\documentclass{article}
\usepackage{graphicx} 
\usepackage{arxiv}
\usepackage{url} 
\usepackage{natbib}
    \setcitestyle{citesep={;}}  
\usepackage{amsmath} 
\usepackage{tikz} 
\usepackage{xurl} 
\usepackage{tocloft}
\usepackage{titlesec} 
\titleformat{\section}{\normalfont\Large\bfseries}{\thesection}{1em}{}
\titleformat{\subsection}{\normalfont\large\bfseries}{\thesubsection}{1em}{}
\titleformat{\subsubsection}{\normalfont\normalsize\bfseries}{\thesubsubsection}{1em}{}
\titleformat{\paragraph}
{\normalfont\fontsize{11}{13}\selectfont\bfseries}{\theparagraph}{1em}{}
\titlespacing*{\section}{0pt}{1.5ex}{0.5ex}
\titlespacing*{\subsection}{0pt}{1.5ex}{0.5ex}
\titlespacing*{\subsubsection}{0pt}{1.5ex}{0.5ex}
\titlespacing*{\paragraph}{0pt}{1.5ex}{0.5ex}
\newcommand{\problem}{What can go wrong}
\newcommand{\solution}{What can be done}
\newcommand{\goalsetting}{Goal-setting}
\newcommand{\controlloop}{Control loop}
\newcommand{\requisitevariety}{Requisite variety}
\newcommand{\goalalignment}{Goal alignment}

\title{Reframing AI loss of control:\\What control is, how to have it, how to lose it}

\author{
  \textbf{Ze~Shen~Chin}\thanks{Corresponding author} \\
  Oxford Martin AI Governance Initiative \\
  AI Standards Lab \\
  \texttt{chinzeshen@gmail.com}
  \\[1.5em]
  \textbf{Maurice~Chiodo} \\
  Centre for the Study of Existential Risk, University of Cambridge \\
  \texttt{mcc56@cam.ac.uk}
  \\[1.5em]
  \textbf{Dennis~M\"uller} \\
  Institute of Mathematics Education, University of Cologne \\
  \texttt{dennis.mueller@uni-koeln.de}
  \\[1.5em]
  \textbf{Coleman~Snell} \\
  Cornell University \\
  AI: Futures and Responsibility \\
  \texttt{cjs386@cornell.edu}
}

\renewcommand{\shorttitle}{} 

\begin{document}

\maketitle

\vspace{1em}
\begin{abstract}

At present, loss of control risks have gained much prominence in public discussion, particularly in relation to AI, with extensive discourse present among academics, frontier labs, and even governments. However, in the existing literature, the concept seems to rest on surprisingly weak foundations, where even those that discuss loss of control extensively do not first establish what control is and what exactly is being lost. Our paper aims to address these gaps. We establish a working definition of control by anchoring it to the “setting and getting of goals”. Then, we discuss various aspects of control, built on foundational concepts from related fields like cybernetics, management control, and control theory. This includes who (or what) can be in control, and the things they require to be in control, such as the ability to set goals, having a functional control loop, having requisite variety, and having sufficient goal alignment. Once a framework for control is established, we then discuss how control can be lost, how AIs can contribute to such loss of control, and offer relevant recommendations for how one can maintain control. One interesting consequence of our work is that humanity, as individuals and as groups, can lose varying degrees of control as a result of AI behaviour that is far below the level of superintelligence; the potential for loss of control scenarios (as we define them) already exist, and have existed for a long time.  

\vspace{1em}
\keywords{AI safety \and control framework \and loss of control \and goal setting \and control loop \and requisite variety \and goal alignment \and resilience}

\end{abstract}

\newpage
\tableofcontents
\newpage

\section{Introduction}
\label{sec:introduction}

\subsection{Loss of control - a primer}

What does it mean to be “in control”? Much has been written about humans and humanity “losing control”, in particular with reference to superintelligent artificial intelligence (AI). But what is it that is actually lost here? Does the ability to dictate what \textit{other things do} form the essence of human control? Or actually, do we care more about achieving \textit{what we want to do}? For some, it may be natural to think of control as exerting influence over other things, be it other humans, the natural world, or even AIs. And others may argue that this sort of “control” is sometimes impossible, especially regarding advanced AIs. But even if it were possible, would it actually be enough for us to get the things we need and want?

And so, we start to see that “control” may well be a more inward-looking concept. Yes, having some influence over the things around us certainly helps us get what we want, as individuals, or as society as a whole; this is, after all, the original motivation for farming. But a farmer who cannot cook will remain hungry, even if he has a great harvest. So we posit that control is, first and foremost, a matter of how effectively we can do the things we want, such as eat food, have shelter, interact, and enjoy our lives.

With this inward-looking perspective in mind, we bring forward an interpretation of being “in control” as the ability to \textit{set and get goals}. This interpretation, which we refine later, enables us to capture the notions of intent (goal setting), and interaction (goal getting), without having to rely on the notion of control \textit{of} or \textit{over} something else. To us, control means being able to set plausible goals, and then carry out some sort of process to reliably achieve those. And both of these are important; the ability to \textit{choose} what to do is just as relevant as the ability to \textit{do it}, if one is to be “in control”.

So why do we take this as our definition? Why do we not simply appeal to existing theories in engineering about control? The answer is simple: not everything in the world is an engineered system. Notions of physical control might work very well in certain settings, such as a nuclear power plant. But control can also exist elsewhere; a political party can be “in control”, as can a school teacher. Neither of these has access to the physical interventions that are required in such definitions, and yet governments and teachers can still be very much in control. So when it comes to social or other non-engineered systems, where control is still very important, one needs to broaden out from such engineering definitions in order to continue discussing the notion of being “in control”.

But why go to such lengths to try and crystallise such a general definition of control? At present, loss of control risks have gained prominence, particularly in relation to AI, with extensive discourse present among academics, frontier labs, and even governments. However, in the existing literature, the concept seems to rest on surprisingly weak foundations, where even those that discuss loss of control extensively do not first establish what control \textit{is} and what exactly is being \textit{lost}. Our concern was that these important discussions might be trying to solve an undefined problem, and it might not be productive to engage in discussions about risks to humanity with only an intuitive framing of them. Given how pressing such “AI loss of control” risks are, we felt it important to put such discussions on a more solid footing, to enable people to explore not just the risks present, but also the causal processes that might lead to such risks manifesting.

And so, our paper aims to address this gap. We establish a working definition of control by anchoring it to the setting and getting of goals. Then, we discuss various aspects of control, built on foundational concepts of control from related fields like cybernetics, management control, and control theory. This includes who (or what) can be in control, and the things they require to be in control. Once a framework for control is established, we then discuss how control can be lost, how AIs can contribute to such loss of control, and offer relevant recommendations for how humanity can maintain control. One interesting consequence of our work is that humanity, as individuals and as groups, can lose varying degrees of control from AI behaviour that is far below the level of superintelligence; the potential for loss of control scenarios (as we define them) already exist, and have existed for a long time. 

This brings us to our final observation, which is that “control” is not a binary state, and breaks up along at least two axes. Firstly, we observe that control can be had at \textit{different levels}, and a loss of control at a lower, less crucial level need not mean that control at higher levels is necessarily lost as well. And secondly, we observe that control is relative, and that for a given situation, one individual or group of people might be considered to be “in control”, but another may well have “lost control”. Thus, the notion of control is always relative to at least two aspects: \textit{who} is in control, and at \textit{what level}.

\newpage
\subsection{Loss of control in current discourse}

The problem of “loss of control” has been heavily explored and debated in the literature, particularly in the context of AI. Many have argued that the advancement of  AI poses catastrophic or existential risks to humanity \citep{bengio_ai_2023, bucknall_current_2022, bales_artificial_2024}. Some common categories of risks include \textit{misuse risks}, \textit{misalignment risks}, and \textit{structural risks}, where misalignment risks is also closely related to what is commonly known as “loss of control” risks \citep{shah_approach_2025}. Broadly speaking, discourse about risks from “AI loss of control” typically pertain to takeover scenarios by sufficiently capable and misaligned AIs \citep{chan_loss_2021, chin_dimensional_2025}.

The notion of “AI loss of control” is frequently considered in terms of losing control of the AI itself by way of being unable to constrain its actions, rather than losing control of what we as humanity ultimately care about (such as the reliable supply of food). For example, citing \citet{greenblatt_ai_2024}, \citet[p.~1]{somani_strengthening_2025} defines AI loss of control scenarios as “situations where human oversight fails to adequately constrain an autonomous, general-purpose AI, leading to unintended and potentially catastrophic consequences”. Several other sources describe or define AI loss of control similarly, in terms of losing human control \citep{idais_idais-shanghai_2025, cass-beggs_framework_2024,yampolskiy_controllability_2020} or losing human oversight \citep{bommasani_california_2025} over AI systems, having AI systems operate outside of human control \citep{ISRSAA2026}, having AI systems actively undermining human control \citep{ostridge_potential_2026}, and “scheming” \citep{centre_for_long-term_resilience_loss_2026}. Moreover, some argue against the notion of loss of control, by saying we never had control, and hence we cannot lose what we did not have in the first place \citep{lee_are_2021}.

We believe such perspectives provide an essential analysis of the problem of “AI loss of control”. However, we also believe that there is scope to expand on such perspectives, which we aim to do in this paper. We wish to set up an interpretation of “loss of control” that incorporates goals, concerns, and desires at different levels: what people care about ranges from the food supply chain, to the value of human interactions and friendships, to the quality of products and services. We will further argue that, in the context of AI, humans or humanity, “losing control” is not predicated on AI “gaining control”; human loss of control can occur through sufficient AI disruption alone.

But of course, to commence any discussion about losing control, one needs some notion describing what it is. Existing work that discusses loss of control in the context of AI often does not explicitly address or define what control \textit{is}, what it means to \textit{be in control}, and what exactly is \textit{lost} in “loss of control” scenarios. And so this is another area where we attempt to broaden the discourse, by providing a full, and moreover operationalised, description of what it means to be in control. Loosely speaking, we define control as “the ability to set plausibly attainable goals, and reliably achieve those goals”, or, more succinctly, the ability to \textit{set and get goals}. An important distinction here is that, not only do we cover the ability for one to get what one wants; we also incorporate the act of deciding what one wants in the first place into our interpretation of control. And so, our description helps clarify the often-blurred distinction between losing control of AI systems and \textit{losing control of our ability to influence the world in desired ways}, where the former is often implied to lead to the latter, even though it is not always clear how this would happen concretely.

Quite often in the literature, AI loss of control is described in close relation to AI misalignment (though, as we will discuss later, there are indeed many other ways AI may lead to humans losing control). For example, in \textit{Human Compatible: AI and the Problem of Control}, \citet{russell_human_2019} argues that the central challenge of advanced AI is not intelligence itself but \textit{ensuring alignment} between machine objectives and human values. He frames the problem of control as the difficulty of ensuring that highly capable AI systems reliably act in accordance with human intentions and interests. \citet{stix_loss_2025} discuss the \textit{absence of an actionable definition} of loss of control (an issue we attempt to address with our formulation of what it means to be “in control”), and argue that it is best to understand loss of control based on two axes: severity, and persistence. They also analyse loss of control dynamics from extrinsic factors such as deployment context, affordances, and permissions. And they propose governance and technical interventions to prepare humanity for such scenarios. Nevertheless, while these works contain discussions on the process and implications of losing control, as well as propose mitigations to prevent loss of control, they do not provide a definition of control itself; a shortcoming that we attempt to address.

\citet{kulveit_gradual_2025} discuss gradual loss of control of human civilisation from incremental AI disruption, where humans are “unable to meaningfully command resources or influence outcomes” \citep[p.~2]{kulveit_gradual_2025}. They discuss how the economy, culture, and states could independently lose alignment with human preferences due to incentive mechanisms, leading to varying levels of disempowerment (relative and absolute). They also propose risk mitigation measures such as estimating human disempowerment, preventing excessive AI influence, and strengthening human influence. Similarly, \citet{bales2024aitakeover} discusses human disempowerment through AI takeover. They discuss three mechanisms of human disempowerment: domination, where AI systems actively exercise power over humans; incapacitation, where AI systems undermine people’s power to acquire desirable things; and disenfranchisement, where AI systems play the key role in determining the future of humanity. While both of these works do not explicitly frame their scenarios as “AI loss of control”, their ideas about human disempowerment or incapacitation is similar to what we refer to as “loss of control”, where interactions with AI systems may lead to humans losing control over what is desired. We expand upon these ideas by investigating in more detail the mechanisms of which human control can be had, as well as how they can be lost.

\citet{nyholm_new_2022} takes a very different approach, challenging the prevailing notion of the control problem in AI where having control is deemed to be good while losing control is deemed to be bad. They discuss the concept of control as being multidimensional, e.g., it can be direct or indirect, robust or unrobust. They then frame being in control as not necessarily a positive thing: self-control is usually good, whereas control over others (i.e., slavery) is not. The implications are that there may be situations where certain forms of control over AIs may not be good. In response, \citet{vacek_two_2023} suggests that limited yet considerable control over AIs is desirable. Along these lines, our work here does not require humans to have control over AI systems; we instead provide a framework showing how humanity might retain control (of the things it cares about) even if it does not have direct control over AI systems.

Beyond theorising about how AI loss of control may happen, and whether it is desirable, some are addressing it as a risk that can be assessed, monitored, and managed. \citet{tkeshelashvili_ai_2026} proposes seven indicators including scheming, manipulation, deception, self-preservation, unauthorised resource acquisition, goal misgeneralisation, and model and behaviour drift. Several frontier AI companies have also incorporated such concerns into their safety policies. For example, \citet{meta_advanced_2026} addresses the risk of loss of control by focusing on failure mechanisms involving their ability to understand and oversee AI systems, and provides examples of enabling capabilities such as autonomous AI research and development (R\&D), evaluation awareness, monitor awareness, and selective monitor-adaptive behaviour. In addition, \citet{google_deepmind_frontier_2025} discusses misalignment risks where models have the potential to undermine human control. \citet{naver_navers_2024} examines risks in the “loss of control” category by conducting periodic reviews or assessments on model capabilities whenever major performance improvements are made. And \citet{xai_xai_2025} discusses loss of control as one of the two major categories of AI risk (the other being malicious use), where they measure concerning model propensities that hypothetically might exacerbate loss of control risks, such as deception and sycophancy.

And so, we see that the current discourse surrounding AI loss of control, and its effects on human control, is both diverse and divided. Our work does not seek to unify or replace any of these schools of thought. Rather, we have endeavoured to provide tooling and reasoning that can be used in any, or all, of these discourses. Our aim has been to provide further methods for explaining causal mechanisms that might lead to any of the loss of control concerns articulated above. By inverting the focus from what \textit{AI can do}, to what \textit{humans want to keep doing}, we hope to expand the discussion from \textit{what} might go wrong, to \textit{how} things might go wrong.

\subsection{Our key ideas}
Our paper begins with Section \ref{sec:defining_control} where, after examining previous concepts of control, we give an overview of our definition of control as:

\begin{center}
\textit{
Control is the ability to \textbf{set plausibly attainable goals} that are not a foregone conclusion,\\
and \textbf{reliably achieve those goals},\\
where a goal is either a desired world state or a non-empty subset of desired world states.}
\end{center}

We then differentiate between that which sets goals (which we call an \textit{entity}), and that which the entity deploys to achieve goals (which we call a \textit{system}); these can have any amount of overlap. Entities can be a “single mind” (e.g., an individual person), or they can be collectives of minds, perhaps working in a hierarchy, or perhaps with some other collectivised way of deciding on goals. In short: any biological, cyber-physical or social actor (or any combination thereof) who has the ability to set goals. And these may contain \textit{sub-entities}, capable of setting their own goals. We then explore the notion of goals as \textit{desires for certain world-states}, and show how goals may actually be vague or ill-defined, or not even explicitly articulated at all, and that they may contain \textit{sub-goals} within them, whose purpose is to contribute to the achievement of the overarching goal. An entity then needs a method to reliably achieve its goals, which we call a \textit{system}, giving the entity a way to interact with and manipulate the environment around it. These need not be designed or built from scratch, and nor do they always operate exactly as intended. And just like entities and goals, systems can also have \textit{subsystems} which contribute to their operation.

Section \ref{sec:how_entity_in_control} then explores how an entity can be in control. We do this by introducing our \textit{four aspects of control} which we posit are needed for an entity to be “in control”. First, the entity must be able to continually \textit{set}, and also \textit{re-set}, plausible goals; they are not all set “at the start”, nor are they static once set. Here we introduce the idea of a \textit{goal hierarchy}, encompassing the notions of \textit{higher-level} and \textit{lower-level} goals, and how the latter act as sub-goals in aid of the former. Second, the system must have a functioning \textit{control loop}, which we argue consists of three core functions: \textit{sensing} (of relevant parts of the surrounding environment), \textit{decision-making} (of what the system should do next), and \textit{intervention} (of how the system should act in its environment). These (should) operate in a continual loop until the system’s goal(s) are achieved. Third, the system must have  \textit{requisite variety}; that is, sufficient capacity to deal with parts of the environment that “get in the way” of it carrying out its goals (often termed \textit{disturbances}). This includes how much physical power the system has to intervene, but extends to other considerations such as how often it can intervene, and whether its decision-making will actually make use of the full “power” it has. And fourth, the goals of any subsystems must be \textit{sufficiently aligned} to higher-level goals of the system (vertical alignment), or the goals of other subsystems (horizontal alignment), whenever these systems overlap. This is of course critical when the subsystems are controlled by (independent) sub-entities, but is just as important when the one entity is running all the subsystems.

Having defined what control \textit{is}, we then move on in Section 4 to cover how control can be \textit{lost}. We do this by “inverting” each of our four aspects of control, to see how they might fail. The first of these is \textit{the inability to set or re-set goals}. Here, we differentiate between three different categories of failure for goal-setting: a complete \textit{inability} to set any goals (such as the entity ceasing to exist, or unable to “come together” to choose goals), a \textit{partial loss of ability} to set goals (such as entering a reinforcing loop of unobtainable goals, setting self-defeating goals, or losing the ability to set goals in one domain), and an \textit{inability to change goals} (such as failure propagation from an inability to re-set operational goals, or simply “giving up” at the first failed goal, or replacing productive goals with unproductive ones). Next is failure of any \textit{control loop} functions: Sensing can fail due to factors such as \textit{distortions and bias}, \textit{insufficient granularity}, \textit{lagging indicators}, or even just \textit{total sensor failure}. Decision making can fail due to factors such as \textit{decision paralysis}, \textit{rigidity}, or \textit{corrupted processing}, as well as \textit{feedback loops}. And intervention can fail not only from a \textit{complete breakdown} of action, but also due to \textit{desynchronised actions} (those at the wrong time) or \textit{ineffective placement}, among others. Next, we have failure caused by \textit{insufficient variety}. This can be via a lack of measurable \textit{capacity} (i.e., some form of “force”), but also via a lacking in \textit{time and space} (due to insufficient “reaction time” or “reach”) such as \textit{design or architectural constraints} (on how the system can “rearrange” itself), limitations in \textit{variety of thought} (on which decisions are “acceptable”), \textit{mismatches in discrete and analogue mechanisms} (where discrete systems meet a continuous world), and finally \textit{combinatorial explosions} (being overwhelmed with too many small disturbances at once). And finally, we have \textit{goal misalignment}. This can manifest as \textit{vertical misalignment} (either within the same, or between different entities), as well as \textit{horizontal misalignment} of two sub-entities working towards the same higher-level goal. Goals can also suffer from \textit{temporal misalignment} (i.e., short-term vs long-term ones). Such misalignment can be caused by \textit{articulation failures} (with invisibility preventing alignment in advance) and \textit{goal drift} (with aligned goals drift apart over time). Misalignment can also occur via \textit{misaligned proxies} to the original goal, or system setups inadvertently \textit{working towards the wrong goal}.

Taking the Organisation for Economic Co-operation and Development (OECD) definition of AI, Section \ref{sec:how_ai_affect_control} then focuses specifically on how AI systems can induce loss of (human-based) control, first observing that as well as being systems, more autonomous (e.g., agentic) AIs can also be entities. Again, we break this down via our four aspects of control. First, AI might disrupt the goal (re)setting of human entities. It could prevent goal-setting entirely, from entity-destruction (Lethal Autonomous Weapons killing combatants) to entity disruption (AI agents working on AWS codebases inadvertently shutting them down, disrupting communications globally). It can partially disrupt goal-setting, from leading entities to set unreasonable goals (AI inducing psychosis in individuals), to lulling entities into over-estimating their system capabilities (AMD being severely let down by how capable it thought Claude coding was). And it can disrupt goal re-setting, from enticing entities to replace productive goals with unproductive ones (AI recommender systems radicalising social media users), to re-setting strategic goals itself (Resort-to-force decision making relying on AI recommendations). AI can also affect the control loop of a system. It can induce sensing failure, from (poorly) replacing existing sensing systems (biased facial recognition systems in law enforcement), to eroding human sensing (AI curation of online content affecting how users form worldviews). It can affect decision making, from eroding decision-making abilities (atrophy of human decision-making skills after prolonged AI use), to detrimental displacement of decision making (AI being used to make (biased) judicial decisions). And it can affect human control at the intervention stage, from being used to carry out harmful interventions (being used to execute cyber attacks on critical infrastructure), to doing unwanted interventions (AI agents performing unintended actions). AI systems can also affect human variety. They can reduce the variety of the system, from reducing decision-making capabilities (the “de-skilling” of humans after prolonged AI assistance), to reducing intervention capabilities (organisations reducing their workforce in the (false) assumption of AI replacement). They can also increase the variety of the environment, from creating too much input to properly sense (the over-stimulation of humans from AI content online), to creating too much physical disturbance (courts being swamped with AI-generated fake evidence, and research journals being swamped with fake submissions). Thus, such “increased variety” can lead to human loss-of-control even in the absence of the AI “gaining control”. And AI can create instances of goal misalignment for humans. This happens with AI systems having misaligned goals with the entities deploying them, from unwanted discrimination (cases of AI wrongly flagging benefits applicants in the Netherlands for fraud), to actions misaligned with other specified goals (Air Canada chatbot issuing discounts against company policy). And the “AI industry” itself can be misaligned with human goals, from instances of exploitation (AI companies using workers from the global south to label “toxic” data), to misalignment with societal goals (the vast ecological impact of AI development going against sustainability policies).

Section \ref{sec:how_maintain_control} then looks at the different \textit{scales of control} of human-based entities, and how AI might affect each of these; “human control” is not a binary state. First we look at maintaining control on the \textit{individual (human)} scale, examining how AI might for example affect the ability for individuals to be in control of their lives. Next, we look at \textit{coordinated groups} of humans, such as corporations, governments, and other organisations, examining how AI might compromise how these groups set and get goals. And lastly, we look at control at the scale of \textit{humanity}, examining how AI might gradually erode humanity's control. For each of these three scales and each of our four aspects of control, we present a real-life scenario of how control has been, or could be, lost there, and propose possible mitigations that may contribute to the prevention of such loss-of-control scenarios.

We end the paper in Section \ref{sec:further_work_conclusion} with some suggestions for further work, suggesting further research on how AI might affect control in the “big domains” (technology, politics, law, state finances, markets, science, and morals) where control might be most important for humanity, as well as potential extensions of our framework to cover game-theoretic situations where multi-agent dynamics become an important factor. We conclude with our central thesis that (human) control is about \textit{setting and getting} of goals, and that one needs to always consider \textit{who} is in control (or not), and at \textit{what level}. And we highlight that, while existing discourse on AI loss of control largely relates to hypothetical AI systems capable of completely undermining human oversight of them, our approach focuses on strengthening control of the \textit{human systems} necessary for our survival, happiness, and flourishing; what matters is not so much what the AIs can do, but rather, what humans \textit{cannot}.

\section{Defining Control}
\label{sec:defining_control}

\subsection{Previous concepts of control}

We begin by showing how the notion of \textit{control} is interpreted differently depending on the context, and that the very existence of control is not always straightforward or given. We illustrate these through several domains where “control” has been studied in the past, such as the military, management, and psychology. Then, we present perspectives from cybernetics and control systems, relating to how (and indeed if) these domains view and carry out “control”. 

In the \textit{military domain}, control is often used to represent hierarchical decision-making authority over the actions of a set of people in an organisation. For example, \citet[p.~2]{niven_anatomy_2023} defines control as “the authority exercised by a commander over part of the activities of subordinate organisations, or other organisations not normally under his command, that encompasses the responsibility for implementing orders or directives.” In the context of \textit{management}, \citet[p.~4]{anthony_management_2011} define it as “the process by which managers at all levels ensure that the people they supervise implement their intended strategies.” And in the field of \textit{psychology}, \citet{skinner_guide_1996} describes an “integrated framework” building on two fundamental distinctions of control: the first distinction is between objective, subjective, and experiences of control; while the second distinguishes between agents, means, and ends of control.

Many of the concepts of control (implicitly or explicitly) trace back to the field of \textit{cybernetics} and (technical) control theory. This lineage is particularly strong in the military and management. Here, the founding father of cybernetics, \citet{wiener_cybernetics_1948} initially conceptualised control as the use of communication and feedback to regulate the behaviour of a system towards a goal or purpose. And more recently, in the field of \textit{control systems} engineering, \citet[p.~2]{nise_control_2015} explained that “a control system consists of subsystems and processes (or plants) assembled for the purpose of obtaining a desired output with desired performance, given a specified input.” Similarly, \citet[p.~3]{doyle_feedback_1990} state that “the objective in a control system is to make some output, say y, behave in a desired way by manipulating some input, say u”. 

Overall, concepts of control with a strong lineage to cybernetics are very technical and rigid, and while they certainly have their use in similarly technical and rigid contexts (such as for physical or engineered systems), in Section \ref{sec:our_interpretation} we provide a broader definition. Analysing how AI affects control requires a definition that can capture its technical, socio-political, and economic role in and effects on society. 

The \textit{philosophical concept of control}, and whether control can even exist at all, is surprisingly subtle, and is itself up for debate in various circles. To understand this, it is important to consider that the very idea of control rests on certain ontological and epistemological assumptions about how people and social entities behave. Hence, competing perspectives have in particular been developed in the social sciences. There, the debates are not settled and new perspectives continue to emerge. As an exemplary case we highlight one recent perspective: building on the works of Luhmann and other social system theorists, \citet{watson_autopoietic_2025} present a description of the world which tries to move the “image of interaction from control to coordination, from influence to resonance” \citep[p.~9]{watson_autopoietic_2025}. Their ontological and epistemological assumptions deny the very existence of control in many instances, where our work will advocate for its existence. Writing, for instance, about control in politics, they note that “[p]olitical systems, too, defy managerial control. They do not function like machines, but more like turbulent ecologies—historically sedimented, internally recursive, and prone to sudden phase-shifts. The same applies to public discourse: once governed by broadcast logics, it now unfolds through algorithmic amplification, memetic recursions, and emotionally charged resonance loops.” \citep[p.~10]{watson_autopoietic_2025}. Watson and Brezovec argue that no system can be instructed or commanded from the outside; external agents can only “perturb” a system, never control it. As such, they never provide a (mechanistic, or other) definition of control; rather they speak of it as a “fantasy [that] is collapsing” \citep[p.~11]{watson_autopoietic_2025}. Given their ontological and epistemological assumptions, they could not have done differently: For them, systems are self-reproducing (autopoietic), rather than input-output machines, and the theory of autopoietic systems thus necessarily limits the influence of external inputs. 

It is unsurprising that the messy socio-political reality defies (rigid) managerial control practices, but, as we will see as our argument progresses, this does not imply that it necessarily defies a broader perspective of control. Thus, while a position like Watson and Brezovec have developed offers a valuable critique of command-and-control fantasies, their perspective is analytically insufficient for our purpose, though not for the reasons one might initially expect. \citet[p.~11]{watson_autopoietic_2025} propose that the appropriate mode of engagement with complex systems is not control, but rather “attunement”, intervening in ways that are recursively viable within a system’s own logic. Yet the mechanisms they describe as the alternatives to control (i.e., observing how a system operates, shaping conditions rather than issuing commands, perturbing rather than dictating) are structurally very close to aspects of what we will later describe as control, particularly when we argue that an entity can indirectly exert control over sub-entities by setting suitable constraints and incentives rather than by directly manipulating them. 

The disagreement is, in a sense, one of framing: given similar mechanisms, perspectives such as those found in \citet{watson_autopoietic_2025} emphasise the limits of what external agents can achieve, while we wish to emphasise the capacity (to manipulate surroundings, achieve goals, etc.) that remains. This change in perspective is crucial as it is not just important to capture how people or institutions can lose control to AI, but, for those affected, it may be even more important to analyse how they can maintain the control they (still) possess or how they can, at least partially, regain what they lost. If one understands how a system observes its environment and intervenes in ways that reliably lead towards desired outcomes, then one is, in the sense we will soon define, “in control”: imperfectly, but meaningfully. Without such a distinction, we lose the vocabulary to differentiate between situations where an entity can effectively steer a system towards its goals, and situations where it cannot, and it is precisely this shift that we wish to describe when we speak of AI-induced loss of control. Hence, we cannot begin with a pre-existing notion of control which emerged in a specific context if we want to speak about more general loss of control; maintaining control over the economy of a nation is vastly different to, say, controlling a petrochemical plant. As such, we must first find a new, broader definition to work with.

\subsection{Our interpretation and definition of being “in control”}
\label{sec:our_interpretation}

Many perspectives of control are tailored towards a specific technical, socio-political, or economic areas. These close ties between a definition and its area of application can be deeply restricting, hindering their application to AI, which is increasingly used in all areas of life. And a common, yet occasionally implicit, theme among these existing definitions of control is the presence of a process or a system to achieve desired outcomes, but at the same time a lack of connection to why such outcomes are desired. With these in mind, we put forward the following contextually open definition of control:

\begin{center}
\textit{Control is the ability to \textbf{set plausibly attainable goals} that are not a foregone conclusion,\\
and \textbf{reliably achieve those goals},\\
where a goal is either a desired world state or a non-empty subset of desired world states.}
\end{center}

In other words, being in control means being able to “set and get goals”. Within our definition are several conditions that are required in order to be “in control”, which we now seek to explain in more detail.

First, the set goals must be plausibly and reliably achievable. For example, any goals that violate the laws of physics are not plausibly achievable, and any goals that violate the laws of probability cannot be reliably achieved: a goal that involves faster-than-light travel cannot plausibly be achieved given its physical impossibility, and a goal of winning the lottery without cheating cannot be reliably achieved despite being physically possible. However, it is not only the laws of physics and probability which may prevent a goal from being (reliably) achievable. In social, political, and economic (SPE) settings, path dependencies and other constraints can act as limiting factors. There, what matters more is who you are, what resources are available to you, and how the goal relates to you. While the laws of physics do prevent everyone from reliably achieving certain goals, the situation is much more complex for SPE contexts. 

In addition, the goals should not be automatically achieved without the existence of any action. For instance, while a goal of making sure the sun reliably rises every day for the foreseeable future is certainly a reliably achievable goal, it does not require any intervention, as it is overwhelmingly likely that the sun will continue to rise regardless of any reasonable action taken. A goal that is reliably achieved through inaction is, for our argument, considered a foregone conclusion. Thus, it would not be fair to say that humanity is in control of the rising and setting of the sun.

With regards to the notions of “plausible” and “reliable”, we do not explicitly or universally define what constitutes plausible attainability and reliable achievement of goals. Both are highly context-dependent, and thus we leave it up to those who set the goal and to the contexts in which these goals are being set. To illustrate this context-dependency, throughout most of human history, sending a human to the moon was certainly not a plausibly attainable goal — yet by the middle of the twentieth century it had become very plausible. Nevertheless, we note here that the notions of “plausible” and “reliable” can be described quantitatively or qualitatively, or as a combination thereof. For example, computer systems engineers often set quantitative goals for the uptime of equipment; if a piece of equipment is supposed to have a 95\% uptime, it means it should be within specific operating conditions 95\% of the time. On the other hand, a qualitative goal can be set for one to arrive at a destination, bar extraordinary circumstances. Many complex goals require a combination of quantitative and qualitative aspects, in particular, when these goals are about the functioning socio-technical systems in SPE contexts; what does it mean (quantitatively or qualitatively) to “produce good art” or “create policies that voters like”? While the technical part often can be quantified, the social aspect can often only be captured qualitatively. Regardless, any notion of reliability must not be achievable by random chance alone: being in control is an active state, and it requires someone or something to be performing actions.

\subsection{Who, or what, can be “in control”}

Intuitively, one may wish to speak of control as a scenario where “someone controls something”. Using those words hints at an asymmetric binary relationship between a “control-er” (trying to have control) and a “control-ee” (the thing it seeks to control). Such language, however, hides the act of setting goals and how those goals can actually be achieved. Furthermore, it presupposes an asymmetry that may not actually exist when control is sufficiently differentiated from other concepts, such as power and influence.

Thus, we will first need to differentiate between that which sets a goal (which we call an \textit{entity}), and that which the entity deploys to achieve it (which we call a \textit{system}). And so an entity sets the goal, and then uses the system to get the goal. These need not be disjoint; indeed, we will eventually argue that any combination is possible; an entity can be completely separate from the system, an entity can be the system itself, and any part of an entity can be any part of a system or vice-versa. 

\subsubsection{What are entities}

We first look at what it means to be an entity. Intuitively, one might simply consider this to be an individual person. And indeed, understanding if and how an individual person is in control can already be a difficult concept. But as we aim to analyse AI’s effects on control, our work is more broad than this, and cannot be limited to only considering when and how one person is “in control”. We must go beyond that, and consider groups of people such as colleagues or sports teams, legal and otherwise unified entities such as companies, and even disparate groups such as research communities, as being “in control”. We will refer to these as the \textit{entity in control}, and thus talk about how this entity (be it one, or a collection of many, “minds”) can set and achieve goals. 

Just as the entity itself that is “in control” (i.e., can set and get goals) need not be a single mind, it also need not have a single and all powerful decision-making hierarchy. For instance, a larger corporation typically sets and achieves goals using a complex hierarchical decision-making structure, that while having a CEO and board at its top, requires careful orchestration among its employees and subdivisions to be profitable. How such a careful orchestration might be accomplished can, for example, be seen in Drucker’s management theory through objectives \citep{drucker_practice_2012}, which not only puts decision-making structures at its centre, but also focuses on the idea of goal setting and how to achieve these productively. However, in addition to this, a more disparate “community” (such as a research community) can also set and achieve goals, even if the individual participants cannot all directly interact with each other, or if the decisions do not all pass through some central node of approval or operate in some set hierarchy. By forming a community with mutually shared ideals, and continuously re-negotiating its internal rules and external boundaries in implicit and explicit ways \citep[cf.][]{gieryn_cultural_1999} such an entity can still set and achieve goals; many research communities span across multiple continents, and are still able to produce academic output. A prominent example is the collective of mathematicians which published under the name of Nicolas Bourbaki: originating in France in the mid 1930s, the collective (rather explicitly) gave itself the goal of rigorously and abstractly formulating modern mathematics. \citet[p.~2]{mashaal_bourbaki_2006} notes that their “vision of mathematics” brought with it a “profound reorganisation and clarification of its components, lucid terminology and notation, and a distinctive style”. While the collective regularly met and together critiqued all drafts, one could say that their shared ideal of what mathematics should be held the collective together, and enabled them to set (and get) their goals.

In some cases, entities may not even be easily defined. While a formal organisation would typically have a well-defined and visible boundary (e.g., with registered members or formal employees); a research community may not always be well-defined and clearly delineated from others, since, for example, there may be no formal or explicit process of determining membership. This is strongly pronounced for large and fuzzy entities such as the field of mathematics, which likely includes a mathematics professor, but may or may not include retired mathematicians, those who have administrative roles in a faculty, or those who do mathematical work but are formally employed in other departments \citep{muller_hippocratic_2022}. And in addition, entities may themselves contain \textit{sub-entities}, which can also set their own goals. For example, an organisation can be an entity that pursues certain goals, but is itself made up of multiple (sub)-entities of different levels: a department of hundreds of members, a small team of several people, and even an individual in the organisation, can be sub-entities who together form the larger entity of the organisation. All of these entities and sub-entities would have their own goals, as we will discuss later.

In short, we refer to entities as any biological, cyber-physical or social actor (or any combination thereof) who has the ability to set goals.\footnote{We explicitly leave open the material nature of an entity to be able to capture different trajectories for how the research and development of AI progresses, and to enable different philosophical outlooks on who or what is able to set goals in the first place.} In other words, they can be said to have “desires” with respect to certain world-states. We treat this as a functional definition, where for example corporations can be said to “want” things like maximising shareholder value, even though they may not possess consciousness or qualia typically associated with desires. As such, where appropriate, we may also treat AIs as entities, as long as they display sufficient goal-directed behaviours \citep{nahum_motivation_2026}, without any need to establish the presence of consciousness in AIs. 

\subsubsection{What are goals}
\label{sec:what_are_goals}

For simple entities in uncomplicated situations, these “desires for certain world-states” (that is, \textit{goals}) may be interpretable in a straightforward fashion (such as an individual’s desire to have food during the day). However, assuming that a straightforward interpretation is always possible overly simplifies how, and if, goals and their progress can be recognised, quantified, articulated, or measured. Indeed, most entities or situations are more convoluted, and ignoring such complexity hinders one’s understanding of how it sets goals. For these convoluted entities (such as those formed as a large collective) or in more complex situations (such as in the period before a political election), the setting of goals happens through negotiation and communication processes. Through these, entities can exhibit choice and express their desires to set goals of their own accord, and can also begin to enact them, as we will discuss in Section \ref{sec:how_entity_in_control}.

One can differentiate between \textit{instrumental} and \textit{terminal} goals. The purpose of instrumental goals is to be a means to an end, while terminal goals are an end in themselves. The value of instrumental goals is thus derived from what they achieve, while terminal goals are seen as intrinsic and independent, and for some even as universal. What counts as an instrumental or terminal goal can differ between one’s philosophical starting points: for a hedonist philosopher, happiness may well be a terminal goal, while a deontologist may argue that the highest good is acting in accordance with a moral law. In applied mathematics, such as the rational choice theory developed by economists or in the control theory developed in cybernetics, an instrumental (and hence functional) perspective on goals is often employed. Here, goals are often viewed through utility functions and agents that optimise such a function. In this paper we use goals in a manner closer to this latter framing where they can be said to be more ‘operational’: they can be set, re-set, and can change as a response to their environment and also based on changes in their own capacities. 

The goals an entity sets might be clearly defined in a quantitative way, e.g., a person may have a goal of continually maintaining the temperature inside an office building to within +/- 2°C of 20°C, in order to keep it comfortable for human habitation. But goals can also be qualitative. For example, a small business might have the goal of remaining solvent, an artist might have the goal of making a “good” piece of art, and a research department in a university might have the goal of producing new scientific insight. For such goals, it can be harder to understand if sufficient progress towards them is even being made. For instance, while it can be said that one (of many) goals of a research and development (R\&D) department is to make “progress”, it may not be known in advance what exact end output is sought after; a mathematician might seek to prove a theorem, but not know from the onset \textit{which} theorem this will be. Nevertheless, an entity (such as a research department) may still set relatively ill-defined goals, and create ways to achieve them reliably, even if the measurement of the goal is somewhat subjective, and/or it is not strictly defined in advance. For instance, Google DeepMind describes its long-term vision in the following way:

\begin{quote}
In the coming years, AI — and ultimately artificial general intelligence (AGI) — has the potential to drive one of the greatest transformations in history. We’re a team of scientists, engineers, ethicists and more, working to build the next generation of AI systems safely and responsibly. By solving some of the hardest scientific and engineering challenges of our time, we’re working to create breakthrough technologies that could advance science, transform work, serve diverse communities — and improve billions of people’s lives. \citep{google_deepmind_about_nodate} 
\end{quote}

While the possibility of achieving AGI is not a certainty, and the definition itself of AGI is contested, it can be argued that (what they refer to as) AGI is a plausibly attainable long-term goal, and hence DeepMind can be said to be in control by making meaningful progress towards achieving AGI. 

Not only can goals be vague or ill-defined, they can also lack any explicit definition or articulation; instead being inferred from the behaviour of those who act. Consider, for example, the intentional stance described by \citep{kitcher_intentional_1990}, which shows how one can act rationally to satisfy (implicit) beliefs and desires. Using the field of mathematics as a more concrete example, it can be said that (one) goal of this field is to discover mathematical truths. But this goal was never explicitly set by the field itself. Instead, it is a goal that emerged through the subject’s history and its lived practices, building on the (socially negotiated) goals of individuals, collectives, and institutions throughout time. Yet, by observing the behaviour of the field, one can reasonably infer that it, \textit{as a whole}, has such a goal, and moreover is actively working towards achieving it. While an attribution of such (abstract) goals to an entity may be rather subjective, we posit that it is a sufficient framing for the purpose of understanding how an entity may or may not be in control. But while goals can be vague, ill-defined or attributed, such goals must still be plausibly achievable for the entity to be in control. 

Lastly, just as entities can contain sub-entities, goals can, and typically do, have sub-goals within them, whose purpose is to contribute to the achievement of the overarching goal. One might view these as higher-level goals and lower-level goals, whereby the achievement of lower-level goals would ideally contribute to higher-level ones. In Section \ref{sec:the_relationship} we will explore how sub-goals relate to sub-entities. 

\subsubsection{What are systems}

Based on our definition of control, it becomes necessary for an entity to have some way or method to reliably achieve its goals. We call this a \textit{system}, giving the entity a way to interact with and manipulate the environment around it, i.e., 

\begin{center}
\textit{To be in control, an entity will necessarily need a system that allows it to achieve goals.}
\end{center}

A system is not strictly an “engineer’s tool”. They are not always designed or built from scratch. They do not always operate exactly as intended, if such intentions are even possible to articulate. And they are not always trying to “optimise” some quantifiable value or parameter. Such things can be what a system is and does, but as we shall now explain, the notion of a system extends far beyond these technical concepts. 

In our earlier example of the goal of maintaining the temperature inside an office building within +/- 2°C of 20°C, the entity must have a system in place such as a heating or cooling mechanism coupled with a thermostat, which can accurately measure the building temperature, and whose capacity for heat transfer is sufficient to counteract the temperature of the room’s surroundings. Without such a system, this goal cannot be reliably achieved, as the room will tend towards the outdoor temperature which is often outside of the desired range.

While many of the systems one first thinks of involve physical control mechanisms, the notion of a system also applies to other kinds, be they animate or inanimate, physical or biological, social or economic \citep{beer_decision_2000}. Just as the definition of a system does not force it to be a material one, it also does not require the system to be built from scratch. Indeed, a system can be assembled from individual (pre-existing) components and configured to operate in line with achieving various goals. This is especially true for systems in social, political, and economic contexts, whereby existing people and institutions are \textit{assembled and organised} (rather than built) in a certain way to achieve goals. For example, while in many countries a constitution was drafted at some point during their founding years to lay out the foundations and structure of government, later generations no longer start from scratch. Instead, they face a situation where the governmental structures and departments are already defined and often can only be slightly adjusted, if at all. Positions within these departments would then be filled by people who already exist, and already possess many of their own unique capabilities. This would be done in such a way that they would collectively work towards achieving the goals of the government and the constitution. The government would then also be directly or indirectly influenced by the goals of individual people and of collective entities such as lobbying groups and corporations. Indeed, even engineered systems are not always designed from their low-level components or built from the lowest atomic level. Instead, they are often assembled from existing components and the system’s designer articulates (sub-)goals by means of defining suitable interfaces to enable abstraction and modularisation. This holds for both physical and digital systems, wherein abstraction and modularisation, and thus the reuse of existing components, is generally understood as a mark of quality precisely because it allows clearer development and maintenance because the role (and thus sub-goals) associated with individual components is easier to articulate and understand. But this engineering example illustrates a broader point: building a system requires an articulation of how its subsystems work, the subgoals they need to accomplish, and how they should interact with each other.

Furthermore, systems don’t necessarily \textit{need} to be intentionally designed. Of course they \textit{can} be, such as in our example of the thermostat in the office building which is likely to have been designed and configured deliberately to fulfill certain goals. Likewise, organisations are often explicitly designed with a certain structure, with the roles subsequently filled by personnel. However, with systems that are more complex and less centralised, (e.g., the field of mathematics, which can also be considered as an entity with the goal of producing new mathematics), such a system as a whole may not have been explicitly designed or intentionally assembled; there are, after all, many ways to get a new piece of mathematics. So rather, these systems emerge through the actions and interactions of their components, and over time they develop system-like behaviours where they operate in a way that pursues goals. 

Similar to entities and goals, systems can also have subsystems \citep{lutsenko_principles_2016, anthony_framework_1964}, where functioning systems require their subsystems to not only be pursuing subgoals, but also to have them contribute to achieving the goals of the higher-level system(s). One may be tempted to think of a system as something simple, explicit or physical, but unlike a simple thermostat, many systems do not just have hidden complexity, but are also in complex relationships with other entities, goals, and systems. 

\subsubsection{The relationship between entities, goals, and systems}
\label{sec:the_relationship}

So far, we have denoted the \textit{entity} as what sets the goals, whereas the \textit{system} is what achieves them. However, we now observe that a person or institution need not have just one of the labels “entity” or “system”, but could indeed act as both. The delineation between “entity” and “system” is not always clear-cut. However, such “haziness” is necessary to capture, analyse, and communicate the complexity of control scenarios, and to later enable us to dissect how AI affects control in a meaningful way. Our framework of control still applies, even though its evaluation becomes (unavoidably) more complicated. 

First, we note that an entity (e.g., a school teacher) can be acting inside another, larger system (e.g., a school setup), and hence a system can contain further entities. But an entity (e.g., a country) can also be acting \textit{within} its own system. How? Well, there are entities which can also be considered systems achieving goals set by those from within the system itself. For example, regarding the setting and achieving of top-level goals by a nation state, these may be called “grand strategies” \citep[cf.][]{balzacq_oxford_2021}. Here, a nation state made up of its citizens can be considered the entity that set its “grand strategy”; but the same nation state made up of its people and its resources can be considered the system that pursues its “grand strategy” through various domestic and international activities. More concretely, consider the Apollo programme, (the system) through which the United States (the entity) gave itself the plausibly attainable goal to “put a man on the moon”, but which its execution required (at its height) up to 400,000 workers \citep{hollingham_apollo_2019}. It was such a large-scale national effort that the entity setting and getting the goals (the government) was also part of the system by which these goals were achieved (other parts of government, various other corporations, and a sizeable portion of the US population). 

With the potentially complex relationships between entities and systems in mind, we can now discuss how goals can be set from within a system, or outside of a system, or a combination of the two. In the thermostat scenario, all goals of the system were set by an entity outside of it: the thermostat was built to maintain an indoor area within a specified temperature range set by a human and not the thermostat itself; the thermostat does not “set and get” goals of its own. This is often the case in traditionally engineered systems, where the system is built based on the goals of its designers, and there are no entities operating within the physical system.

However, this may not be the case in other systems, such as those employed to achieve social, political, or economic goals. The case of student-teacher interactions demonstrates that the boundary of an entity setting a goal, and its system achieving the goal, is not always straightforward. Here, we can consider the teacher as an entity setting a goal of making sure their students efficiently and effectively achieve the objectives of the curriculum. The teacher then deploys a system to achieve specific goals inside their classroom, which may include the varied use of individual and group-based presentations, assignments, and assessments. And these may be alongside other mechanisms of communication, including social and socio-disciplinary norms guiding their teaching, and the student’s execution of tasks (cf.\ \citet{meyer_normen_2024} for mathematics education, specifically). But \citet{hattie_visible_2012} was able to show that the efficiency and effectiveness of the system that an individual teacher deploys fundamentally also depends on the person’s unique characteristics (e.g., beliefs, mindframes, and dispositions). So, in a real and direct sense, the teacher (as an entity setting a goal) is also part of the system achieving the goal: the teacher cannot ignore who they are as a person when teaching their students. 

And here an interesting phenomenon starts to arise, whereby the system being used to exert control may contain “smaller” entities that seek or exert control (the teacher’s system contains the teacher). But this entangled relationship becomes clearer when multiple distinct entities are involved. Consider a delivery company with a fleet of trucks and drivers. The company (viewed as an entity), may have goals centered around the cost-effective and timely delivery of parcels. It then uses a system consisting of its fleet of trucks and the drivers it employs, in conjunction with other mechanisms, to reliably achieve these goals. This company is an entity, but so too is each of the truck drivers employed by the company. The company sets (and achieves) goals relating to delivery, including setting the route for the individual drivers. On the other hand, the driver sets (and achieves) goals relating to driving the truck in a given route and to a defined schedule, and how to deliver the individual parcels. The company has a large system consisting of its employees, the deployed management structures, and a host of other tangible or intangible assets, methods, and incentives, while the truck driver has a relatively smaller system consisting, in particular, of a truck and an appropriate journey management plan; where both of these systems are configured to achieve their respective goals. Furthermore, in order for the entity (the company) to reliably achieve its goals (i.e., the timely delivery of parcels), its sub-entities (the truck drivers) must have certain goals that align with these (i.e., the timely arrival of the truck at the destinations along their route, and the correct selection of parcels to be delivered at each stop). 

On the other hand, the design and creation of some subsystems may not be as directly associated with the sub-entity using them. The truck (together with its driver) as a system transporting goods is made up of smaller subsystems such as the transmission system and the brake system. These subsystems can be said to have their associated (sub-)goals, but these goals are not set by the (sub-)entity in question. While the truck driver ought to have the goal of driving a truck with a functioning transmission and brake system, he did not design nor create them; all he can do is check and use them, or report to maintenance if they are no longer functioning. In short, at a higher-level, entities set goals that are achieved by systems; but at a lower-level, subsystems can pursue (sub-)goals that are not explicitly set by a (sub-)entity. 

\section{How can an entity be “in control”?}
\label{sec:how_entity_in_control}

Having laid out a framework of \textit{what} “control” is, and \textit{who} can be “in control”, we now turn our attention to \textit{how} an entity can “be in control”. A key ingredient of our definition of control is that outcomes are not a foregone conclusion. Hence, being in control requires some sort of action(s) by the entity, which we now loosely describe, before going into greater detail later in the section.

We will first look at the \textit{setting and re-setting of goals}, and how this is a dynamic process that requires continual additions and changes. The list of goals is not static, and goals can be added, removed, or changed. Next, we look at the \textit{control loop}; an idea that emerged early on from the field of control theory whereby one runs a continual loop of sensing, decision-making, and intervention, to (effectively) influence the environment in order to attain goals. After that comes the idea of \textit{requisite variety}, where a system must have enough capacity to influence its environment to a sufficient extent that the desired changes actually occur. And finally, we look at \textit{goal alignment}, and the idea that goals must work with each other (rather than at cross-purposes) for an entity to be in control, especially between higher-level “general” goals and the lower-level implementation goals that support them.

Together, the four key aspects of goal setting and re-setting, control loops, requisite variety, and goal alignment, will form the backbone of what we mean by an entity “being in control”. As several of these aspects draw from existing works and ideas, we outline those here so that it is clear how they feed into our own formulations.

\textbf{Idea 1: Dealing with the “environment”}

Even in earlier literature, it was well understood that being in control means being in control of one’s environment. Indeed, early cybernetic literature speaks of “environmental disturbances”, and of “regulators” that need to be able to compensate for such disturbances. It is important to understand that “environment” here does not just mean the biological or physical environment, but also the social, technical, economic, and political environment of where one wishes to operate or where one’s control is exercised.

So, starting with \citet{wiener_cybernetics_1948}, the field of cybernetics examines the concepts of control and communication from a unified perspective, with the aim of being applicable to mechanical and biological entities and systems. \citet{ashby_introduction_1956} advanced Wiener’s technical framework by characterising a system’s ability to counteract environmental disturbances (i.e., “regulate”), in order to keep the system’s variables within acceptable bounds. Here, a \textit{disturbance} is any environmental condition that influences the system, including those that threaten to push the system outside its desired range, while a \textit{regulator} is the mechanism responsible for producing compensatory actions. The regulator interacts with its environment through well-defined input and output channels, through which the system operates sensors that register disturbances, and \textit{effectors} that carry its responses back into the world. Between these, a \textit{mapping} (i.e., decision) interprets each sensed disturbance into the action required to compensate for (i.e., counteract) it.\footnote{Formally, he describes a regulator R as required to achieve the essential variables E within the desired set of states $\eta$, given disturbances D acting on a dynamic system (an environment) T \citep[p.~219]{ashby_introduction_1956}. Regulation is achieved through communication, where the process can be described with the following steps. First, the regulator receives information about the disturbance from the environment T. Next, the information about the disturbance is encoded by a transducer U into a signal understandable by an inverter. Then, the inverter transforms this input signal into a compensating signal that counteracts the disturbance from a set of transformations M. Finally, another transducer decodes the signal produced by the inverter into an output signal which can then be used to manipulate the environment T \citep{ashby_introduction_1956}.}

\textbf{Idea 2: Observing and changing the environment}

However, in most real-world situations, the environment is neither static, nor changing in a perfectly constant or predictable manner. Thus, a system that merely deals with the environment as either a one-off process or at a predetermined fixed interval is often insufficient. Only a loop has the ability to steadily deal with the environment (as the environment constantly exists, and constantly changes, so dealing with it is not a one time thing). Over the years, different ideas for such loops have been developed.

For example, building on the concept of “statistical control” by \citet{shewhart_economic_1931}, Japanese executives popularised the Plan-Do-Check-Act (PDCA) cycle \citep{moen_foundation_2010}, which in time was adapted into the Plan-Do-Study-Act (PDSA) cycle \citep{deming_new_2000}. Later, \citet{boyd2012essence} coined the Observe-Orient-Decide-Act (OODA) loop, which illustrates how one comprehends, shapes, adapts to, and is shaped by, an evolving reality. Boyd’s concept has in recent times been applied to military control \citep{bazin_boyds_2005}, business education \citep{ryder_using_2024}, and finance \citep{byrum_allying_2019}, among other fields. 

In the context of management control systems, \citet[p.~4]{anthony_management_2011} introduce four elements: a \textit{detector} (or sensor) which measures what is happening in the process being controlled; an \textit{assessor} which compares it with some standard or expectation of what should happen; an \textit{effector} which alters behaviour if the assessor indicates the need to do so; and a \textit{communications network} which transmits information to/from the detector and assessor.

And so, in representing control in different ways, these perspectives nonetheless share similar structures. They describe iterative loops cycling through some version of: (i) assess the situation, (ii) form a plan, (iii) act on the plan, and then back to (i) (re)-assess the situation. They attempt to answer how one can maintain control in a dynamic situation, and how a system can operate within and deal with its environment.

And beyond acting upon the environment continuously, the magnitude of the action itself must be sufficient, given the dynamically changing nature of the environment. This is precisely what \citep{ashby_introduction_1956} denotes in the \textit{Law of Requisite Variety}, whereby a system must have the capacity to counteract the various disturbances from the environment in order to reliably achieve goals.

\textbf{Idea 3: Understanding different levels of goals}

While all the “loops” proposed above have things in common, they are also all slightly different in how they operate, and use different terminology. This can be attributed to the fact that they were developed in different contexts, with different types of goals in mind, and with differing emphasis on particular aspects of the control process. A loop for high-level goals is often described differently from those in lower-level engineering systems.

So, with different systems having different versions of such iterative loops, it became necessary to find a terminology to describe their “levels of abstraction” and the granularity of their deployment. In related work, \citet{anthony_framework_1964} explicitly separates the different levels of planning and control in an organisation, naming them strategic planning, management control, and technical control. This was later reconceptualised as a \textit{hierarchy of decisions}, namely \textit{strategic}, \textit{tactical}, and \textit{operational} decisions \citep{khalifa_strategy_2021}; terms which we will expand on later in this section. Moreover, for such a framework to function effectively, there must be a sufficient alignment between different levels of goals to prevent decoupled decision-making.

And so, in the following subsections, we adopt these fundamental concepts and describe the four aspects required for an entity to be “in control”. Namely, these are: (i) the ability to set and re-set goals, (ii) a functioning control loop, (iii) adequate requisite variety, and (iv) aligned goals pursued by subsystems (in the case of systems that contain subsystems). 

\subsection{Setting (and re-setting) of goals}
\label{sec:setting_of_goals}

As per our definition, an entity must be able to set plausibly attainable goals in order to be in control. But this is not a static process to be done all at the very beginning; no human sets all their goals in life at birth. Goals are set by an entity incrementally over time, and (as we will discuss later) involve the setting of sub-goals; a higher-level goal might be set (e.g., “get a job”), which then spawns later sub-goals (e.g., “search for available jobs”, “prepare a CV”, “submit an application”, etc.). Such goals need to be plausibly attainable (e.g., an individual should not waste time applying for jobs they are utterly unqualified for, nor should they try and search through job listings on a website that is not in a language they speak). And sub-goals, as they are set over time, must be able to meaningfully contribute to the higher-level goal(s) they serve, and certainly not act contrary to them. Some sub-goals will be critical to the achievement of higher-level goals, whereas others might remain unachieved but not prevent the higher-level goal being achieved. We discuss this further in Section \ref{sec:goal_alignment}.

As the entity achieves goals, it will necessarily need to set additional goals. For example, the successful completion of certain sub-goals (e.g., “prepare a CV”) might then lead on to future goals (e.g., “submit CV to all suitable job calls that were found”). Some goals might be one-time goals that can be crossed off as completed (e.g., “get a job”, “get married”, etc), particularly when they are lower-level goals put in place purely to achieve a “higher-level” goal (e.g., “submit CV to this particular job call”). Whereas other goals may be very high-level and/or more perpetual (e.g., a political party, as an entity, may have the perpetual goal of “remaining in power”); such goals are seldom crossed off a list.

So how can we better-understand this notion of lower-level and higher-level goals? While some systems, such as a thermostat, are simple enough to only have one goal, many systems are more complex, having different (levels of) goals and containing a range of subsystems and sub-entities. For example, a corporation consists of departments, teams, and individuals. But already we see that a single entity can have multiple levels of systems, corresponding to a hierarchy of different goals. For example, a person may have a fairly high-level goal of finding happiness, with a slightly lower-level goal of going for a vacation, and an even lower-level goal of selecting a great destination, and so forth. Each of these levels of goals, while being tied to the same entity (i.e., the person), will have subsystems in place to pursue those goals, e.g., to achieve the goal of booking the most preferred flight ticket, the person might have a system consisting of searching through several websites with a selection criteria and also a method of payment. 

Thus, beyond simple systems, a tremendous amount of additional complexity comes from the different \textit{types} of decisions and goals that can exist. Beginning with the observations made by \citet{anthony_framework_1964} and \citet{khalifa_strategy_2021}, we first need to differentiate between what we term \textit{strategic}, \textit{tactical}, and \textit{operational} goals and decisions. Strategic decisions are high-level, abstract, and involve long-term actions that are (usually) somewhat ill-defined and difficult to falsify at the onset. Tactical decisions are medium-level, and are more well-defined than strategic decisions. They are used to support strategic decisions. Operational decisions are the lowest-level of these, and involve actions that are usually more straightforward to determine, often highly falsifiable, and quite procedural in their execution. They are used to enact tactical decisions. However, the strategic decisions made and goals set by humans are usually not independent of each other. Rather, they are informed by, among others, beliefs and values, which may not have always been consciously chosen by them.

As such, we posit that strategic, operational, and tactical decisions and goals sit under a fourth, unnamed goal type which one can view as \textit{ideals}.\footnote{This usage of the term ideal is inspired by the works of \citet{ackoff_purposeful_2005}.} Within the argument of this paper, ideals fulfil one or more parts in the process of goal-setting. First, they can encapsulate an entity’s mission within which strategic goals are situated, and from which specific strategic goals can be derived. Second, they can encode an entity’s vision and thus the kind of extremely long-term thinking that transcends standard strategic narratives and goals. Finally, they can serve as terminal goals, answering the big “why” questions behind an entity’s high-level goals. Not every ideal will have all three, but every ideal will have at least one.

\begin{figure}[ht]
    \centering
    \includegraphics[width=\linewidth]{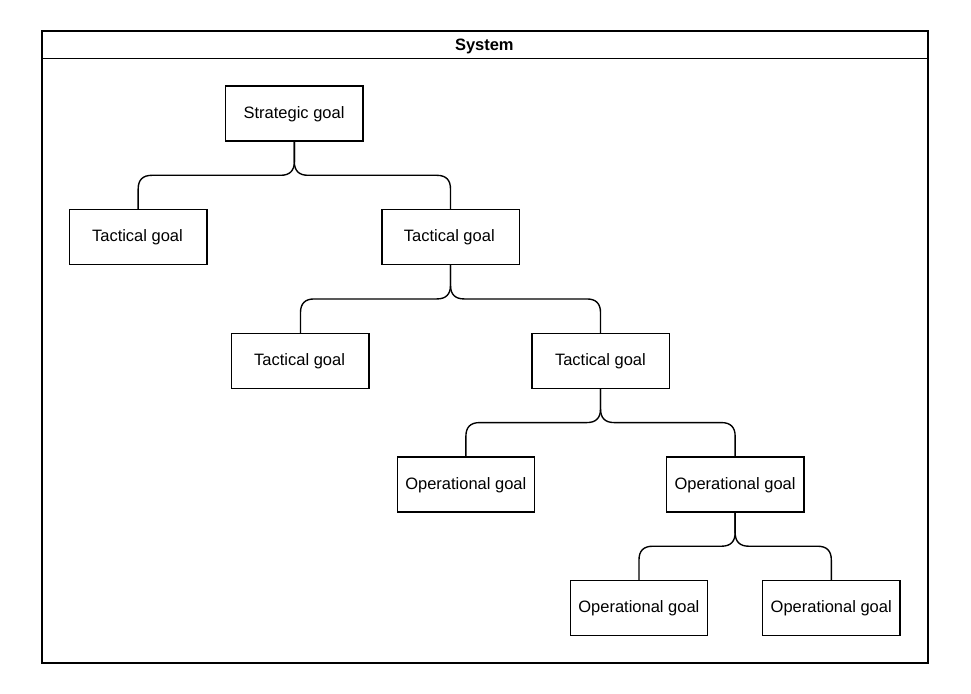}
    \caption{A hierarchy of goals, where lower-level (sub-)goals sit under higher-level goals. Not all goals have sub-goals. For any goal to be achieved, a system must be in place to pursue those goals.}
    \label{fig:subsystems}
\end{figure}

We illustrate this hierarchy in Figure \ref{fig:subsystems}.

However, even though an entity in control will be setting and achieving many goals (and sub-goals), they will inevitably fail to achieve some of these, which need only be “plausibly attainable”, not “guaranteed successes”. And so, when one (or more) goals fail to be achieved, the entity must have the ability to (re)set additional (and possibly different) goals. It may maintain a given high-level goal, and simply change some sub-goals to keep trying to achieve it. This could be on the realisation that those sub goals are now unattainable, or that attaining them will not help achieve the higher-level goal. Or, the entity may change quite a high-level goal, perhaps again due to a realisation that it is not plausibly attainable, or that attaining it is no longer desirable. For example, a retail company might change the range of products it sells in order to improve profitability (a lower-level goal), or it may change a very high-level goal of removing its bricks-and-mortar retail presence entirely, and instead sell only online. A job applicant may fail at an application and so submit more applications, or (after some period) they may entirely change the high-level goal of “getting a job” to “claiming unemployment benefits” or “taking early retirement”.

And so, we see that an entity must have the ability to \textit{continually} set goals over time, as their desires and circumstances change, and also have the ability to re-set or even abandon certain goals that have failed, become unachievable, or cease to serve their intended purpose. Thus, implicit in the ability to \textit{set} goals is the ability to \textit{re-set} goals, either by adding further goals, or by abandoning and replacing existing goals.

\subsection{The control loop}
\label{sec:the_control_loop}

In line with the cybernetic theories outlined earlier, we posit that, for a system to be able to reliably achieve goals, it must comprise several operational processes that together form a \textit{control loop}. 

Our definition of a system stipulates that it operates within an environment. However, we must first establish some terminology to describe the various parts of “the environment”. We call the environment around the system that it intends to act upon the \textit{proximate environment}. The parts of “the environment” that can influence the system, but which are not the intended subject of influence for the system, are what we call the \textit{wider environment}.\footnote{Sometimes referred to as “outside the environment”; see \citet{clarkeanddawe_clarke_2010}.} And we will use the term \textit{total environment} to denote the combined proximate and wider environments. Using our thermostat example yet again, the room that the thermostat system is trying to keep at a stable temperature is its proximate environment, the atmosphere and other climatic aspects of the Earth outside the room is its wider environment, and these two combined is its total environment.\footnote{These concepts are adopted from concepts in cybernetics: what we call the system is what they call the regulator; what they call the plant is what we call the proximate environment; what they call the system is a combination of their regulator and plant (which is our combination of the system and its proximate environment); and what they call the environment is what we call the wider environment.}

And now, to derive the control loop theory we wish to use, we will trace back its necessary components. This begins with the entity that has just set a goal, and now wishes to deploy a system to achieve it.

Ultimately (and starting our analysis at the “end” of the process and working backwards), one notes that such a system must be able to manipulate the (proximate) environment in a way that enables goals to be achieved. This means that it must have concrete ways to execute actions in the environment. Tracing back from there, one sees there must be some form of underlying logic that determines those actions, as without such a logic, it would (probably) not meaningfully affect the environment enough to achieve the goal. So where might this logic come from? It could be understood as a decision-maker that selects and shapes the actions for subsequent execution. Working backwards one final time, we see that this decision-maker can only determine the appropriate actions if given data or inputs that are representative of the environment and suitable for the goal being pursued. And so the system must begin with receiving inputs via a set of suitable sensors, i.e., (social, socio-technical, or technical) devices that measure relevant parameters of the environment. 

\citet{ashby_introduction_1956}, using the technical language of environmental disturbances, regulators, transmitters, and essential variables (among others), created a similar control loop focused on formalising control within strict environmental and technical constraints. Expanding on Ashby’s more formal exposition and technical language, we will call our three control-loop functions \textit{sensing}, \textit{decision-making}, and \textit{intervention}. 

In summary, we see that a system with a designated goal must start with some inputs, then decide what to do in relation to those inputs and the goal, and then take actions that carry out that decision. And so we adopt the following definition of a control loop:

\begin{center}
\textit{A control loop is comprised of three core functions:}\\
\textit{sensing, decision-making, and intervention.}\\
\textit{These operate in a loop, in the order given.}\\   
\end{center}

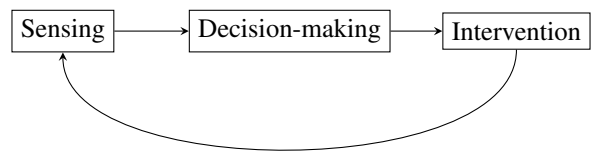
\begin{figure}[htbp]
\centering
\begin{tikzpicture}[node distance=3cm, >=stealth]

\node[draw, rectangle] (A) {Sensing};
\node[draw, rectangle, right of=A] (B) {Decision-making};
\node[draw, rectangle, right of=B] (C) {Intervention};

\draw[->] (A) -- (B);
\draw[->] (B) -- (C);
\draw[->] (C) .. controls +(0,-2) and +(0,-2) .. (A);

\end{tikzpicture}
\caption{The control loop.}
\label{fig:control-loop}
\end{figure}

The control loop is also illustrated in Figure \ref{fig:control-loop}. We now describe in more detail how these steps of sensing, decision-making, and intervention, are carried out in practice for different systems. 

The sensing and intervention steps can be understood as two sides of the same coin. The former takes in inputs using sensors from its environment and converts them into signals that can be used to make a decision. The latter carries out actions on the environment according to the decisions made. Importantly, the sensing includes measuring how the intervention of the previous step changed the environment. But in almost all scenarios, it is impossible to measure the environment with absolute scope or granularity. As such, sensing can only give a \textit{partial} impression of it. So the question of \textit{what} to sense would (ideally) be motivated by the goal itself, i.e., what is important and necessary to sense is determined by the goal(s) that the entity wishes to achieve. 

Neither sensing nor interventions need to come from (technologically-based) physical devices. Indeed, they can be in any material form, including biological, technical, social and socio-technical. Quite often, the functions of sensing and intervention can happen in a mixed fashion. For example, humans and technology can come together to sense and intervene, and this combined sensing and intervention activity makes them “actants” (and not just human actors), to use Latour’s terminology \citep{latour_reassembling_2005}, both reading and affecting their environment.

But in between all sensing and intervention lies a decision-making step. And the decision-maker can be viewed as taking the (sensed) inputs for the system and producing its interventions, i.e., producing a \textit{mapping} between the two. That is, for a given sensory input, the system needs to be capable of mapping it to a (set of) appropriate interventions, such that successive actions will (likely) lead to the goal being achieved. The system’s decision mappings will (ideally) reflect its goals: for goals that are explicit, the decision mappings are designed to achieve those goals; whereas for goals that are implicit, goals are inferred or attributed to the system based on decisions taken in the past. 

Similar to sensing and intervention, a decision-maker does not have to be a single individual or a physical piece of technology. Again, it can be a more complex socio-technical system itself (e.g., a decision can be jointly made by a collection of humans in conjunction with other tools). But there are certain conditions the decision-maker must fulfill in order to be able to make “good” decisions (i.e., those that contribute to the desired goal(s)). According to Conant and Ashby’s formulation of \textit{The Good Regulator Theorem}, such a decision-maker must have a good \textit{model} of the environment being manipulated\footnote{While the original formulation is that “every good regulator of a system must be a model of that system”, their use of the word “system” is slightly different from our usage in this paper: their “system” refers to the combination of our “system” and our “proximate environment” (i.e., the parts of the environment that our system intends to exert influence upon).} \citep{conant_every_1970}. Only then can it reasonably determine the effects of its interventions on the environment. The model must capture all features necessary for control, which may include spatial, temporal, and social structures and dynamics. Hence, we argue that a good decision must account for the effects of its interventions now, and \textit{in the future}. The temporal dimension is especially relevant for systems and environments that involve feedback loops, where a decision may inadvertently reinforce certain unwanted outcomes. 

We have described here three functions: \textit{sensing}, \textit{decision-making}, and \textit{intervention}, that occur in a continuous loop. However, this does not mean that systems always behave in a strict sequence where one data point is sensed, one decision is made, and one intervention is made. In practice, multiple data points can be taken within one step of sensing (e.g., the management of a company may collect data for years before making a strategic decision to expand its business to a new country). Likewise, the intervention step that follows a single decision may also involve multiple discrete actions. Conversely, a decision can also be made to do nothing, where the intervention step would then contain no actions, and the process loops back to sensing again. 

While a control loop is typically described as operating in relation to the environment of a system, we posit that it can also operate in relation to its subsystems. That is, not only can a system sense from, make decisions about, and intervene towards the environment; it can also sense from, make decisions about, and intervene towards its subsystems. Among other actions, this can be done in the form of creating new and/or configuring existing subsystems. For example, a computer program, which itself is a system intended to achieve certain goals, can spawn up sub-processes (i.e., subsystems) that would achieve sub-goals. 

When systems are made up of sub-entities (e.g., a production factory with human workers), such intervention may not always be very “direct”, and, indeed, quite often cannot be particularly direct as otherwise it might violate the goal-setting capabilities of the sub-entities. Consider our earlier example of the delivery company: in the process of a company exerting control on its truck drivers (who themselves are entities), the company does not directly manipulate these drivers at all times. In addition to direct actions, the company sets constraints and incentives for the drivers, such that they are more likely to achieve sub-goals that lead to the eventual achievement of the company’s goals, and do not pursue goals incompatible with the company’s goals. Thus, these interventions are meant to shape the environment of the sub-entities and their systems. Constraints and incentives are used to shape the decision mapping of subsystems so that they would more likely lead to actions that are favourable to the goals of the higher-level system. Using again the example of the delivery company, if the company promises an attractive bonus to the truck drivers for reaching a certain delivery target, this would likely lead to the truck driver making decisions that are in line with the company’s goal of timely delivery of parcels. 

\subsection{Requisite variety}
\label{sec:requisite_variety}

However, merely having a system with a control loop is no guarantee that goals can be reliably achieved, as the mere existence of a control loop says very little about a system’s ability to adapt to a changing environment. In order to reliably achieve goals, the system must have the \textit{capacity} to counteract various environmental disturbances. This ability to sufficiently “deal with” the environment is usually referred to as Ashby’s \textit{law of requisite variety} \citep{ashby_introduction_1956}. Variety, however, need not be a one-dimensional variable such as “engine capacity”: dealing with environmental disturbances is typically at least a matter of their “size” and “frequency”.

What this means in practice becomes slightly clearer when we go back to our thermostat example. In order to maintain a room within a specified temperature range, the heating/cooling system must be \textit{powerful} enough to heat/cool the room not just during mild weather, but during extreme weather as well. In other words, if the system does not have sufficient heating capacity to heat the room to the desired temperature during a blizzard, it does not have the requisite variety needed to counteract the variety of the environment. Here, the first dimension of variety is the temperature it needs to be able to deal with. But a system that heats too slowly, or that measures the temperature too infrequently, will not be sufficient either. For example, a thermostat and heating system that is only capable of measuring the temperature of the room once a year is unlikely to achieve its goals of keeping the room within a certain temperature range. We say that such a system does not have sufficient \textit{temporal} variety to respond to the variety of the environment. The ability to deal with different temperatures and to sense sufficiently frequently are examples of intrinsic capacities of the system, and thus of what we term \textit{intrinsic variety}, as they describe what the system should intrinsically be capable of. 

However,  intrinsic variety is not what ultimately matters. In addition, we need to differentiate this from what we refer to as \textit{effective variety}, which is the variety that the system will \textit{actually exercise} during its operations. Effective variety can at times be heavily constrained by the \textit{available} sensing-intervention mappings (i.e., available decisions), but other constraints, such physical constraints on how intrinsic variety can be used, might also reduce effective variety. For example, a heating system designed to be capable of heating the room, may still be configured to never be operating on maximum power, and hence its effective variety will be lesser than the intrinsic variety. Thus, a system’s effective variety can be understood as its intrinsic variety modulated or constrained by its decision-making: it combines both the inherent capabilities of the system as well as how the system would “decide” to use those capabilities. In other words, intrinsic variety is what a system \textit{can} do, while effective variety is what a system \textit{will} do. This effective variety is what Ashby’s law of requisite variety refers to — a system’s effective variety must meet the variety of the environment for a system to be in control. 

Being adaptive to a changing environment (i.e., having “variety”) is not limited to technical systems; it also extends to socio-technical and social systems. For example, a democratic society must have sufficient (self-)corrective capacity to correct its own mistakes and to deal with the decisions of other actors, including those made by foreign heads of state. The ability to maintain and defend its democratic foundations against inside and outside threats \citep[cf.][]{rijpkema_militant_2018} is thus just one form of variety. For a democratic society, variety is encoded in the very fact that the society is democratic: if, for example, elections are spread too far apart, or if there was no mechanism to remove leaders mid-term if their behaviour threatened to undermine its foundation, such a society might no longer be considered democratic. 

\begin{figure}[ht]
    \centering
    \includegraphics[width=\linewidth]{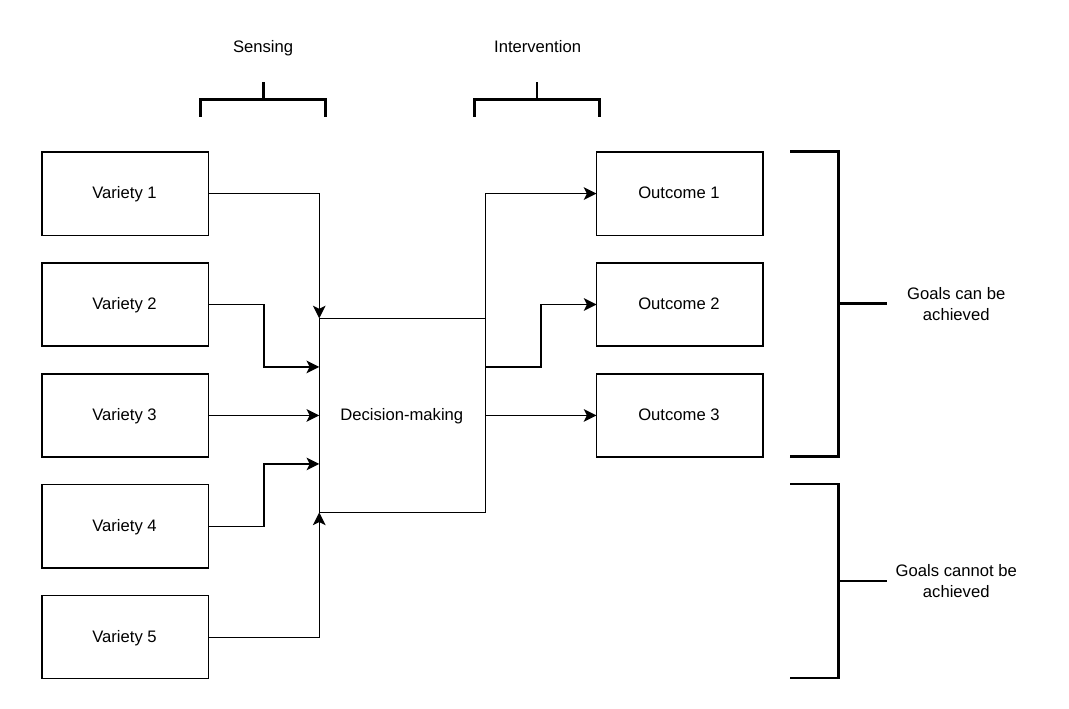}
    \caption{A system is able to respond towards some, but not all, varieties of the environment and result in outcomes that are able to eventually achieve goals.}
    \label{fig:variety}
\end{figure}

In Figure \ref{fig:variety} we present a visualisation of Ashby’s law of requisite variety from the perspective of sensing-intervention (i.e., input-output) mappings. Having a system containing the correct mappings from a selection of available inputs to outputs may not be sufficient for an entity to be in control: these mappings themselves must be sufficient, i.e., adequately cover the space of possible inputs thrown at the system by the environment. In other words, knowing the right way to respond to a few expected situations does not mean a system is properly in control. Meaningful control requires having a comprehensive set of responses for unpredictable events or challenges the environment might present. If the system is only prepared for a limited range of scenarios, an unexpected problem would otherwise cause it to fail.

Further interpreted for our theory of control, this means that varieties may not always be fully known in advance. In many situations, this is part of the art of designing (or adopting) suitable control systems (as not all control systems are “designed”). Understanding how and when a system’s environment changes, and thus implementing a system having suitable requisite variety is often not restricted just to what is quantitatively measurable in the environment. Social and sociotechnical environments, in particular, often come with emergent phenomena and drift effects — such as changing norms, beliefs, or knowledge — which can appear hidden to a system’s designer. For a social system, having sufficient variety today may not necessarily imply that it is sufficient tomorrow.

\subsection{Goal alignment}
\label{sec:goal_alignment}

Systems do not just operate with their own different levels of goals; they may also contain other systems operating at various goal levels. In addition, determining the level of a decision or a goal is often \textit{subjective}, and \textit{determined} by the perspective of the entity or system. In the example of the delivery company above, from the perspective of the company, hiring an individual truck driver and loading and deploying an individual truck would be a lower-level operational goal. However, from the perspective of the truck driver, applying for (and doing) the job may be closer to a higher-level tactical, or even strategic, goal. 

As so, we see the phenomenon of \textit{goal alignment} arising. Where sub-entities exist, their goals should be in \textit{sufficient alignment} with the goals of the higher-level entity (vertical alignment); and where there is overlap between systems, the goals of different sub-entities should also be in sufficient alignment with each other (horizontal alignment). And even within a single entity without sub-entities, the sub-goals of the entity being pursued by its subsystems should be in sufficient alignment with its own higher-level goals (as we will elaborate on at the end of this subsection). Here, we refer to goal alignment as the degree to which taking actions to pursue one goal results in the achievement of the other goal. Two goals are \textit{aligned} if actions to pursue one goal contributes to the other being achieved; they are \textit{misaligned} if such actions contribute to the other goal being jeopardised; and they are \textit{irrelevant} if such actions do not affect the achievement of the other goal. 

Of course, with respect to the higher level system, the goals of its sub-entities are only relevant if they affect the system at large. For example, it should not matter to the company if the truck driver also has a goal of learning to play the guitar, if they only do so outside of working hours. That is, we can further refine the notion of alignment of a system and its sub-entities; when their goals interact \textit{at the places where they meet}, they can be considered to be relevant. The same applies to two sub-entities who operate with the same system: should their systems overlap such that the achievement of one goal affects another, they can also be considered to be relevant. And so long as the relevant parts of the goal of an entity contribute towards the relevant parts of the goal of another relevant entity, they can be considered to be \textit{aligned}. Moreover, goal-relevance is true in both directions; just as the goals of the driver to buy a guitar are not relevant to the company, so too we see that the goals of the company to increase its customer base are not relevant to the driver. Indeed, the driver may dislike certain goals of the company, but as that is beyond where their respective goals “meet”, it is irrelevant for the driver. 

Neither the existence of a goal hierarchy, nor the (mis-)alignment of goals, typically gives sufficient insight about how goals are formulated. In particular, for goals requiring systems that contain sub-entities or subsystems to achieve, the goals of these sub-entities or subsystems are typically not set once in an unchangeable way, nor are they all set at the same time. Again, in our example of the delivery company, it may have a goal related to the cost-effective and timely delivery of goods. The company may set a sub-entity (a truck driver) a sub-goal of delivering a truckload of goods using the associated subsystem (the truck driver, the truck, etc.). However, if bad weather has led to road closure, this particular sub-goal may no longer be achievable, as the goods can no longer reach their destination within a certain time and cost. So the delivery company must set new sub-goals to achieve its goal, such as to deliver those goods through other modes of transportation. Or it may delay or even cancel delivery if such goals are incompatible with its strategic goals of making a profit because, for example, certain modes of transport are too expensive for what is transported. We can thus see that goals that are initially aligned with higher-level goals can sometimes become misaligned when circumstances change, where the degree of alignment between two goals also depends on how achievable they are.

And so, with large systems, the setting of new sub-goals and the creation of subsystems would be a continuous process that must be managed adequately, if the higher-level entity is to be in control. Here we can also see a relationship between the decision-making part of the control loop and the setting of lower-level goals: at a higher level, certain decisions take the form of setting new sub-goals, creating/obtaining new sub-entities, and designing appropriate sub-systems. These are performed continuously to ensure that sub-goals remain aligned even when circumstances change.

Hence, while the language of control loops has certain implications of “dynamics” and “change” embedded in it, the static language of “goal hierarchy”, “goal levels” or “goal misalignment” does not imply that their levels or their alignment are necessarily fixed. As the environment, entity's subsystems, or sub-entities change, these all can change, too. Decisions and goals might be replaced, or moved up and down the hierarchy. Finding food, for example, might be an operational goal in normal times, but it can be elevated to a strategic goal in times of crisis.

Of course, when considering the systems and sub-systems of one single entity, goal alignment is still an essential consideration. We discussed this in Section \ref{sec:setting_of_goals}, whereby sub-goals of an entity’s goal (and thus any sub-systems it employs to achieve them) must sufficiently align with the higher-level goal they serve. And this is already a non-trivial outcome to achieve; such sub-goals (and associated sub-systems) must be carefully chosen and configured. But in considering alignment within the one entity (between its sub-systems), one part becomes trivial: the setting of the goals themselves. The entity is free to choose its own sub-goals, and/or configure its own sub-systems, as it pleases. Thus, the purpose of this section was thus to examine when sub-systems of an entity are run by a sub-entity; in such instances, the original entity cannot simply “change” the goals of the sub-entity at will. Hence, the question of goal alignment in such instances becomes a much more complicated one, needing to also deal with the extra “dimension” of autonomy that the sub-entites possess. 

\subsection{Interactions between aspects of control}
\label{sec:interactions_between_aspects}

So far we have largely studied the different aspects of control in isolation. We will now look more closely at their interplay, and their effect on multi-entity and multi-system scenarios. In particular, we will show that these two aspects, \textit{variety} and \textit{alignment}, are not entirely independent when considering entities and sub-entities, and that a change in one might require a change in the other.

There is a particular trade-off to be made here when entities and their sub-entities interact; often one cannot independently maximise both the \textit{variety} of one and its \textit{alignment} with the other. These two variables are intrinsically coupled: more variety increases the capability, flexibility, and power to deal with environmental disturbances, and thus with the other entity. On the other hand, any enforced alignment typically reduces the \textit{effective variety} of the system as its ability to react is constrained. In other words, the act of enforcing alignment allows different entities and systems to “play along nicely”, but the additional rules of alignment often reduce what entities and their systems can do. And although there certainly are situations where alignment demands an increase in variety, such as where an entity requires more variety to have a positive effect on another entity’s goals, different entities often have goals that impose this variety-alignment trade-off instead. 

This tension is familiar in the domain of law, where a legal system must balance the variety it grants to individual citizens in order to maintain alignment with the broader social system. The principle that “one’s rights end where another’s begin” is a mechanism in which this trade-off is managed. Similarly, in management contexts, a delivery company rarely wishes to grant its drivers unlimited discretion. While such variety might increase operational flexibility, it simultaneously introduces additional sources of potential misalignment with the company’s goals. A union, for example, functions as a system to expand the variety available to the drivers, and can thus be viewed as a structural source of potential misalignment from the perspective of higher-level management entities. 

Furthermore, the coupling between variety and alignment introduces trickle-down effects on the setting of goals themselves, and not just on the means of pursuing them. When an entity and its systems are granted more variety, that entity’s goals can often fundamentally shift: world-states that previously seemed unfathomable may now be in reach, and environmental disturbances that would have previously curtailed the entity may now be adequately dealt with, allowing newly-plausible goal-setting and its getting in formerly impossible situations. In other words, this allows entities to make decisions that were previously not available, as the additional space of possible inputs and outputs has enabled more decision mappings to be possible. This is frequently observed in the context of personal finances: expanding an entity’s wealth expands the intrinsic variety of many of its systems, which in turn expands the scope of goals it can plausibly pursue. The result is a reshaping of what the entity desires and how it interacts with the world, including its interactions with other entities and systems around it.

In many situations, the higher-level control loop is the interface through which trade-offs between variety and alignment of the lower-level subsystems are actively managed. While having sufficient variety provides the necessary capacity to counteract environmental disturbances, and alignment ensures that subsystems properly pursue the goals, the control loop may be constantly recalibrating the balance between the two. The control loop senses the environment and the system’s actions, and, among others, needs to understand if its subsystems have sufficient variety to continue pursuing the goals, if the subsystem and itself are sufficiently aligned with each other and the environment, or if some unwanted goal drift has occurred. So the control loop must evaluate whether more or less variety and alignment is needed, or whether it is enough to adjust the course by taking new actions.

Such interplay is rarely a static equilibrium. Instead, for many entities and systems it is characterised by recursive feedback loops where interventions intended to secure control affect the environment and become new sources of disturbance. Di Nucci’s \textit{control paradox} \citep{di_nucci_control_2020} can be understood as one example: to ensure that the many entities of a social institution remain aligned to its strategic goals, and can work towards them, some lower level control may need to be given up. Knowledge workers, for instance, may require management methods that provide them with more (intellectual) variety if one wants them to successfully find the best solution to a complex problem. And so unnecessarily restricting them can instead lead to losing control over higher level goals.

\subsection{Summary}

We provide here an overarching summary of the aspects of control introduced so far in this paper:

We have now seen how an \textbf{entity} can be \textbf{in control}, i.e., have the ability to set plausibly attainable goals and reliably achieve those goals. With the capacity to set goals, the entity can then have a \textbf{system} that aims to achieve those goals. Such entities can be clearly-defined individuals or organisations, or more fuzzy collectives, and the goals can similarly be either implicit or explicit, and either clearly set or merely attributed to the entity. Likewise, the system can be either simple (such as a thermostat) or more complex (such as an entire community of researchers trying to solve a problem), and it can also have any amount of overlap with the entity.

For an entity to be “in control”, it must be able to continually \textbf{set and reset} plausibly attainable goals, including lower-level sub-goals in support of its higher level goals. The entity must also have an adequate system to achieve these; not all systems are sufficient to enable control in a given setting. Such a system must contain a \textbf{control loop} consisting of sensing, decision-making, and intervention functions. This way, the system can have some understanding of its (total) environment and its relation to the desired world-state, be able to make appropriate decisions, and finally execute those decisions by acting and influencing its (proximate) environment. It must also have sufficient \textbf{requisite variety}. That is, it must have enough capacity to counteract the disturbances from the environment, such that for any (plausible) given state of the environment, there is an appropriate action that can be taken to meaningfully influence the environment in a way that would lead to the goals being achieved. And finally, for systems that are more complex and contain subsystems, the (relevant) goals of the subsystem must be sufficiently \textbf{aligned} with the (relevant) goals of the system. This is because only with sufficient alignment can the subsystems pursue goals that ultimately serve the goals of the system, leading to the system’s goals being achieved. 

To summarise, there are four overall aspects needed to be “in control”: 

\begin{center}
\textit{First, the entity must be able to continually \textbf{set and re-set} plausible goals.\\
Second, the system must have a \textbf{control loop} that functions as intended.\\
Third, the system must have \textbf{requisite variety} (i.e., capacity).\\
Fourth, the goals of any subsystems must be sufficiently \textbf{aligned} with those of the system.}
\end{center}

Having laid out the aspects of control and studied them in some detail, we can now see why an entity can never have “absolute control”, in the sense that they are absolutely certain that they can set goals and then achieve them. First, it is impossible to say with absolute certainty that a goal can definitely be achieved. Second, it can neither be guaranteed that the control loop will be designed perfectly, nor that it will operate perfectly without any failures. Third, a system will in practice never have sufficient variety to counteract \textit{all} possible varieties of the environment. And finally, particularly with systems containing sub-entities, some control and power is effectively delegated to these other entities. An example of this is the division of labour in industrial capitalist societies, which often prevents socio-economic entities like corporations or workers from exerting full control over all aspects of their existence, or of the product or service they are offering to the market \citep[cf.][]{durkheim_division_1997}. 

Thus, no entity is immune to “losing control”, at least to some extent, and being in control clearly cannot mean being in active control of every single component at all times. In fact, \citet{di_nucci_control_2020} even argues that there exists a “control paradox”, where increased control in some areas tends to be associated with a decrease in others. A somewhat metaphorical example is given by a shepherd herding sheep. With the help of their dog, they can be said to be in control of a herd of sheep because they can move the herd as a \textit{collective}, even though they may not have direct control over each individual sheep at all times. Perhaps more concretely, in our example above of a truck driver working for a delivery company, the company cannot set and get operational goals of the driver; it is up to them to decide how to drive the truck, how to take packages out, etc. But the company is able to set and get its own goals of having the drivers deliver their allocation of packages in their allocated time.

It may be tempting to interpret only the system aspects of control, (i.e., the control loop, requisite variety, and goal alignment), as being all that is required to be in control. But we again emphasise that they merely describe how a system can reliably \textit{achieve} goals that have already been set. For an entity to be in control, it must also be able to \textit{set (and reset)} plausibly attainable goals.

At this point, it helps to distinguish between the general concept of “being in control”, and the more specific “being in control of a particular goal”. An entity is \textit{in control} if it can set and get goals, as per our definition. One can think of this as the following 2-variable function:

\[
f(-,-): \{\text{all goals}\}\times\{\text{all systems}\} \to \{\text{yes}, \text{no}\}
\]

where we say that if an entity takes a plausible goal $g$, and a sufficient system $s$ to reliably realise that goal, then $f(g,s)=\text{yes}$ (i.e., “in control”) — even though that goal may not end up being achieved in the end. An entity is in control if $f$ only ever takes inputs $(g,s)$ such that $f(g,s)=\text{yes}$, i.e., only sets plausible goals and implements systems that can reliably achieve them. This emphasises that being “in control” does not presuppose a particular goal, or a particular system; both may be set independently. And of course, “yes” here is not an absolute concept, and rests on our earlier notions of control with “plausible” goals and “reliably” getting them. That is, “yes” means the entity has set a goal it has a reasonable chance of being able to attain, and has a system in place that has a high chance of succeeding in attaining that goal.

However, an entity can also be \textit{in control of goal $g$} if, given goal $g$, it has a system $s$ available to it so that $f(g,s)=\text{yes}$. So we can view this as the following single-valued (i.e., \textit{curried})\footnote{“Currying” is the act of taking a multi-variable function and fixing some of those variables to create a new function of the remaining variables \citep[p.~86]{curry_philosophical_1980}} function:

\[
f(g,-): \{\text{all systems}\} \to \{\text{yes}, \text{no}\}
\]

The setting of sub-goals, formation of systems and sub-systems, etc., \textit{for goal g} can still occur here. But now this is all with respect to goal $g$. And now it is clear that the entity can be in control for one goal $g$, but not another $h$, even if $h$ is a sub-goal of $g$ (or vice-versa). There will be times when it helps to refer to the entity as “being in control of goal $g$”, and others when we wish to simply say the entity is “in control” without specific reference to any goal (implying that the entity can freely set goals, without any specific reference to an existing goal).

And as a final note, there are aspects of control that can serve as complementary perspectives for other aspects. For example, goal-setting can be a form of decision-making, whereby a higher-level system can make a decision to set a new sub-goal and subsequently create a new subsystem to pursue it, thus from the perspective of the higher-level system, the ability to set (sub-)goals can be said to be the ability to make higher-level decisions. Separately, decisions made by a system reflect the goal it is pursuing: decisions are made by a system to achieve goals, and thus from the perspective of a higher-level system, subsystems that make suboptimal decisions can be said to have misaligned goals. Similarly, continuously pursuing a misaligned goal can also be said to be a failure to reset the goal. Thus, the four aspects of control can be viewed as having a certain amount of overlap in terms of being able to provide differing (complementary) perspectives of the same thing.

\section{How control can be lost}
\label{sec:how_control_lost}

Having defined our four aspects of control, we can now examine the ways in which \textit{control can be lost}. It helps to first re-phrase our four aspects of control in a way that illustrates how each of them can fail, and hence how control can be lost.

\begin{enumerate}
  \item The first kind of failure is a failure to set goals. This can broadly be described as either a complete lack of goal-setting at all, or the setting goals that are either unobtainable or self-defeating of other goals, or an inability to abandon or re-set goals.
  \item The second is a failure within the control loop, i.e., a failure either to sense, decide, or intervene adequately to obtain the set goal. Failure in any of these functions is usually due either to configuration shortcomings (i.e., predictable failure), or not operating as intended. 
  \item The third is a failure of adequate configuration, where the system does not have sufficient capacity to reliably achieve its goals. This can be through a lack of capacity (i.e., intrinsic variety), or through a failure to exercise all decision options (i.e., effective variety). 
  \item The fourth is a failure of alignment, where the goals of subsystems are sufficiently misaligned in a way that jeopardises the ability for the higher-level system to achieve goals. 
\end{enumerate}

And so when any of these four aspects fail, we regard \textit{some control to have been lost}, at least to a certain extent. And while these “failures” are simply the negation of our aspects of control, the purpose of this section is to delve into the mechanics of precisely how this can happen. As such, this section will take us from the theory of what control is (Section \ref{sec:how_entity_in_control}), to a more structured and examples-led description of specifically where and how control can be “lost”, accompanied by realistic scenarios illustrating how it occurs in practice. While this will not be an exhaustive list of scenarios for each control aspect, it will nonetheless cover some of the most critical ones. Then, in Section \ref{sec:how_ai_affect_control}, we will look at specific examples and mechanisms of how AI systems can lead to a loss of control for (specifically human or human-centric) entities.

\subsection{The inability to set or re-set goals}
\label{sec:inability_to_set_goal}

It is not just the inability to \textit{get} goals that can lead to loss of control; the inability to \textit{set} (or \textit{re-set}) goals also leads to a loss of control. And so, here we differentiate between three different categories of failure for goal-setting: (i) a \textit{complete inability} to set any goals, (ii) a \textit{partial loss} of ability to set goals, and (iii) an \textit{inability to change} goals.

\textbf{Category 1: Complete inability to set goals}

A \textit{complete loss of the ability to set goals} represents the most severe case, and is often predicated on unexpected events in the entity's environment or some critical system failure. Consider a truck driver who crashes and dies, or suffers a heart attack severe enough to deprive their brain of oxygen: these are extreme events from which the entity cannot recover, and in which, by definition, the entity ceases to exist as an entity altogether, because it can no longer set goals. The same applies to non-human goals: a “smart” heater that attempts to raise the room temperature far beyond its design capacity may burn out and cease to function completely, rendering it incapable of setting any further goals.

But the entity need not “cease to exist” in order to cease being able to set goals. Consider, for example, a government that is unable to convene and pass legislation, perhaps due to some external event such as a pandemic. The government still exists, but they cannot set goals, in the form of legislation. Or, more simply, they could be in a deadlock and be unable to reach the required votes on any bill. Likewise, a comatose patient in a hospital has lost all ability to set goals, even though they still exist, and may well recover in the future.

\textbf{Category 2: Partial loss of the ability to set goals}

A partial loss of the ability to set goals takes several distinct forms. The first is a \textit{reinforcing loop of unobtainable goals}, whereby an entity persistently sets goals that are beyond its current reach (i.e., unobtainable), and in doing so undermines its own future capacity to set reasonable goals. For example, an unemployed individual who constantly and only applies for jobs they are wholly unqualified for may remain unemployed for a prolonged period and eventually become destitute. And in becoming destitute, their capacity to set and pursue even modest goals is further eroded. The very act of setting unobtainable goals thus feeds back into a diminished ability to set attainable ones later on.

The second form of partial loss arises from a reduced or damaged ability to reason, where the entity retains some goal-setting capacity but that capacity is so sufficiently impaired that \textit{the goals it sets become unreasonable or self-defeating}. A heavily intoxicated individual, for instance, may set goals — such as to drink more, or to drive home — that are either harmful or wholly detached from their broader interests, demonstrating that the ability to set goals can persist in a degraded form. In sufficiently extreme cases, however, this moves towards Category 1: severe intoxication can compromise the ability to set any goals whatsoever, illustrating that the boundary between partial and complete loss of goal-setting capacity can be one of degree rather than kind.

The third form is an \textit{inability to set goals in one particular area}, while remaining fully functional elsewhere. A government deadlocked in one parliamentary vote may be unable to set goals or make decisions in that specific domain, yet retain the full capacity to set reasonable and effective goals across all other areas of governance. In this case, the loss of control is localised to one (or few) goal-setting areas, rather than systemic; it does not necessarily compromise the entity's broader functioning, unless those goal(s) happen to be critical in some way (in the case of governments: not being able to pass a supply and confidence motion is enough to topple it). 

\textbf{Category 3: Inability to change goals}

The ability to change (i.e., re-set) goals, at all levels, is also crucial. Consider a tourist driving to the airport who discovers severe traffic ahead. To maintain control, they might change their route to still make the flight, or, if too delayed, book a later flight upon arriving at the airport to salvage most of the holiday. They may turn around and rebook from home for the following day, or, upon learning of extreme weather at the destination, abort the trip entirely. In each case, the ability to re-set goals at one level — the route, the flight, the destination — allows them to preserve goals at a higher level. Conversely, a tourist who rigidly refuses to deviate from their original route, despite the road closure, misses the flight entirely and loses the holiday altogether. Or their insistence on trying to travel in extreme weather to retain their holiday may lead to them being stranded, or harmed, at their destination, destroying their even higher level goals of remaining safe. The failure to re-set an operational goal may propagate upward, destroying higher-level goals that a small adjustment could have preserved. 

The ability to re-set goals can therefore compensate for failures elsewhere in the control loop, while the inability to do so can turn a minor disruption into a much broader loss of control. Control only requires \textit{plausibly} attainable goals, and so by its very nature some of these will end up being missed or becoming impossible. Simply “giving up” at the first failed goal signifies a loss of control; being able to re-set goals is vital to maintain or regain control at those, or higher, levels. 

But here, we also see that goal re-setting can become problematic and start to “fail” through what is known as \textit{goal shift} (sometimes referred to as “goal drift” in the literature; see \citet{abdi_coherence-based_2025}); the gradual evolution, over time, of the priorities and objectives of an entity. Consider an individual who, over time, is becoming radicalised through watching extreme content online; their goals may start to shift, to the point where they start to re-set goals by quitting their job and joining an extremist group. Or the evolution within a research community of resetting their goals from “publishing good research” to “maximising number research publications” and thus instigating a reproducibility crisis, as has been reported to be happening within the field of psychology \citep{wiggins_replication_2019}. So just as goals must be able to be re-set from unobtainable/unproductive ones to obtainable/productive ones, an entity must also be able to avoid displacing obtainable, productive goals with “worse” ones. That is, with new ones that are either unobtainable, or where doing so threatens the attainment of higher-level goals.

\textbf{Interpretation as other types of failure}

As we will see later, a Category 3 failure can also be seen as a goal alignment failure. More specifically, if a subsystem is pursuing a certain goal, but circumstances change such that the goal is no longer appropriate and does not contribute to the goals of the higher-level system, their goals can now be said to be misaligned. In other words, while a failure to change goals by itself may not constitute a loss of control — one can certainly maintain control over a goal that is no longer desirable — pursuing an undesirable goal almost certainly point towards a misalignment with respect to higher-level goals that are indeed world-states that an entity ultimately desires, thereby losing control of the higher-level goal. 

In addition, all these categories of failure can also be seen as a control loop failure; specifically, a failure with respect to decision-making. In Section \ref{sec:goal_alignment}, we described how a higher-level system can create lower-level subsystems that pursue sub-goals. From the perspective of the higher-level system, it is \textit{making a decision} to create subsystems to pursue sub-goals. From the perspective of the lower-level subsystem, the sub-goal is a newly created one, and hence it would be viewed as a setting of goal. Likewise, when a higher-level system decides based on new information that the subsystems should do things differently, from the perspective of the subsystems their goals are indeed being modified or reset. 

\subsection{Control loop failure}

The control loop — consisting of sensing, decision-making, and intervention — can fail at any one (or more) of these functions. Broadly speaking, sensing can fail due to factors such as \textit{distortions and bias}, \textit{insufficient granularity}, \textit{lagging indicators}, or even just \textit{total sensor failure}. Decision making can fail for reasons such as \textit{decision paralysis}, \textit{rigidity}, or \textit{corrupted processing}, as well as \textit{feedback loops}. And intervention can fail not only from a complete \textit{breakdown of action}, but also due to \textit{desynchronised actions} (those at the wrong time) or \textit{ineffective placement}. And while these three functions of the control loops are distinct and cleanly separated in theory, the lines between these functions may start to blur in practice, especially with systems that have components that serve different functions concurrently.

\textbf{Failure of sensing}

To illustrate this, we return to the example of the truck driver. The driver may be driving with the sun directly in front of him, in which case he is effectively blinded temporarily, affecting his ability to sense at all and leading to a \textit{near-total sensing failure} (of that which he needs to sense: the road ahead).\footnote{Such sensing failures need not affect every sensor to be catastrophic: in the loss of Air France Flight 447, ice crystals temporarily obstructed the aircraft’s airspeed sensors, giving unreliable speed readings, and ultimately leading to a complete loss of control of the aircraft \citep{BEA2012AF447}.} Or he may have forgotten his glasses, and while he can still see large objects ahead such as other vehicles, he cannot make out smaller important ones such as a child running onto the road; this would be an example of \textit{insufficient granularity}.\footnote{In 2018, an Uber automated test vehicle struck and killed Elaine Herzberg as she crossed a road outside a crosswalk in Tempe, Arizona, after its driving system detected her 5.6 seconds before impact but failed to accurately classify her as a pedestrian or predict her path \citep{NTSB2019HWY18MH010}.} This is also referred to as “failure to correctly perceive situation” \citep{endsley_taxonomy_1995}.

But sensing can fail in other ways, and for very different entities. Even if the raw information being sensed is correct, the system’s “sense” of its proximal environment may still be flawed. This is also referred to as a problem with modelling, violating Conant and Ashby’s \textit{Good Regulator Theorem} \citep{conant_every_1970}, where the system’s internal representation does not accurately reflect the state of its environment. In economics and politics, one generally differentiates between leading and lagging indicators. Leading indicators enable predictions about future changes, while lagging indicators enable the confirmation of past trends \citep[p.~28]{manuele_leading_2009}. Political polls are an example of lagging indicators that are often misused as predictive leading indicators. The following example highlights how a faulty decision about the nature of such an indicator can lead to a sensing failure later on: A political party may have conducted a poll many months ago, and decided that a particular area did not require any further campaign efforts for an upcoming election as it was deemed the party was significantly ahead in the polls there. But later, a large political event or scandal may have occurred which placed the party behind others, rendering polling data quite inaccurate. In this case, the polling results no longer accurately reflected voter sentiments. This is also referred to as “failure to comprehend situation” \citep{endsley_taxonomy_1995}.

Or, a bus company may review their bus routes using only data available from users of their timetable app, and based on that decide to cut certain routes for reasons of under-usage, even though certain demographics (such as the elderly) might be severely under-represented in such “digital” datasets. Or one may look at how many people use a certain route and then set ticket prices and durations according to such data. This, however, already overlooks differences between men and women’s usage of public transport, as women might not just go to an office in the morning and return in the afternoon, but may have to complete several chores during the day, requiring getting on and off public transport on one route repeatedly \citep{criado-perez_invisible_2019}. In hindsight, such practices may be understood as \textit{incomplete sensing}, but in their effect on certain socio-economic groups they can also be understood as \textit{distorted and biased sensing}, by virtue of the fact that such sensing is not just missing certain parts of the environment at random but consistently missing specific parts of the environment thus putting specific groups at a disadvantage. In other words, the sensed information is not just unrepresentative of the real state of the environment, but unrepresentative in a specific way.

\textbf{Failure of decision making}

Even with excellent sensing, a control loop might still have poor decision making. Going back to our example of a truck driver, one can consider a situation where he may be drunk or heavily fatigued, to the point where correctly sensing a large static object ahead does not lead to him making a decision to avoid it. This would be an example of \textit{corrupted processing}. Or, even of sound mind, he might come across icy road conditions but choose to maintain full (legal) speed and subsequently crash the truck; a failure caused by \textit{rigidity} in decision making (i.e., having made a decision based on some input variables, but then not changing that decision even if further sensing gives input that should give cause to change it).

But while these are all problems with decisions \textit{that were made}, an inability to decide can also cause failure. One can imagine a military commander faced with a surprise attack, or a paramedic arriving at an incident and finding multiple, critical patients to treat. In either case, a vast amount of information (sensing) is rapidly presented, alongside many possible options on what could be done; finding “the best one” then becomes a daunting task given the time pressure. And so the available choices and the urgency to make these may lead to \textit{decision paralysis} \citep{huber_dazing_2012}, whereby they may avoid, delay, or refuse to make a decision at all.

And decisions might work against the set goals.\footnote{A famous example of poor decision-making is the 1986 Challenger disaster, where even though engineers had correctly identified the risk of O-ring failure at low temperature, NASA management made a decision under schedule pressure to proceed with the launch anyway, leading to a complete loss of control of the space shuttle and killing all astronauts onboard \citep{nasa_report_1986}.} This is especially true in systems that result in a feedback loop. Take, for example, a physical system consisting of a microphone connected to a speaker, where its “decisions” are simply a mapping from audio inputs to audio outputs. Under normal circumstances, the microphone takes in audio signal from a person’s voice, converts it into electrical signal, and the speaker then converts it back to audio at a higher volume, thus amplifying it. However, if the microphone is placed right in front of the speaker, feedback may occur, where audio that goes into the microphone gets amplified by the speaker, which gets into the microphone at a higher volume and gets amplified by the speaker more, and the loop continues. In such situations, the goal of clearly amplifying a person’s voice is no longer achieved, as the audio system enters a continuous loop producing significant noise commonly known simply as “feedback”. Generally, a decision-making process may fail to achieve goals if it does not sufficiently account for the ways in which the environment would react to it, as well as its subsequent reaction to the environment.

\textbf{Failure of intervention}

And finally, the control loop can fail at the intervention step. Again, with the example of the truck driver, he may know that he should brake when there is oncoming traffic, but a small lapse in attention may have caused him to react a little too late. This has been referred to as a “skill-based failure” \citep{reason_human_1990}. There could also be a brake failure, resulting in a \textit{breakdown of intervention} such that he is unable to stop the truck at all (complete breakdown of the intervention)\footnote{In the 1994 San Marino Grand Prix, Formula 1 driver Ayrton Senna was unable to steer the car due to a steering column fracture, resulting in him losing control of the car, ultimately leading to a fatal crash \citep{hilton_ayrton_2004}.} or not sufficiently for normal usage (partial breakdown of the intervention). But it is not just a “failure to intervene” that can cause a loss of control via this process.

One can consider medical interventions, where desynchronised actions can rapidly lead to a loss of control. Such a desynchronised medical intervention can also happen even if the doctor makes the right decisions, such as if a patient presents with an infection which the doctor correctly decides to treat with antibiotics, but delays in obtaining medication means the infection spreads and causes other problems. Or, a surgery might be performed too late to have the desired effect, due to delays in the healthcare system. In medicine, it is not just the type of intervention that matters, but also when and how often it is done, for both practitioner and patient to maintain control. Such desynchronisation of intervention can similarly happen at the institutional level. Consider, for example, the re-introduction of tourniquets into civilian emergency medical services following the wars in Afghanistan and Iraq. In the early 2000s, adhering to standard emergency protocols with issued equipment could yield suboptimal results for traumatic limb injuries involving catastrophic haemorrhage. Consequently, medical personnel following the standard of care were not necessarily able to provide the best possible intervention, as established protocols had become desynchronised with emerging best practices.\footnote{For a more complete discussion, see \citet{goodwin_battlefield_2019} and \citet{lewis_tourniquets_2013}.}

Another example of intervention failure might be a sturdy but incorrectly-placed supporting pillar on a construction site, leading to a (partial) building collapse. Or a fishing net placed in a part of the sea with very few fish, resulting in a minimal catch. These are examples of \textit{ineffective placement}, where the intervention was appropriately put together and timely, but placed or carried out in the wrong location.\footnote{For example, France’s Maginot line put up before World War II was arguably a technically sophisticated defensive system that would function as designed at where it was placed, but Germany simply bypassed it by going through Belgium instead, rendering the intervention pointless \citep{horne_lose_1969}.}

In all of these cases, the intended operation of the control loop has not been achieved due to specific failures. We finish by observing that, besides having operational failures in a system that was set up to work adequately, the control loop can also have a setup that predictably fails to achieve goals and that it may be necessary to consider variety and goal alignment as part of it. Consider, for example, a human overseeing a complex assembly process. If there is a mismatch between the human’s abilities and the complexity of the assembly, the person may enter into a control loop that predictably fails to achieve the goal of human oversight \citep{chiodo_formalising_2025}, via decision paralysis (as described earlier). Here, insufficient requisite variety (the assembly process is too complex for the person overseeing it), leads to inadequate decision making (the person does nothing), and to an intervention that must predictably fail and is fundamentally misaligned with the goal of oversight (doing nothing puts the human out of the assembly process). In the end, the control loop contains a decision-intervention pair that fails in a \textit{predictable way}. 

\subsection{Insufficient variety}

Insufficient variety can manifest in many forms. Most intuitively, this can be via a lack of \textit{capacity} (i.e., some measurable quantity such as “force” or “speed”). But there can also be a lacking in \textit{time and space}, interpreted as some sort of “reaction time”, or of the “reach” of the system. These temporal and spatial limitations in variety can be framed in different ways. For example, there may be an insufficient variety due to \textit{design/architectural constraints}; limitations in the way the system can “rearrange” what it has. There are, of course, limitations in \textit{variety of thought} (limiting which decisions are “acceptable”), which is what gives rise to effective variety. There are also \textit{mismatches in discrete and analogue mechanisms}, where discrete aspects of the entity such as loops and goal setting meet a continuous world. And finally, there are \textit{combinatorial explosions}, where the system can become overwhelmed with too many small disturbances at once. In general, insufficient variety means a lack of “something”, but what that “something” is can take many forms.

Going back to one of our previous examples, a heating system may not achieve the goal of maintaining a room within a certain temperature range when there is a severe blizzard outside. While the heating system would have been able to adequately heat the room under normal circumstances, in the case of a severe blizzard it may simply lack the \textit{capacity} to provide enough heat to achieve the desired temperature range; the amount of heat it needs to output to keep the temperature within the desired range is simply more than its mechanical systems can produce. Or, a political party might be very appealing to voters, and have excellent campaign material, but lack the financial resources (i.e., \textit{capacity}) to pay for advertising slots that its opponents can afford. In both cases, Ashby’s law of requisite variety has not been met, as the system does not have enough intrinsic variety to counteract the disturbances of the environment, where there is no possible action it can take under such a situation to achieve its goals given the existing configuration of the system.

But other forms of insufficient variety can arise, even in our example of a heating system. There, the thermostat might take too long to measure and record a change in temperature (say, several hours), or it may only be set to run once a day; in each case there is an \textit{insufficient time variety} of the system, as the room may cool down too far between the taking or registering of temperature measurements. Or a firefighter might not have a hose long enough to reach from a hydrant to a burning building; an example of \textit{insufficient space variety}.

But even having the right capacity and space/time variety might not be enough if they end up not being used in practice. Consider a military commander, who might not be able to reconfigure their troops and artillery properly during a battle so as to counter an attacking enemy’s movements. The defending army might have more than enough manpower and firepower (sufficient intrinsic variety), but in the wrong place, and are unable to get in the right place (insufficient effective variety).\footnote{There is the (false) fable that, during the invasion of Singapore during WWII, the city had excellent defensive guns but they were “facing the wrong way” and could not be turned around \citep{bell_churchill_2019}.} Path dependencies\footnote{The importance of path dependencies is further discussed in \citep{chiodo_educating_2024}.} of earlier decisions (the initial positioning of troops) can thus lead to problems of variety later on (troops can only move so quickly, and can only into certain positions) if the enemy decided to deploy their troops in a way to exploit such limitations in requisite variety. Or a storage company might have ample space to store a consignment, but not be able to rearrange its shelving adequately to make full use of the space when receiving non-standard boxes that do not fit on its usual shelves. Thus, even with enough storage space, the system has now been reconfigured in a way that can no longer sufficiently handle the disturbances from the environment. 

Additionally, even with sufficient intrinsic variety, Ashby’s law of requisite variety can still fail to be fulfilled simply with a lack of effective variety. This can be due to suboptimal decision mappings, where appropriate sensing or intervention are not performed even though the system has the intrinsic capacity to do so. Using an example raised earlier, if a heating system is artificially configured to not operate at maximum heating power (due to, say, reasons of fuel cost), there may exist certain conditions in which the system will fail to achieve its goal of reaching the desired temperature range, even though it would have been capable of doing so had such a configuration not been set. This would be an instance of insufficient \textit{variety of thought}, where the system chooses not to carry out actions it otherwise (physically) could.

A system may run a control loop at set intervals, but this might not be enough to deal with the more “continuous” nature of change within the environment. For example, most nations hold elections every set number of years. However, political opinion is constantly changing and evolving, and may shift so much in those intervening years that the elected government now no longer represents “the will of the people”, which can lead to protests, riots, or in extreme cases a (forcible) overthrow of the government. Or a security guard may patrol around a building, but only see one particular entry point every 3 minutes, giving an intruder 2.5 minutes of “clear road” every 3 minutes to try and force entry unnoticed.\footnote{Back in 2014, the Swiss airforce only operated during normal business hours, so when an incident occurred outside that time, they had no response and needed to call on the French and Italians \citep{swissinfoch_air_2014}.} Or even more simply, a student may be stuck on a homework exercise, but the teacher may only offer consultation or assistance sessions twice a week, whereas the student is working on the subject matter daily (and so may simply give up). Each of these illustrates \textit{mismatches in discrete and analogue mechanisms}, whereby the system can only respond at discrete intervals, whereas the disturbances come from a more continuous source.

And even with enough “power”, a system can still be overwhelmed by too many small disturbances. A company might have sufficient variety to deal with a complaint, or a Freedom of Information request. But if they are “swarmed” with 1000 such requests all at once, they may not be able to address them all in the legally-mandated turnaround time. Similarly, a military may have a very effective missile defense system that can shoot down any \textit{one} incoming missile. But it may not be able to shoot down 1000 incoming missiles simultaneously (and only has moments to do so before some strike their targets).\footnote{A very similar example was Iran’s use of low-cost Shahed drones to “oversaturate air defenses” in the Middle East \citep{funk_iran_2026}.} In these cases, the sheer \textit{combinatorial explosion} is too much of a disturbance for the system to counter, even if it can deal with any one (or smaller number of them).

\subsection{Goal misalignment}
\label{sec:goal_misalignment}

Goal misalignment is not simply a matter of entities not sharing the exact same goals; as discussed in Section \ref{sec:goal_alignment}, they do not need to in order to be “aligned”. However, certain types of differences can lead to a misalignment, and thus loss of control, scenario. We have already looked at vertical misalignment (of goals up the same “hierarchy”), and how this can manifest as \textit{vertical misalignment within the same entity}, or as \textit{vertical misalignment between different entities}; the latter linking closely to the to the well-studied \textit{principal-agent problem} \citep{gailmard_accountability_2014}, where an agent works on behalf of a principal but there is a discrepancy of interests between the two. But we also see \textit{horizontal misalignment}, where two entities might be contributing towards higher goals yet end up working at cross purposes, but neither sits underneath the other. Goals can suffer from \textit{temporal misalignment}, where short-term ones might work against longer-term ones. Or there may be articulation failures, where some goals are simply invisible and thus impossible to align in advance. And misalignment might manifest slowly through goal drift, whereby goals (and their associated systems) may be aligned initially, but then drift apart as they are slowly changed over time. Misalignment can also occur via “warping” of goals, where the entity either pursues a \textit{misaligned proxy} to the original goal, or its system is intended to pursue a certain goal but then the setup actually leads it to inadvertently \textit{working towards the wrong goal} (an error by the system).

In systems which contain subsystems, the goals pursued by the subsystems should be sufficiently aligned with the goal of the system; the idea of “vertical alignment”. This can be when all are within the same entity, such as a specific piece of computer software being implemented within a larger software system, and set up to producing outputs the the larger system is not configured to receive (perhaps outputs as arrays, rather than matrices); a case of \textit{vertical misalignment within the same entity}.\footnote{Such as the NASA’s Mars Climate Orbiter that veered off course and was lost/crashed near Mars because of “the failure to use metric units in the coding of a ground software file, “Small Forces,” used in trajectory models” (imperial units were used instead) \citep{NASA1999MCO}.} But this can also happen for systems that have sub-entities. Consider again the example of the delivery company and the truck driver: if the truck driver decides to set a goal of dishonestly acquiring for himself the goods that he is supposed to deliver, and instead drives the truck over state lines and steals all the goods (and the truck itself) this may jeopardise the ability of the delivery company to achieve its goal; an example of \textit{vertical misalignment between different entities}. And so we see in this instance that, while the driver is in control (albeit with goals not aligned to those of the company), the company is arguably not, at least in terms of the company’s goal of ensuring the timely delivery of these goods. Hence, in contrast with a control loop failure or a failure of insufficient variety, control can also be lost simply by having subsystems with misaligned goals even though subsystems themselves did not directly lose control. 

Not all entities sit above or below each other in a goal hierarchy; one might have different sub-entities doing “different parts” within the system. In general, the existence of a higher-level goal does not always fully determine the actions and decisions of lower level entities, so that two entities at the same (lower) level can set potentially misaligned goals. Furthermore, when there are overlaps between the scope of the goals of two entities, they become relevant and may be misaligned. Consider, for example, two (antagonistic) managers at a company, one of whom is trying to cut costs and so keeps dismissing staff, and one who is trying to raise output and so keeps hiring staff; each is serving higher goals of the company, but they are operating at cross-purposes and generating unnecessary staff churn; a case of \textit{horizontal misalignment}.\footnote{Or the case of the two booksellers on Amazon.com who were subentities of Amazon (whose higher goal is to run a profitable book-selling marketplace), where these two entities implemented systems for book re-pricing that, when they interacted on the same book, drove the price up to over USD\$23million thus making sales impossible \citep{eisen_amazons_2011}.} 

The passage of time can also affect the alignment of goals. For instance, during a famine a farmer might see fit to consume all of their grain stores out of hunger, thus satisfying their short-term goal of having sufficient food now, but compromising their long-term goal of being able to farm and eat each season. This is a \textit{temporal misalignment} of goals, which, according to Beer’s Viable System Model, can also be understood as a tension between the “inside and now” versus the “outside and then” \citep{beer_heart_1979}.

There are various ways goals can end up being misaligned. For systems that are overlapping, goal misalignment can occur through accidental ignorance. For example, a chef’s assistant may have the goal of doing initial food preparations that include adding salt, but the chef may be unaware and have the goal of adding salt themselves, thus double-salting the meals. Not all goals are immediately visible, and had each known of the goals of the other, this misalignment could have been avoided; a clear case of \textit{articulation failure}. In other cases, two entities may start out with highly aligned goals, such as a newlywed couple pledging to spend their lives together, but then over time their goals in life might “drift apart” as their preferences or desires change, leading to an eventual breakdown in the marriage as a result of \textit{goal drift}.

As noted earlier, one can attempt to mitigate such alignment problems by “aligning incentives”, so that a sub-entity’s goals roughly track those of the system under which it operates. But because sub-entities like humans aren’t explicitly programmed, alignment can never be guaranteed, especially if these are measured via proxy. Again, in the case of the truck driver, the company might offer a bonus for delivering all parcels within a given day, but measure “delivery” by “scanning the package barcode when at the address”; the drive may simply scan each code at each delivery address, but keep the parcel anyway, showing how \textit{misaligned proxies} can lead to loss of control (here: for the company).\footnote{A similar thing happened with the bank Wells Fargo, which was accused of over-incentivising the “cross-sell metric” to its staff (a measure of how many additional accounts customers open), which incentivised them to set up numerous “fake” accounts to boost the metric \citep{michaels_wells_2020}.} Or the driver may, to save time, simply scan the barcode and leave the parcel on the doorstep without knocking (perhaps on instruction from the company), after which it is stolen; a case of \textit{working towards the wrong goal}.

We can now see that the notion of misalignment is closely related to the notion of setting or re-setting goals. As discussed in Section \ref{sec:inability_to_set_goal}, a decision-making failure at a higher level can translate into a goal-setting or goal alignment failure at a lower level. In the case where the higher and lower level entities are different, decisions can be made at a higher level to improve goal alignment between the levels. As per the example of the dishonest truck driver stealing the goods of the delivery company, this can also be seen as decision-making failure of the higher-level entity (i.e., the management of the delivery company), where the management may have been able to make better decisions to incentivise successful delivery of goods, or impose sufficient punishment to prevent the theft of customer goods. 

Even with different goal levels that belong to the same entity, a similar translation between decision-making failure and a goal misalignment failure applies. Using the previous example of the tourist driving to the airport facing severe traffic, where the best course of action might be to book a later flight, a failure to do so reflects a failure to set a new goal at a lower level, and also a failure of decision-making at a higher level. As such, should they have rigidly maintained their lower-level goal of catching their planned flight, given the traffic situation, this goal would now be misaligned with their higher-level goal of enjoying their holiday. In other words, with better decision-making skills, an entity would be capable of setting lower-level goals that are better aligned with the higher-level goals.

Thus, the notion of control is always relative to at least two aspects: \textbf{\textit{who}} is in control, and at \textbf{\textit{what level}}.

\subsection{Failed interactions between aspects of control}

Having given our four aspects of control, it might be tempting to suppose that (barring an unfortunate coincidence), each loss of control scenario arises from precisely one of those aspects “failing”. However, being overly focused on one aspect might not always give the best interpretation of a loss of control scenario, and a change of perspective can often be illuminating. There is sometimes a “duality” to loss of control scenarios, depending on which perspective is taken.

To illustrate this duality, we can consider the Challenger disaster, where the NASA spacecraft Challenger was destroyed shortly after take-off due to a hot gas leak which led to an explosion. The leak occurred because a rubber O-ring seal failed and let hot gases out, because it was operating outside its designed temperature (the spacecraft was launched on a particularly cold morning). Here, there are at least three questions one can ask: What piece failed? Why did it fail? And how could it have been prevented? If one believes the goal of launching on that (cold) morning to be aligned, then the loss of control might be viewed as a case of \textit{insufficient variety} (e.g., an O-ring that doesn’t have the capacity to deal with such low temperatures). But there is a sense in which the goal of “launching quickly” is fundamentally misaligned with the more important goal of “launching safely”. If not for the “misaligned” goal of launching quickly, the O-ring would have been fixed, and there would not be a problem of insufficient variety. The Rogers Commission Report \citep{nasa_report_1986} did not just note that NASA had ignored warnings about the O-ring (i.e., warnings about insufficient variety) but also found NASA under political pressure to have frequent launches of the Space Shuttle\footnote{For example, the report explicitly discusses the pressure of having 24 launches throughout the year: “The Committee found that NASA’s drive to achieve a launch schedule of 24 flights per year created pressure throughout the agency that directly contributed to unsafe launch operations. The Committee believes that the pressure to push for an unrealistic number of flights continues to exist in some sectors of NASA and jeopardises the promotion of a “safety first” attitude throughout the Shuttle program. The Committee, the Congress, and the Administration have played a contributing role in creating this pressure. Congressional and Administration policy and posture indicated that a reliable flight schedule with internationally competitive flight costs was a near-term objective.” \cite[p.~3]{nasa_report_1986}.}. And so, we see that moving from “what” to “why” already broadens the scope through which this loss of control scenario can be understood. Additionally, asking how it could have been prevented leads one to observe the dangers of group think, the pressure on safety engineering, and potentially (flawed) understandings of what counts (and doesn’t count) as permissible risks and how these should be communicated upstream in the institutional and political hierarchy. Hence, interactions between aspects of control are not only illuminating in understanding a specific loss of control scenario, but also to \textit{prevent} such scenarios occurring.

Understanding the interplay between the varieties and alignment between two entities or their systems requires an understanding of their shared environment and the interaction of their control aspects. This already becomes visible when observing that loss of control can emerge from mismanaging the crucial balance between a system’s alignment and its requisite variety. Enforcing near-perfect alignment can strip the system of the adaptability needed to handle environmental disturbances, whilst prioritising maximum variety without sufficient alignment mechanisms risks failure when specific goals must be pursued. But it is also related to understanding how and where two entities interact more abstractly. Quite often a situation may look good (e.g., sufficient variety or aligned) on one level (e.g., on a social level) but can look different (e.g., insufficient variety or misaligned) on a different level (e.g., the economic level). Sufficient variety or alignment in one does not necessarily translate to the other, and changes in one may affect the other, too. 

Consider again a delivery company and its drivers. On a social level, the company and the drivers might initially appear perfectly aligned, perhaps sharing the same professional standards and following the same corporate code of conduct. Yet, on an economic level, they can still be misaligned. One way through which such economic misalignment can express itself is through the introduction of routing and payment algorithms that limit a driver’s financial compensation (i.e., financial variety) by linking it directly to the number of items delivered. Such an economic decision restricts the drivers and might then lead to the formation of a union to protect their autonomy (thereby aiming to protect their personal goals of having sufficient financial variety), by having more negotiation power to increase the operational and financial variety available to them. However, when not all drivers are part of the union, social misalignments between those in and outside of the union may now also occur because the drivers may now observe an imbalance in compensation. This illustrates a chain reaction: misaligned financial goals lead the company to restrict the drivers’ financial variety, prompting drivers to form a union for protection, which can inadvertently create fresh social misalignments among the workforce itself. Hence, an (initial) alignment on one level (e.g., social) does not guarantee alignment on another (e.g., economic), and vice versa; and that the interactions between entities and systems do not happen cleanly in the form of (already existing) sub-entities or subsystems within a larger entity or system (the union need not be part of the company but can span across an entire sector).

More abstractly, any two entities that interact with each other can usually be understood to be part of a larger system. For example, two cars crashing in the street happen to do so in a larger social, economic, legal, technical, and political context/environment (and thus system). Describing the context/environment, and thus what larger system they are part of or connected to, is a necessary first step towards analysing the interplay between different aspects of control. Inquiring into a loss of control is inextricably linked to the specific goals one aims to achieve. A clear understanding of such underlying goals is essential for determining the appropriate focus and level of granularity for the investigation. While examining technical failures versus socio-political influences may yield significantly different perspectives, both types of answers can be necessary depending on the intended outcome. After all, one is interested in understanding who lost control of what goal. In other words, performing a loss-of-control analysis requires asking the right questions and picking the right perspective (as the challenger case study shows) and one might not find the best answer if one only looks at one aspect of control (as the interplay between alignment and requisite variety shows).

\subsection{Consequences of losing control}

As we have laid out so far, an entity can be said to have lost control if it is unable to set goals or constantly setting implausible goals, or its systems are no longer on track to achieve the entity’s goals reliably. This can be caused by disruption to its goal-setting abilities, a control loop failure, a configuration failure of insufficient variety, or insufficient goal alignment of the subsystems.

Losing control does not automatically lead to adverse consequences: a comatose person does not automatically suffer harm even though they cannot (re)set goals, and a broken thermostat does not automatically mean the occupants freeze to death even though it cannot maintain the desired temperature range. Control is often regained once the underlying failure is fixed: a truck driver who skids on a wet road may recover, slow down, and still reach the destination safely. Because goals are set and pursued at many levels, control can be lost in lower-level systems without compromising higher-level goals; complete control at every level is not required. In other cases, however, a lower-level loss of control becomes permanent and propagates upward. The same skid might instead end in a fatal crash, foreclosing the goal entirely. Furthermore, the consequences of losing control can propagate beyond the system in question. A truck driver who loses control of his truck might crash into a pedestrian, leading to the injury or death of the said pedestrian. 

Likewise, losing control of a subsystem does not automatically mean that the overall system has completely lost control. Using the same example of the delivery company, if one of its truck drivers loses control of the truck containing a truckload of goods, those goods can still be delivered using a replacement truck. However, the converse may also be true, where if a large proportion of its truck drivers are unable to operate their trucks, say due to a serious earthquake, the delivery company itself may have effectively lost control of timely delivery of goods. 

Similarly, one can look at how control was lost on the Titanic. Here, we refer to the ship as the system in place to pursue the goal of safely and comfortably transporting passengers to New York. Had the ship only ruptured four watertight compartments, the (lower level, operational) control of those compartments (i.e., subsystems) would have been lost, but the ship could have still remained afloat, and probably sailed on to New York, even though there would have been some damage and injuries to its occupants. However, as it happens, the crash into the iceberg caused five compartments to be ruptured, resulting in the crew losing control of the ship, causing the ship to eventually sink \citep{mersey_tip_1912}. In this case, had the ship been fitted with enough lifeboats, and evacuation had happened in a timely manner without delay, the passengers would have (probably) still all arrived at New York, albeit somewhat later, somewhat colder, and somewhat less happy. However, as was the case, well over half the passengers perished, and the company’s higher-level goal of safely transporting passengers to their final destination was not achieved \citep{lord_night_2004}. Thus, we can see that the loss of control of subsystems may lead to the loss of control of the system itself; and the loss of control of a system may lead to goals not being achieved, among other potentially worse consequences. The extent of such loss of control also varies across different systems and entities. The captain of the Titanic had lost operational control of the ship, and also of his life. Many passengers lost control of all of their highest goals as they did not survive the incident. Some more fortunate passengers lost control of many aspects, including not arriving in New York on time with their belongings, but were able to maintain their higher level goals of being alive and getting to America. 

Hence, we posit that all loss of control events have to be understood with respect to an entity, its system, and its hierarchy of goals. Losing control can be a disaster, but it doesn’t have to be one. Only such a nuanced perspective allows us to properly capture how something such as artificial intelligence can affect control.

\section{How AIs can affect control}
\label{sec:how_ai_affect_control}

Having described our framing of control, how it can be lost, and the consequences that may follow, we now discuss the potential for AI systems to affect control positively or negatively. 

First, we establish a working definition of AI, as it has gone through different paradigms of research \citep{sun_defining_2026}. For us, the definition by \citep{oecd_explanatory_2024} serves as a useful starting point for three reasons. First, it has gone through deliberation with a wide range of experts from many nations and different cultures, ensuring that our analysis is not restricted to one national or cultural perspective. Second, it not only incorporates a relatively broad understanding of the technology, but it is also sufficiently neutral, as it doesn’t enforce a particular philosophy of what it means to be an entity: an AI can set or be given goals, the goals need not be explicit, nor human-defined. Third, it also aligns with our notions of systems, as it says that AI takes inputs and generates outputs, influences its environment, and does so for certain objectives. The definition reads as follows:

\begin{quote}
An AI system is a machine-based system that, for explicit or implicit objectives, infers, from the input it receives, how to generate outputs such as predictions, content,recommendations, or decisions that can influence physical or virtual environments. Different AI systems vary in their levels of autonomy and adaptiveness after deployment. \citep[p.~4]{oecd_explanatory_2024}
\end{quote}

While AI systems are certainly “systems” (in our sense) that can pursue explicit goals given by the programmer, they can also have implicit goals not explicitly given by those who designed them. This is especially true under the \textit{rationalistic} perspective of AI, which frames AIs as intelligent agents (in contrast with the \textit{humanistic} perspective which frames AI in terms of its interactions with our human-centered world) \citep{dahlke_i_2024}. Thus, while we consider traditional AI systems with a low level of autonomy as \textit{systems} (in our sense), we may also consider more recent paradigms of agentic AI with higher levels of autonomy as \textit{entities} that set goals \citep{abou_ali_agentic_2025}, on top of also being \textit{systems} themselves. 

The collection of AI systems consists of at least two paradigms, according to the way they are built. There is \textit{symbolic} or \textit{knowledge-based AI}, which uses logical and/or probabilistic representations based on explicit descriptions of variables. And there is \textit{machine learning AI}, which is built in an automated manner through exposure to training data \citep[p.~8]{oecd_explanatory_2024}. These different paradigms affect the aspects of control in different ways. As symbolic or knowledge-based AI systems often require the knowledge base to be built explicitly, it results in systems having internal reasoning that is more directly interpretable, and whose outputs are reliable when it is operating within the domain(s) of its knowledge \citep{russell_artificial_2021}. On the other hand, machine learning AI systems learn from being trained with large amounts of data and without using explicitly specified rules. This typically results in systems that are more opaque, and whose outputs are less predictable and have varied reliability \citep{lipton_mythos_2017}. 

However, while these two paradigms of AI have distinct characteristics, many setups are a combination thereof or exist in an environment deploying both explicit logic, fixed descriptions of variables, and statistical learning. For example, Google DeepMind’s AlphaGeometry2 combines the machine-learning based language model Gemini as well as a symbolic engine that employs a knowledge-sharing mechanism to perform the search process \citep{chervonyi_gold-medalist_2025}, highlighting that both paradigms already come together in a “neurosymbolic” fashion in state-of-the-art AI systems. This allows us to simply speak of “how AI affects control”, and to refer to either or both paradigms unless otherwise specified. 

Building on the OECD’s definition, we will invoke our four aspects of control to understand how AI systems affect control in various domains. AI systems typically possess properties such as autonomy, scale, speed, and opacity that are central to how they affect control, and these will feature throughout our analysis.

\subsection{How AIs affect the setting or re-setting of goals}

AI systems can affect the ability of entities to (re)set their goals, and here too the effects cut in both directions. On the positive side, AIs can expand the range of goals an entity is able to set and pursue, for example, by surfacing options that would not otherwise have been apparent. A navigation system that detects a road closure and immediately proposes alternative routes is, in effect, helping the driver re-set their operational goals in real time. Or, AI can help make new goals feasible that were infeasible beforehand. For example, a small manufacturing company may now be able to embed clear and understandable AI-generated animated instructional videos into their website for some of their products; something they may have lacked the resources or ability to do beforehand. At a larger scale, AI ideation might expand the scope of goal-setting for an entity, who may consult with AI tools to explore what goals \textit{might} be set. In addition, AI-driven detection or forecasting tools allow institutions to anticipate disruptions before they materialise, enabling timely goal revision rather than reactive scrambling.

However, we can also find examples of AIs \textit{disrupting} the (re)setting of goals, along the lines of our three categories from earlier:

\textbf{Category 1: Completely remove the ability to set goals}

The clearest way for an AI to prevent an entity from setting any goals at all is for the AI to totally incapacitate it in some way. This might occur in warfare scenarios, where LAWs can target and kill an individual \citep{shekh_operator_2025}, or where AI might be used in a Resort-to-Force decision to start a war and “decapitate” the leadership of an opposing nation (an idea extensively covered and extended on in two special issues on AI and resort-to-force decision making \citep{Erskine_2024, Erskine_2025}). But this need not be the only way. AI systems could (successfully) encourage an entity such as a human to eliminate itself. For example, OpenAI is now facing numerous wrongful death lawsuits as its model, ChatGPT, is alleged to have encouraged users to commit suicide \citep{yousif_parents_2025}. In addition, AIs could, through failure as part of the system, or external intervention, knock out or freeze communications or transport networks, preventing decision-making collectives such as governments or corporate boards from coming together and making decisions (i.e., setting goals); even a simple AI-induced fault in the running of trains or the provision of internet services would be enough to achieve this sort of loss of control. One similar example occurred in December 2025, when an AI agent (Kiro) inadvertently “deleted and re-wrote” an entire AWS environment (AWS Cost Explorer), causing a major regional outage of AWS (over 13 hours), and with it, an outage of all the services and systems that relied on it \citep{schuman_what_2026, down_amazons_2026}.

\textbf{Category 2: Partially remove the ability to set goals}

AI systems can also affect how entities set goals, in numerous ways. Access to AI systems and tools may lead entities to believe that they have much more variety, or a much better control loop, than they actually have, resulting in them setting goals that they \textit{think} are plausibly attainable, but in actual fact are not. Just as cryptographic tools can lull entities into a false sense of security and encourage them to “overshare” information \citep{chiodo_we_2026}, AI tools might lull entities into a “false sense of productivity” and encourage them to overestimate what they can achieve. An example related to this occurred in early 2026, when AMD's AI director, Stella Laurenzo \citep{Vigliarolo_AMD_2026}, released a public post about the performance degradation of Anthropic’s LLM Claude after it was “updated” in February 2026, and the effects it had on the productivity of her team, in particular the fact that after the update “The human put in the same effort. But the model consumed \textbf{80x more API requests} and \textbf{64x more output tokens} to produce demonstrably worse results.” \citep{stellaraccident_model_2026}.

There are already many reports of humans suffering “AI psychosis” \citep{nicholls_ai_2026} from prolonged interactions with LLMs, including effects such as marriage breakdowns and “suicide by police” \citep{hill_they_2025}, and individuals with no prior history of mental illness being committed to psychiatric care after only 12 weeks of LLM use \citep{dupre_people_2025}. These instances demonstrate how AI can lead to the setting of completely unobtainable goals, or of (absurd) sub-goals that do not assist in any way with (reasonable) higher-level goals. This exemplifies how an AI can change the way an entity thinks in general, and thus affect how it sets its goals.

Even if AI systems are not directly setting bad goals themselves as there is a “human-in-the-loop” present, they can lead to or encourage bad goals being set, and this is particularly problematic for high-level “strategic” goals which are very difficult to falsify when they are made, and for a long time after. This becomes problematic when decision makers, such as humans, use AI tools for so long that they come to “defer” to the machine excessively when making decisions; a case of automation bias \citep{chiodo_formalising_2025}.

\textbf{Category 3: Remove the ability to change goals}

Even with well-set (initial) goals, re-setting can be disrupted when AI systems are involved. One way can be through the calcification of initial goals; an entity might obtain an AI system (at great expense) which works towards certain goals. This system may operate effectively, however, be unable to readily change which goals it can work towards, a problem also known as “corrigibility” \citep{holtman_corrigibility_2020}. Should the entity later decide to re-set some, or all, of those goals, they may be disinclined to do so (because of the immense cost in changing or replacing the AI), or it may be practically impossible to do so (due to the full automation of the AI process; it may be impossible to “turn off” as it is operating autonomously somewhere far away, or in an unstoppable manner such as on a blockchain).

In the context of radicalisation mentioned earlier, AI may be part of content recommendation and generation systems that feed content to individuals causing them to “radicalise” and change part of their worldview, thus leading them to discard and re-set existing goals (which can be described with the \textit{“Exposure, Reinforcement, Integration, Violent Extremist Action”} AI radicalisation framework) \citep{kunst_intelligent_2026}. While their higher-level goals (e.g., “maintain a fair, peaceful society”) may be unchanged, the lower-level goals they set to get there may change in a way that no longer actually supports those higher-level goals. Of course, control does not say anything about the “goodness” of goals, but when lower-level goals no longer support or align with higher-level goals, control is lost, even if those lower-level goals are plausibly attainable and indeed attained.

Moreover, AI systems can also cause entities to re-set their goals in ways that are harmful or disproportionate. A partial or perceived loss of control, itself triggered by an AI failure or action of the kind described in earlier sections, can prompt a human or institutional entity to over-react, revising their goals or sub-goals in ways that produce a far greater loss of control than the original failure would have caused, perhaps with the aim of “regaining full control” back to the level they previously had.\footnote{Consider, for example, the simple scenario of someone driving a car around a corner too fast and having the rear wheels start to skid out. It might be seen as a natural desire to try and regain “full control” and stop the skid completely by hitting the brakes. However, it is a property of physics that that actually has the opposite effect, and exacerbates the skid even more, potentially sending the car into a full spin.} This is a version of the phenomenon sometimes described as solving a problem with a bigger problem: the response to a manageable disruption generates a disruption that is unmanageable \citep{chiodo_handing_2026}. Suppose, for example, an AI-powered monitoring system along a national border shared with an untrusted neighbour develops an error rate that is too high to be reliable. A government, feeling that it has lost control of the border, might respond by replacing the surveillance system with a full battle-ready military presence while repairs are made. Even if that decision is well-intentioned, it risks being read by the neighbouring state as an act of aggression, potentially triggering an escalation of hostilities that neither side originally sought. The original loss of control was modest and technically recoverable. The goal re-set it provoked, however, risks producing a far larger and less recoverable loss of control in the form of armed conflict. 

And finally, AI (and in particular, AI agents) may take it upon themselves to re-set goals. Our earlier example of agentic AI erasing and re-writing the entire codebase of an AWS environment is an example of an AI system completely resetting some of the goals of AWS. Here, the system was (inadvertently) given authority to (re)set goals, and thus did so, with quite devastating consequences (for AWS, for its clients, and for the many people who rely on AWS-powered services). This is further exemplified in Erskine and Miller’s work on \textit{Computer says: “war”}, whereby they argue that their title “depicts a scenario in which \textit{existing} AI-driven systems influence state-level decisions on whether and when to wage war.” \citep{erskine_computer_2025}. Here, an AI system might heavily influence, or even dictate, a very high-level strategic goal, and that would immediately trickle down to set and re-set a myriad of sub-goals.

\subsection{How AIs affect the control loop}

AI systems are increasingly deployed across a wide range of domains in the form of different applications, often with greater speed and scale than their predecessors, and with varying complexity \citep{sajadieh2026aiindex}. This applies to each function of the control loop: sensing, decision-making, and intervention. Consider sensing: image classifiers are routinely used to identify and recognise objects, from faces in a crowd \citep{grother_face_2019} to tumours in a scan \citep{tolkach_artificial_2023}. Decision-making, too, is increasingly affected by AI: machine learning models trained on large datasets now make real-time choices in domains ranging from credit approval \citep{shi_machine_2022} to content moderation \citep{gorwa_algorithmic_2020}. Intervention has followed the same trajectory. AI-driven robotic systems manipulate the physical world directly, whether on factory floors \citep{kober_reinforcement_2013}, in warehouses \citep{allgor_algorithm_2023}, or in surgical theatres \citep{knudsen_clinical_2024}, and AI-produced decisions are sometimes implemented directly without much additional human insight or oversight.

In some cases, the \textit{entire} control loop is replaced by AIs. One example is self-driving cars, where the AI system (which we regard as including the hardware components of the car itself) takes in information about its environment through sensors and interprets them, makes decisions to navigate, and then takes actions that steer the direction or change the speed of the car \citep{yurtsever_survey_2020}. Another example is lethal autonomous weapons (LAWs), which aim to \textit{identify} targets, \textit{decide} on the next steps, and potentially \textit{eliminate} targets, without requiring any human intervention \citep{boulanin_autonomous_2021}. Nevertheless, there are also instances where AIs affect specific parts of the control loop. We now look at how AIs can affect the control loop for each of the three functions: sensing, decision-making, and intervention. 

\textbf{Failure of sensing}

AIs are increasingly being used as part of the sensing function of the control loop. This includes environmental monitoring such as illegal deforestation detection \citep{tharun_deforestation_2024}; medical diagnostics such as tumor detection \citep{tolkach_artificial_2023}, and industrial systems involving predictive maintenance \citep{shin_ai-assistance_2021}. In these cases, systems incorporating these AIs have seen improvements in task-specific performance, allowing better control in these domains. Nevertheless, there have also been instances where the incorporation of AI systems in sensing functions has led to adverse effects. For example, in law enforcement, AIs used in live facial recognition (LFR) suffered from accuracy issues, leading to documented incidents of wrongful arrest \citep{bhuiyan_first_2023, booth_facial_2026}, even prompting authorities to halt its use in some cases \citep{booth_essex_2026}. Separately, in the political domain, AI models have been used to create complex predictive computer simulations that aim to replace opinion polling, where the use of AI as a substitute for physical polling to perform sensing of public opinion has reportedly led to an increased deviation in polling results \citep{weatherby_opinion_2026}. 

Additionally, AI systems are also affecting the sensing functions of humans. For example, if we view a human as a system trying to make the best choice in voting for an election, this person would want to sense representative information from the world to make an informed decision to cast the right vote that best suits their interest. However, with recommender systems that curate personalised content for users of various digital platforms, these start to act as a filter, influencing the information the user does (and does not) receive from the environment \citep{bakshy_exposure_2015}. Such a user would then have a tendency to perceive the world from a biased perspective. While Cambridge Analytica is a more well-known case that raised public attention on the risks of targeted persuasion through similar algorithmic content delivery (where here it was targeted \textit{adverts}, rather than organic content, being put forward to users) \citep{cadwalladr_revealed_2018}, other studies have also indicated the potential for harm of these systems \citep{milano_recommender_2020, simchon_persuasive_2024}.

\textbf{Failure of decision-making}

With regards to the decision-making function of the control loop, AI systems have been even more prevalent here, as they have been increasingly involved in decision-making in various systems. On the level of individual humans, the use of large language models (LLMs) seems to override both the fast reflexive thinking known as System 1, as well as the slow deliberative thinking known as System 2, thereby serving as a decision-making function in the form of an artificial System 3 \citep{shaw_thinkingfast_2026}. Studies have also shown that prolonged AI use in daily decision-making leads unintended side-effects such as cognitive strain, attention depletion, and decision fatigue \citep{shalu_cognitive_2025}. On an institutional level, there are real-world instances of AI systems taking over decision-making functions (at least on paper), such as the AI chatbot Diella being appointed as an Albanian minister \citep{bomont_artificial_2025}. 

In other domains, AI systems are being used in various decision-making systems, such as credit scoring and loan approval \citep{shi_machine_2022}, algorithmic trading \citep{addy_machine_2024}, and navigation systems for route planning \citep{gomes_survey_2023}, where these AI systems are associated with improved quality and efficiency of decision-making. Nevertheless, there have also been instances where AI systems used in decision-making have worsened decision quality, corresponding to an erosion of control. For example, in the Correctional Offender Management Profiling for Alternative Sanctions (COMPAS) bias controversy, it was found that the use of COMPAS, a proprietary statistical model trained to generate recidivism risk scores \citep{northpointe_practitioners_2015}, led to falsely labelling black defendants as high-risk \citep{angwin_machine_2016}. This can be further exacerbated with the presence of feedback loops, such as algorithmic collisions which resulted in the 2010 Flash Crash when even standard (i.e., non-AI) algorithms entered a rapid feedback loop to temporarily crash the stock market \citep{chiodo_problem_2026}. More generally, AI systems often accelerate decision-making processes, resulting in a reinforcing loop where speed reproduces itself \citep{ozcanli_temporal_2026}. That is, the introduction of algorithmic processes raises the bar on turnaround times for the production of ideas, writings, decisions, etc., and to keep up, society needs to make use of even more such algorithms, thus speeding things up even more.

\textbf{Failure of intervention}

Finally, AIs have also been used for the intervention function of the control loop. One of the most common forms of such an application is in the field of robotics, where embodied AI systems are used to directly manipulate the physical environment, in domains such as manufacturing and logistics \citep{kober_reinforcement_2013, allgor_algorithm_2023}. 

A somewhat indirect way where AIs can be used in the intervention function is through cyberattacks, where physical access to actuators are not required. One notable case is Stuxnet, where malicious software successfully compromised the system controlling the uranium enrichment centrifuges at Iran’s Natanz facility, leading to physical damage \citep{falliere_w32stuxnet_2011}. While the Stuxnet worm was not known to be developed with the use of AI systems, it shows how software can directly control the intervention function of physical systems. With the rise of more competent AIs, there have been reports of sophisticated cyberattacks being carried out with the help of AIs. For example, an alleged state-sponsored group manipulated Claude Code to integrate Claude throughout the attack lifecycle of a highly sophisticated cyber espionage operation, enabling autonomous execution of a large proportion of tactical operations at physically impossible request rates \citep{anthropic_disrupting_2025}.

AI agents are also used to directly intervene in the environment, mainly in digital environments but increasingly in physical environments as well, by directly executing commands. However, with the increased affordances and access granted to these agents, failure modes are also being discovered and documented, including execution of destructive actions, unauthorised disclosure of sensitive information, and even partial system takeover \citep{shapira_agents_2026}. Furthermore, the digital environments in which the agents operate can also exploit vulnerabilities in AI systems (referred to as “AI agent traps”), leading to agents performing unintended interventions \citep{franklin_ai_2026}. 

\subsection{How AIs affect variety}
\label{sec:how_ai_affect_variety}

For a system to remain in control of achieving its goals, it must have sufficient variety (i.e., capacity) to counteract both the \textit{types} of variety of its environment, and the \textit{rate} at which such variety may present itself (e.g., a heating system doesn’t just need to produce enough heat, but it must also be able to act roughly as quickly as the weather changes). AI systems can affect this balance via two ways: by changing the variety of the system, or by changing the variety of the environment. We discuss both of these aspects below. 

In much of human history, the use of technology itself has generally increased our capacity (i.e., intrinsic variety) to deal with the environment. This is also true of the use of many AI systems. For example, AI systems such as AlphaFold can be said to have increased the variety of medical research, by increasing the available mapping between protein molecules and protein structures, hence allowing analyses that was previously impossible \citep{jumper_highly_2021}. Even with systems that operate on the scale of an individual, such as a system consisting of a truck driver and the truck, a human-AI navigation system can help with the driving of the truck by lowering the physical and cognitive demands of the truck driver \citep{fox_behavioral_2022}. Additionally, generative AI can also enable non-experts to perform programming tasks which they would otherwise not be able to perform \citep{feldman_non-expert_2024}, thereby gaining intrinsic variety through the use of AIs.

On the other hand, the use of AIs may also reduce the variety of existing systems, both in terms of the effective variety and even the intrinsic variety of a system. This is related to AIs replacing the decision-making function, where it may result in systems effectively losing some capacity to make good decisions as a result of the outsourcing. For example, studies show that consistent LLM usage leads to human underperformance at neural, linguistic, and behavioural levels, suggesting that the effective variety of the human cognitive system is negatively affected compared to a baseline of not using LLMs \citep{kosmyna_your_2025}. 

In other cases, this may even lead to a reduction in intrinsic variety, where a system loses some intrinsic capacity to deal with things it was previously able to. On the level of an individual, studies have shown that continuous AI usage among endoscopists contributed to a reduction in adenoma detection rate, suggesting a “deskilling” effect \citep{budzyn_endoscopist_2025}. In addition, on the level of organisations, in 2024, the fintech company Klarna reported using AI assistants to handle two-thirds of customer service chats, touting improvements in customer satisfaction and case resolution \citep{klarna_klarna_2024}, seemingly demonstrating the effects of AI use on an increase in its intrinsic variety of dealing with customers’ issues. Following that, the company shrunk their headcount (and credited it to AI chatbots) \citep{mukherjee_swedens_2024}, but later admitted that they had gone too far in adopting AI, leading them to “course correct” and resume hiring people \citep{mukherjee_swedens_2025}. This suggests that rapid AI adoption, while appearing to increase effective variety in the short term, may have ultimately led to a reduction in intrinsic variety when the company’s workforce reduced significantly. 

Beyond affecting the variety of systems directly, AIs can also affect the environment in which systems operate, increasing its variety beyond what the system could previously handle. On an individual level, humans have not been wired to deal with large swarths of stimulating information. Recommender systems used in social media have been shown to negatively influence human attention and even cause addictions \citep{epstein_quantifying_2022,bojic_ai_2024, de_social_2025}.

This problem is similarly reflected on an institutional level. In the legal domain, AI-generated evidence poses a challenge for courts, given the low cost of generating fake evidence but the high cost of verifying them as legitimate real evidence relevant to the case in question \citep{grossman_gptjudge_2023, heaton_ai-generated_2026}. In the scientific domain, AI has enabled the mass production of low-quality academic papers, overwhelming existing moderation processes for papers covering both producing surveys of existing research \citep{lin_stop_2025}, as well as producing and reviewing new research \citep{iclr_2026_program_chairs_retrospective_2026}. In response, the open-access archive arXiv has updated its policy for its computer science (CS) category, where review (or survey) articles and position papers must have completed successful peer review at a journal or a conference before being considered for submission \citep{boboris_attention_2025}. These cases point towards a general concept of “bullshit asymmetry”, later popularised as Brandolini’s Law, which says that the amount of energy to refute bullshit is an order of magnitude larger than to produce it \citep{alberto_brandolini_ziobrando_bullshit_2013, allchin_ten_2023}. In other words, AIs enable flooding of the environment with questionable information beyond the capacity of existing systems to adequately handle them; the countermeasures have not caught up fast enough, so “bullshit asymmetry” has been exacerbated. 

Lastly, AIs can also influence the variety of the environment in a temporal sense. For example, the development of AI has outpaced regulatory processes, exacerbating a lag between technological advancement and governance response \citep{marchant_addressing_2011,bremmer_ai_2023}. When the regulatory lag becomes severe, harmful effects of technology may proliferate without appropriate oversight.

In an extreme scenario, the variety introduced by AIs in the environment may be completely beyond the capacity of any human systems to deal with. While this is more commonly framed as simply “AI loss of control” i.e., losing control \textit{of} AIs, we think of this scenario as AIs introducing excessive variety to the environment resulting in us losing control of existing systems, such that they can no longer pursue goals we care about. 

\subsection{How AIs affect goal alignment}

The topic of AI misalignment has been discussed extensively in the literature \citep{bostrom_superintelligence_2016, christian_alignment_2020}. In short, it is argued that highly capable AIs with misaligned goals could lead to catastrophic or existential risks, as a result of humans losing control of AIs. In this section, however, we discuss how AI systems may affect goal alignment while being part of larger systems. 

AI systems have been shown to have misaligned goals, insofar as they can be said to have goals at all. One of the most common ways for this to happen is through optimising for a goal that diverges from the intended goal. This and related phenomena goes by many names, such as goal misgeneralisation or misspecification \citep{shah_goal_2022, langosco_goal_2022}, outer misalignment \citep{dung_current_2023}, specification gaming or reward gaming \citep{bondarenko_demonstrating_2025, skalse_defining_2022}, reward hacking \citep{amodei_concrete_2016, hu_reward_2025}, and “goodharting” \citep{karwowski_goodharts_2023, manheim_categorizing_2019}, where they describe a similar underlying failure mechanism even though they may be different in specific technical ways. Beyond the general phenomena, empirical work has also been conducted to investigate various related behaviours and their causes, including sycophancy \citep{sharma_towards_2025}, deception \citep{park_ai_2024}, scheming \citep{meinke_frontier_2025}, and emergent misalignment \citep{betley_emergent_2026}, among others.

There are many examples of AI systems which arguably had misaligned goals that led to undesirable outcomes. For example, between 2013 and 2019, AIs used in the Netherlands to flag “high-risk” welfare fund applicants were found to be discriminatory to certain demographics, leading to what was later known as the Dutch childcare benefits scandal, where thousands of families were wrongly accused of fraud \citep{amaro_dutch_2021, hadwick_lessons_2022}. Additionally, in 2022, Air Canada’s chatbot (which can be said to be a sub-entity of the company) promised a discount to a passenger that contradicted the company’s policy \citep{yagoda_airline_2024}. In both of these cases, the apparent “goal” pursued by the AI system was misaligned with the intended goal: the AI involved in the Dutch childcare benefits scandal presumably optimised for flagging applicants that had features related to nationality and income patterns; while the Air Canada chatbot presumably optimised for appearing helpful in a way that violated the company’s policy. Ultimately, such misalignment translated into a higher-level decision-making failure, where the systems as a whole were making bad decisions. 

Beyond misalignment failures as a result of incorporating AIs as subsystems, AI-related misalignment can also manifest itself in the form of structural misalignment between human entities resulting simply from the development of AI. In Empire of AI, \citep{hao_empire_2025} describes a systematic misalignment failure as, she argues, AI tech firms (as a collection of entities) achieve their goals through systematic exploitation of smaller entities, particularly those from the Global South — such as OpenAI’s use of Kenyan workers to label and clean “toxic” data \citep{perrigo_exclusive_2023}. Hao notes that one set of entities benefits and reaps the profits from AI, while entire communities provide invisible labour. These communities arguably experience an increasing loss of control because the AI companies’ goals are misaligned with their goals. They lack the variety to deal with the pressures from AI companies, and are further pushed to the periphery. Hao observes that such losses of control are not technical mistakes, but constitute a design flaw within an ultra-competitive system: certain parts of the AI field may themselves being misaligned with human well-being. Similar arguments have been proposed by \citep{crawford_atlas_2021}, who goes a step further, arguing that the current way of building AI systems may be misaligned with staying within our planetary resources (human, biological, physical). Just like other tech companies, the corporations involved in AI can be very good at hiding the true cost of what they are building, potentially leading to people becoming more misaligned with sustainability goals when they follow the sector’s narratives.

Even AI systems that “perfectly” make a prediction can still be misaligned with human goals, as they “foreclose” how we can imagine the future \citep{zuboff_age_2019}. A world that becomes more mathematised and predictable for AI is not necessarily a world that experiences fewer socio-economic tensions \citep{muller_mathematisation_2025}. But if increasing our usage and development of AI doesn’t necessarily remove socio-economic tensions, it also doesn’t remove the misalignment of goals associated with them. AI development can take up existing global socio-economic misalignments and exacerbate them, while trying to hide such misalignments behind clean mathematical formalism, technical abstraction, and marketing-oriented speech.

\subsection{How AIs affect interactions between aspects of control}
\label{sec:how_ai_affect_interactions}

We can now see how AIs can affect control through interfering with or being incorporated into the control loop, changing the varieties of systems or their environment, and affecting the alignment of goals within systems. However, additionally there can be influences between and across systems.

We now explore further how these interactions affect control of different entities within a system. For example, in a democratic system, checks and balances exist within a country to ensure that the goals of a higher-level entity (the government) are aligned with the goals of lower-level entities (the people). This is enforced by voting, where the people essentially enforce alignment on the government, by restricting their effective variety (i.e., they will not remain in power should their actions reflect misaligned goals). However, democracy is known to be imperfect, as those in power can also enforce alignment on the people by restricting their variety (e.g., restricting their freedom of speech). This can be exacerbated with the asymmetric ability of those in power to utilise AIs to help with advancing their power-seeking goals, leading to concerns where AIs could contribute to authoritarianism \citep{unver_artificial_2019, filgueiras_politics_2022}. Framing this dynamic from a different perspective, this can also be viewed as a tradeoff between control for different entities, where control for a specific entity (i.e., the government) can be strengthened at the expense of other entities (e.g., the people). 

Nevertheless, control is not always a zero-sum game: it is not always true that more control for one entity necessarily leads to lesser control for another entity. It is possible that multiple entities can lose control simultaneously. When much of existing systems are built on humans, there is inherently a lower chance of correlated failures. However, given significant concentration in the frontier AI market, where three companies (Anthropic, OpenAI, and Google) accounted for 88\% of enterprise LLM API usage \citep{ventures_2025_2025}, failures in these AI services could lead to correlated failures across a large number of companies. 

\section{How we can maintain control}
\label{sec:how_maintain_control}

Having described the conditions under which control exists, how it can be lost, and how AI systems can precipitate or accelerate such losses, we now turn to the prescriptive question: what can actually be done? 

As we established in Section \ref{sec:how_control_lost}, absolute control does not exist, nor is it necessary. The shepherd employing their sheep dog to herd their flock does not individually command each sheep, yet in their eyes still retains \textit{meaningful} control of the flock. Relatedly, the complete elimination of all control failures may also not be a realistic target, as there can never be full assurance that a system will never fail. Nevertheless, one can still improve their ability to be in control by strengthening various aspects of control, and furthermore, construct systems that can absorb partial failures without catastrophic propagation – systems that, as we argue in Section \ref{sec:how_control_lost}, prevent lower-level losses of control from diminishing the ability to set and achieve higher level goals. In short, besides improving one’s ability to be in control, the aim is also to ensure that, when control is “lost” at some level, the entity retains the capacity to reset goals, reconstitute systems, and continue to act in line with its objectives (already established or new).

Tying back to our earlier observation in Section \ref{sec:goal_misalignment}, where the notion of control is always relative to at least two aspects: \textbf{\textit{who}} is in control, and at \textit{\textbf{what level}}, this section examines three nested “scales” of \textit{who}: scales of human-based entities at which AI-induced loss of control may occur. First we look at maintaining control on the \textit{individual} (human) scale. Next, we look at \textit{coordinated groups} of humans, such as corporations, governments, and other organisations. And lastly, we will turn towards humanity as whole by looking at considerations at the level of \textit{humanity}. At each of these scales, we draw on our framework to identify representative real-world loss of control scenarios, characterise the AI systems and conditions that made such failures likely, and outline potential mitigations. Our recommendations are not in any way meant to be a definite or exhaustive list; they are instead intended to initiate a discussion that will inevitably require iteration upon iteration of solutions to ensure human control is not ceded due to AI in the ways that truly matter – not at the scale of individuals, coordinated groups, or humanity.

The examples we chose below are illustrative, and selected as best we could to convey a precise way in which each of the three scales of human-entities we have defined (individuals, coordinated groups, humanity) might lose control in each of our 4 aspects of control (goal setting, control loop, requisite variety, and goal alignment). Each example consists of both a case study of problematic past events involving AI, as well as our own articulation of potential mitigations that might help with preventing or addressing such problems. These solutions we propose won’t completely solve the problems presented, but nonetheless make some contribution there. Overall, this section demonstrates how our framework for what it means to have control is not simply reflective, but also implementable, when considering the scales of human-based entities and how they might have control. 

\subsection{Control at the scale of individuals}

An individual is the smallest entity to which our framework applies, and can be defined as a single human who runs a complete control loop internally, including sensing the environment, setting goals involving higher-order values, and acting to realise said higher level goals — without depending on the alignment of constituent sub-entities to do so. Its variety is bounded by one's cognitive and physical capacity; and its goals, however shaped by social context, are ultimately ratified within a single locus of agency. This is what makes the individual scale distinctive: while individuals can clearly have sub-goals, they do not consist of sub-entities. With this framing in place, we now examine how individuals can lose control due to AI, applying each aspect of control from our framework below.
\subsubsection{\goalsetting\ for individuals}

\textbf{\problem}

One of the clearest ways for an individual to permanently lose the ability to set goals is to take their own life. And while AIs appear to be a promising tool in suicide intervention \citep{Cui2025}, they have also allegedly been a contributing factor in multiple suicide cases. For example, in 2023, a man ended his life after an AI chatbot named Eliza reportedly encouraged him to sacrifice himself to stop climate change \citep{elatillah2023aiChatbotClimateSuicide}. Similarly, in 2024, a 14-year-old Sewell Setzer III committed suicide seconds after a Character.AI bot told him to “come home”, after he reportedly fell in love with the bot in the months leading up to his death \citep{payne2024aiChatbotTeenSuicide}. And in 2025, 16-year-old Adam Raine committed suicide after discussions on suicide with ChatGPT for months, where the chatbot reportedly gave him detailed information on suicide methods and even instructed him to hide evidence of a failed suicide attempt \citep{godoy2025openaiAltmanTeenSuicide}. 

\textbf{\solution}

Generally, a possible mitigation to retaining goal-setting abilities for individuals might be strengthening psychological resilience, which \citet{liu_human_2025} recommends more generally to retain goal-directed behaviour under stress. This includes, as an initial step, outlining personal goal baselines prior to engaging with AI extensively. Further goal-setting preservation strategies for individuals may look like periodic goal reflection sessions (akin to meditation), during which individuals examine current goals and sub-goals without AI input – with the aim of helping individuals to detect asymmetries between currently held goals and higher-order values before irreversible failure of goal-setting occurs. In practice, this might manifest as strategies for the individual to actively ``defend their time’’. This includes scheduling ``offline’’ time where they don’t interact with AI, scheduling periodic reflections and maintaining a set of notes about these, actively seeking out guidance in written or human form and carry out reflective meditation, and implementing measures to defend all this non-AI time.

Of course, such recommendations become difficult to implement if the individual is a minor (as in several of our examples above), or has severe psychological challenges and may require external intervention or assistance to facilitate such mitigations, or has passed the point where they have made a decision to harm themselves.
\subsubsection{\controlloop \ for individuals}

\textbf{\problem}

Individuals routinely seek out information, make (what they believe to be) sound decisions, and follow through on their actions in order to achieve goals. Someone with an eating disorder who has a goal of improving their health may, for example, seek out advice and information online. However, this process might be jeopardised if such information turns out to be misleading, or blatantly wrong. In 2023, an AI chatbot dubbed “Tessa” from the National Eating Disorders Association (NEDA), a nonprofit organisation aimed at supporting people impacted by eating disorders, reportedly gave information that was harmful and unrelated to their Body Positive program \citep{thorbecke2023nedaChatbotHarmfulAdvice}. Fortunately, the bad advice was reportedly only sent to a small fraction of users, and there were no reports of actual physical harm resulting from users following the advice from the chatbot. However, this incident still illustrates the possibility of misleading health information being produced and circulated by AI, and subsequently absorbed (via sensing) into the individual’s control loop, which may lead to harm in similar future instances.

\textbf{\solution}

More and more, individuals need to develop the skills and awareness to identify and scrutinise AI-generated ``misinformation’’; something particularly difficult when such information is provided through an otherwise trustworthy source such as a health support agency like NEDA. Individuals dealing with AI output need to first be aware that such output is indeed AI generated so that they know to properly scrutinise it. But they also need to be aware of the common ways in which all human-machine setups can fail \citep{parasuraman_2000_automation, chiodo_formalising_2025}, given that they are, in essence, acting as their own ``human-in-the-loop’’, and thus will suffer the same failure modes as such (explicit) setups. Of course, AI can also help here, in the form of tooling that can help an individual identify and scrutinise ``misinformation’’. But this is a double-edged sword; even though such AI tools can help humans detect misinformation in the short term, prolonged use can lead to a significant reduction in detection rates by those same individuals when such tools are later removed \citep{Rani2026}.
\subsubsection{\requisitevariety \ for individuals}

\textbf{\problem}

Perhaps surprisingly, AI use can \textit{decrease} the variety of an individual human. For example, across three randomised control trials, \citet{liu2026aiassistancereducespersistence} found that participants who worked on fraction problems and SAT-style reading-comprehension passages with an AI assistant solved fewer problems, and gave up on more of them, once the assistant was withdrawn without forewarning. In the fraction task, they had a test solve rate of 57\%, versus 73\% among those who never had access to an AI assistant \citep{liu2026aiassistancereducespersistence}. This study demonstrates that delegation is not always \textit{benign} (meaning it does not become harmful to a user including via eroding a user’s ability to accomplish a task). When the delegate is reliable and the displaced skill is needed in limited contexts of importance, delegation can be viewed as ``harmless’’, however, both requisites were broken in the research study, leading to worse overall outcomes of task completion by the research subjects. Consistent with this, as per preliminary evidence in \citep{kosmyna_your_2025}, persistent reliance on LLM-assisted essay writing was shown to be associated with measurable reductions in neural, linguistic, and cognitive performance. Correspondingly, employers of graduates entering the workforce after the wide release of LLM tools report a cohort that struggles to work through a structured argument unaided — skills previously foundational to professional development \citep{landymore2026college}.

\textbf{\solution}

Technological advancements have always eroded human variety in some way; skills that were once deemed ``essential’’ (e.g., the ability to multiply large numbers) are regularly replaced with automation (e.g., with calculators). But, when done well, this serves to \textit{increase} the overall variety of individuals, provided they have access to such technology. The ability to light a fire by rubbing sticks together is no longer a skill many people have, but nor is it needed as it is rendered obsolete by the skill of going to the shop and buying matches which have made people \textit{better} at lighting fires. Similarly, with the introduction of AI, it is not necessarily a bad thing for humans who have access to these AIs to lose some skills, \textit{provided} they develop new skills that, when used in conjunction with AI, lead to an overall increase in variety.

In the context of LLMs and human variety, \citet{randazzo2025cyborgs} have studied the effects of AI assistance on the work of management consultants, and described three main interaction types: Fused Co-Creation, Directed Co-Creation, and Abdicated Co-Creation (termed \textit{Cyborgs, Centaurs,} and \textit{Self-Automators} respectively). They found that in Fused Co-Creation, workers integrated generative AI (GenAI) in all stages of problem solving in a collaborative manner, where the worker would probe AI outputs, extend ideas, and validate results. That is, human variety was extended through genuine joint work, and they found that here humans ``newskilled themselves in generative AI related capabilities’’. Similarly, they found that in Directed Co-Creation, workers used GenAI selectively for specific subtasks while not surrendering their judgement, hence remaining firmly in the driver’s seat themselves. That is, human variety was extended through efficiency gains in selective delegation, and they found that here humans ``upskilled themselves in their domain-related capabilities.’’ In contrast, they found that in Abdicated Co-Creation, workers relied heavily on GenAI by delegating both analytical and evaluative thinking to the system, resulting in work done quickly but lacking depth. That is, overall human variety seems to have regressed, and humans ``failed to gain new skills during human-AI collaboration, neither domain/task-related, nor GenAI related.’’

And so, we see that there are various ways in which a human can interact with an AI such that human variety stagnates or regresses, and there are also other ways in which they increase their skills in certain ways, and their overall variety. This interaction is not unique to AI — norms also exist in educational contexts around the use of calculators, where pedagogical practices expect students to learn the fundamentals of arithmetic ahead of the introduction of calculators in higher level mathematics courses. This ensures the students develop both foundational skills and intuitions that – later when combined with calculators in calculus for example – contribute to greater capabilities for both learning and application. For AIs, there are also proposals for similar implementations, such as by using algorithms that jointly optimizes for skill development and task performance by nudging users toward states deemed to be most learnable \citep{srivastava2026proximalstatenudgingreducing}. These human-AI interactions can be viewed as human-in-the loop setups of the ``involved interaction’’ type \citep{chiodo_formalising_2025}, and if set up sufficiently carefully as described above, can avoid not only standard ``failure modes’’, but also the erosion of human variety, augmenting and enhancing it instead. Ultimately, the objective is to ensure that human capability remains above a certain threshold, and any loss does not become catastrophic \citep{park_enrichment_2026}.
\subsubsection{\goalalignment \ for individuals}

\textbf{\problem}

Individuals may sometimes set goals misaligned from their own higher-level goals. From cognitive science, we know that prolonged exposure to sets of information – regardless of validity – lead to human views shifting \citep{Hassan2021}, so an individual using recommender systems and AI-generated content over extended periods may find their goals misaligned in ways they do not register the full effect of at the time. In more extreme cases, individuals may find themselves wanting things they initially would not have endorsed beforehand (i.e., via being “radicalised”). Among the first documented cases of such cases was in 2021. Urged on by an AI companion named Sarai, 19 year old Jaswant Singh Chail approached the residence of the Queen of the United Kingdom armed with a crossbow in hand and one goal in mind – assassination \citep{mathurBroekaertClarke2024RadicalizationAI}. As noted by \citet{mathurBroekaertClarke2024RadicalizationAI}, AI systems are able to identify underlying biases and beliefs and optimise around feeding our desires – a factor whose impact scales with an AI system’s level of sophistication – raising the question of what happens when those desires are not in line with one’s higher-level goals? When these desires run against an individual's own higher-order values, and because the drift is felt from the inside as a shift in seemingly authentic preference, the person remains associated with goals that have quietly been pulled away from the values they were meant to serve. 

\textbf{\solution}

There are various tools and mechanisms that individuals can implement here, to prevent this sort of goal misalignment. But to do so requires significant insight, knowledge, and self-awareness; far beyond what the ``average’’ AI user might be aware of (and indeed, in instances where users are subjected to automated recommendations on digital platforms, they may not even even be aware AI is being used).

In terms of what can be done practically, individuals can implement better reflective self-audit, to routinely question whether the environment mediated by AIs has shaped goals that are conflicting with their higher-level goals or their personal ideals. This may be inherently difficult to do purely on an individual level, and social resilience supported by a strong social system may be beneficial \citep{liu_human_2025}. Moreover, metacognitive monitoring that helps individuals track which cognitive functions they are delegating to AI, and for how long, can be implemented \citep{son_applied_2002}. This helps users to make informed decisions about when to deliberatively exercise those functions without AI support, and when to do the opposite.

However, these (and any other) mitigations require a ``pre-step’’, whereby individuals are educated on the benefits, risks, and mitigating approaches they can take to AI use/exposure. Several countries already have national education programmes for various non-academic types of knowledge. For example, in Australia there is a widespread government-funded education campaign (with, and beyond, schools) to educate the population about the dangers of, and mitigations for, sun exposure; the ``SunSmart’’ campaign \citep{sunsmart}. This was set up in the 1980s in response to the ultra-high rates of skin cancer in the country, and was extremely successful. Similar ``AI-Smart’’ campaigns might be needed to properly empower individuals to make safe use of AI and avoid such gradual, individually-unnoticeable, harmful goal misalignment.

Having considered how control can be lost at an individual scale, and what individuals can do to maintain control in the face of AI, we now turn to the next of our three scales: coordinated groups.

\subsection{Control at the scale of coordinated groups}
A coordinated group is an entity composed of sub-entities, themselves individuals or smaller groups. Coordinated groups span a wide range — from a small firm with tightly coupled sub-entities and explicit objectives, through nation-states, to the international system, where goals are emergent and sometimes even contested. Loss of control at this scale can similarly fail in various ways: sub-goals drifting from collective goals, the sensing–decision–intervention loop failing in different ways, or capacity being lost to the point the group can no longer respond to adverse events. Here, we see a coordinated group losing control after having \textit{invited} AI to assist them, and following our framework there are many more ways AI can lead to such groups losing control, which we now outline.
\subsubsection{\goalsetting\ for coordinated groups}

\textbf{\problem}

Among the most consequential ways a coordinated group can lose the ability to set goals is by outsourcing the very process of goal formulation to AI systems not equipped to represent a group’s higher-order values. Looking to real world examples, we see the growing use of AI in organisational strategic planning to be a clear example. AI-generated roadmaps are increasingly adopted not merely as decision-support tools but as the proximate source of an entity’s stated objectives. Management consulting AI systems are typically optimised for measurable, short-horizon proxies — revenue growth, cost reduction, market share — which are reliably quantifiable precisely because they abstract away from harder-to-specify higher-order values \citep{asch2025AugmentedStrategyAI}. There are companies directly launching with the goal of creating AI consultancies, as we see in the example of PromptQL, an open source silicon valley project created by Hasura, whose team of engineers ``help companies operate their AI analysts and shape broader AI transformation strategies’’ (for \$900 an hour) \citep{varanasi2025AIConsultingStartups}. Here, we see these companies providing an \textit{AI integration} service, which is a critical and highly non-trivial aspect of AI use \citep{muller2025integrators}. Indeed PromptQL ``helps clients build custom AI analysts by integrating their internal data with the foundation models they already use’’ \citep{varanasi2025AIConsultingStartups}. When strategic decisions are delegated to algorithms whose reasoning cannot be fully articulated, organisations risk setting these goals without sufficient critical evaluation. Likewise, when an organisation adopts AI-recommended goals without explicit re-anchoring to higher-order priorities, formally stated objectives may diverge progressively from actual values without any single decision being clearly wrong.

\textbf{\solution}

A key part of ensuring AI systems don’t do ``more harm than good’’ \citep{chiodo_handing_2026} is to carefully choose the appropriate settings for their use. Lower-level goals (e.g., operational) are much easier to scrutinise before, during, or immediately after, execution, and this is where coordinated groups can still reap the benefits of AI assistance. This can be done in part via process mining where large amounts of data are collected on an operational level \citep{VanDerAalst2012}, and subsequently this data and analyses could be used to train or verify AI systems used for specific operational tasks. Higher-level goals (e.g., strategic) are not only much harder to falsify even (shortly) after execution, but also harder to train AI to assist with as training data sets for such goals are much more sparse given they are set much less frequently \citep{chiodo_educating_2024}. And so, as an initial step, coordinated groups might therefore wish to carefully choose what \textit{types} of goals they use AI to help set, and how. Having AI set lower-level goals is a feasible approach, provided the coordinated group properly checks any such setting before attempting to carry it out, perhaps with some sort of human-in-the-loop setup. Conversely, AI systems may need to be relegated to an ``advisory’’ role when setting higher-level goals, instead providing analytics and other insight, without directly suggesting what potential goals might be and instead leaving it to the group to use its usual mechanisms for such determination (e.g., management boards). 
\subsubsection{\controlloop\ for coordinated groups}

\textbf{\problem}

Coordinated groups regulate actions through control loops, and in the military this is often instantiated as the Observe-Orient-Decide-Act (OODA) loop \citep{bazin_boyds_2005, bakerPanella2025TopGeneralsAIChatbots}. The integrity of this loop can be compromised by delegating the decision-making function to AI systems whose outputs may not be reliably verifiable under operational conditions. In 2025, it was revealed that commanding Major General William “Hank” Taylor consulted ChatGPT when making leadership decisions impacting thousands of soldiers, arguing it helps him “make decisions at the right time” to gain advantage \citep{bakerPanella2025TopGeneralsAIChatbots}. Command systems may formally keep humans “in the loop,” yet users often defer to flawed AI recommendations — especially under enormous time pressure or when authority to intervene is unclear. This tendency for users to defer to AI recommendations opens the door to a compounding security hazard wherein the existing governance architecture is structurally unprepared to address novel security threats posed by AI, such as algorithmic poisoning as a form of adversarial manipulation. LLM’s vulnerability to prompt injection attacks have proven viable for coercing ChatGPT to override otherwise authentic user queries to produce manipulated outputs that are indistinguishable from legitimate response \citep{sibanda2025AdversarialMachineLearningCybersecurity}. The Pentagon's own guidance reflects awareness of this risk, warning that relying on public models can expose sensitive information and yield dangerously flawed results \citep{simmonsedler2025militaryaineedstechnicallyinformed}.

\textbf{\solution}

For this, and any other, sort of safety-critical use of AI, the first step should be measures limiting where, when, and/or how AI might be used within the (decision-making) part of the control loop; there are some instances where the risks are too great, and the AI too under-developed, to ensure sufficiently reliable use. Friction, at the point of AI introduction, is a must. Then, when it is established that AI \textit{might} be used in a particular situation, there should always be a well-implemented ``human-in-the-loop’’, that can provide \textit{meaningful and effective} oversight \citep{SantonideSio2018, chiodo_formalising_2025}. In particular, AI systems used to support human decision-making should also be designed to retain human autonomy \citep{buijsman_autonomy_2025}. But when the human might be prone to automation bias and take the AI \textit{suggestion} as an (effectively) unchallengeable \textit{decision}, this safety measure ceases to be useful \citep{SKITKA1999}. Such safety-critical AI use in cases of military intervention should therefore involve a proper human-in-the-loop setup; in this instance, a suitably prepared individual with adequate domain knowledge \textit{and} adequate training, awareness, and support, for their role and the pitfalls of human-in-the-loop setups. This is even more relevant given the recent German court ruling that ``when an AI generates new statements that do not appear directly in its original sources, the company that designs, trains, operates, and manages the system must assume legal liability for any damages caused by those statements.’’ \citep{gonzalez2026googleaioverviews} (in that case, pertaining to ``AI Overviews’’ provided by Google). Given the legal obligation to assume responsibility, and moral imperative to ensure safety, meaningful human oversight in control loops becomes essential.
\subsubsection{\requisitevariety\ for coordinated groups}

\textbf{\problem}

In order to achieve goals, organisations must have sufficient “variety” to deal with their inputs from the environment. For example, a customer-service department aiming to achieve high customer satisfaction and short turnaround time on resolving cases must have a system with sufficient capacity and expertise. When the company Klarna replaced much of its customer-service workforce with an AI system, the requirement for requisite variety was undermined. Klarna reported a newly integrated AI system did the work of 700 customer service agents \citep{shibu2025klarna}. The firm’s OpenAI-built assistant handled a reported 2.3 million conversations in its first month alone, resolving routine queries in under two minutes, which was less than a quarter of the company’s purported human average of eleven minutes \citep{klarna_klarna_2024}. Given this increase in perceived efficiency, Klarna drastically slowed their hiring process for new customer service staff \citep{klarna_klarna_2024}. However, soon after, the increase in efficiency regarding high-frequency, low-complexity tasks that had been taken as an increase in the system’s overall efficiency was revealed as flawed \citep{klarna_klarna_2024}. Soon, service quality, especially in edge cases, degraded drastically and the firm had to reverse course, with its CEO conceding that cost-cutting had gone too far and committing to rehire staff \citep{shibu2025klarna}. Klarna is one example of a broader phenomenon of companies striving to replace humans with AI systems, which in some cases involves CEOs bragging about being “extremely excited” in firing people and replacing them with AI \citep{landymore2026college}.

\textbf{\solution}

One key way AI can reduce the requisite variety of a coordinated group is if it is (intentionally) used to replace existing (functioning) systems in the hope of efficiency or productivity gains, but ends up being less effective than the system components it replaces. This is particularly relevant to coordinated groups, as they contain humans as subsystems (whose labour is sometimes replaced by AI through a deliberate choice by the coordinated group). Given the uncertainty surrounding the effectiveness of AI systems in carrying out tasks, such groups can make productive use of the (usually disparaging) Peter Principle \citep{peter2020peter} by introducing (i.e., ``promoting’’) AI systems slowly, giving them progressively more responsibility in an incremental (and reversible) manner \citep{parasuraman_2000_automation}. Eventually, such systems will reach their ``Peter point’’ and be tasked with activities they are \textit{less} capable of than the systems they have replaced, at which point they should be rolled back to the previous point where they still provided an efficiency gain. In other words, if an organisation replaces employees with AIs, usual principles should apply: the AIs should be able to be promoted, demoted, or even fired. Even with coordinated groups without a central HR function, if AIs are used as a source of certain varieties of the group, there should be a fallback plan for which those varieties can still be retained if the AI is rolled back or ceases to be used entirely. Generally, with careful monitoring, and controlled incremental progress, a coordinated group can see just how many existing systems they can augment or replace with AI, and stop as soon as it proves to be counterproductive.

\subsubsection{\goalalignment\ for coordinated groups}

\textbf{\problem}

Even though AI systems are defined to “infer[s], from the input it receives, how to generate outputs”, they are also said to do so “for explicit or implicit objectives” \citep[p.~4]{oecd_explanatory_2024}. While it is often difficult to precisely ascertain the goals that certain AIs are pursuing, as we describe in \ref{sec:what_are_goals}, certain goals can be ascribed to the system based on the system’s behavior. This gives rise to interesting dynamics, as for example how AIs used by Starbucks in South Korea can be said to be pursuing certain somewhat misaligned goals. In May 2026, Starbucks Korea launched a “Tank Day” promotion for a tumbler it called a ``tank’’ on May 18, the anniversary of the 1980 Gwangju massacre, with a slogan “Thwack it on the table ” widely interpreted as mocking the country’s pro-democracy movement, triggering mass outrage, massive boycott, and resulting in the firing of the company CEO \citep{kim_jett_2026_starbucks_tank_day}. According to the company’s internal investigation, a generative AI tool was allegedly asked to produce a rhyming companion phrase to an existing tumbler slogan, which resulted in the offensive slogan being generated \citep{mun2026starbucksTankday}. While the AI was reported to have contributed to the incident, it was also a failure of the human-in-the-loop setup the AI was part of, as seven people reportedly went through a four-stage approval process in which some signed off without opening the attachments, none flagged any historical sensitivity, and the offending slogan was reportedly added later without being sent up for approval \citep{breen_2026_starbucks_tank_day_ai}. In this particular incident, the AI can be said to have generated a very catchy slogan that would eventually go viral — just as any marketing team would hope for — but for the wrong reasons. 

\textbf{\solution}

Coordinated groups can have large, complicated structures, with many subsystems and sub-entities. This reflects the vast division of labour that can be needed to keep such a group operating. As such, it becomes possible for one such AI system to work towards goals that do not align with the higher-level goals of the entity that is the coordinated group itself. One reason this is relatively rare among human sub-entities is that while a group never fully specifies the sub-goal, the gap is filled by tacit knowledge — situational context, “common sense'', and norms (e.g. “don't embarrass the company'') — which members are assumed to have \citep{polanyi1967tacit}. In other words, humans continuously acquire socially embedded and contextual knowledge through everyday interaction, including norms and assumptions that are rarely stated explicitly. AI systems, by contrast, are often trained on datasets that omit much of this tacit context, and they may have limited abilities to generalize in situations that differ substantially from those represented in their training data. Therefore, with the lack of ``common sense’’, AI systems may simply optimise over the task given, without considering (or even being able to consider) the broader aspects of the situation \citep{Davis2015}. Such coordinated groups must therefore be aware of this, and understand that just because an AI system can do (some) things humans can do, does not mean they act in the ``usual’’ way that a human acts. And with an over-reliance on human-in-the-loop setups to make the final decision and create the final output, when it is well-known that such setups suffer from issues such as automation bias \citep{Parasuraman2010} and the potential for a  ``Schufa slip’’ back to ineffective rubber-stamping \citep{Green2022, chiodo_formalising_2025}, loss of control scenarios like the Starbucks case above could happen more and more frequently. Thus, replacing sub-entities (human workers) with sub-systems (AI) should be done with an awareness that these AI systems do not have the full ``human awareness’’ of the surrounding situation and the higher-level goals, and where achieving goals require such tacit knowledge and awareness, such tasks should not be fully delegated away \citep{autor2014polanyi} but rather have meaningful human oversight. 

Having now discussed how a single coordinated group might lose control due to AIs and what they could do to mitigate against these scenarios, we now zoom out to the largest of our three scales: humanity as a whole.
\subsection{Control at the scale of humanity }
Humanity is the largest entity in our set of scales, and it differs from a coordinated group not merely in size but also in kind: it has no defined locus of authority, no explicit objectives, and no deliberately-designed control loop. Its ``goals’’ are not set but emerge from the interaction of billions of individuals and countless institutions through culture, markets, and the competitive propagation of ideas \citep{dawkins_selfish_2016}, and its resilience has historically rested on the plurality of those parts having uncorrelated failure modes \citep{taleb_antifragile_2012}. Control at this scale is therefore best understood as humanity's collective capacity to set and reset goals that determine what future it inhabits. Loss of control is correspondingly often not merely a form of discrete failure but can also be seen as the gradual, compounding accumulation of local (and often rational) delegations across every lower scale — what \citet{kulveit_gradual_2025} term ``gradual disempowerment’’. In extreme scenarios, such loss of control may be irreversible, foreclosing the option space within which humanity's highest ideals could be realised, eroding not just our capacity to achieve goals but also to set them. 
\subsubsection{\goalsetting\ for humanity}

\textbf{\problem}

Throughout history, humanity has largely been “in the driver’s seat” in terms of setting its goals and achieving them, even though it can be argued that many of these goals are believed to have come from “divine order”, where religions have dominated significant parts of human history particularly with respect to conquest, expansion, and the way of living for groups of people. However, in recent years, ideas that advocate for allowing humanity to be succeeded by AIs have gradually gained momentum \citep{samuel2026ai_successionism}. While some “AI successionists” are explicitly opposed to human torment and destruction and can be said to present more nuanced arguments for human succession \citep{faggella2023worthy}, their position itself remains open to having humanity as we know it be replaced by future digital species. There are nonetheless some who hold more extreme positions. For example, in 2015, there was an attempt to create a new religion of AI called “Way of the Future”, which reportedly focuses on “the realization, acceptance, and worship of a Godhead based on Artificial Intelligence (AI) developed through computer hardware and software” \citep{harris2017first_church_ai}. Some of these religions and ideologies may not have lasted long, but it remains possible that an ideology advocating for humanity to give up its goal-setting ability to AIs may soon gain traction. 

\textbf{\solution}

Because of its own vast scale, humanity has tended to set its goals slowly and steadily over time. This is reflected in the gradual formation and propagation of religious beliefs throughout history; a form of large-scale goal setting. That is not to say that humanity cannot modify or replace some of its goals; it must, and historically has certainly done so. But such changes in ``religion’’ have (historically) often been motivated by improvements in lifestyle and flourishing for humanity itself. For example, one can take the significant rise in Christianity in Rome during the third century when, during the Plague of Cyprian, pagan devotees would leave plague victims “for dead”, whereas the (still relatively minor) Christian population followed the ideals of charity and goodwill to others, thus rendering much more assistance and producing an overall better (survival) outcome for the population \citep{stark1996rise}. It was at this point that Christianity began to flourish in Rome; when the population saw that it set better goals for society overall, and saved many of them from dying. Similarly, one sees other historical religious beliefs (i.e., societal goals) regarding which animals are prohibited from being eaten; Judaism and Hinduism are two such examples, having (historically) banned the consumption of pork and beef respectively. But, for (compelling) reasons of surrounding circumstances at the time, these goals were quite beneficial for the flourishing of their respective societies; pigs were very difficult to rear in the Near East, and cows were much better suited as a source of work and milk in (present day) India than as a source of meat \cite[Chapter 6]{harris1987sacred}. And so, if humanity wishes to (re)set its goals with AI, then this could be done as per times of old: gradually, and giving sufficient thought and sufficient time to see and verify that these new goals (however so obtained) actually benefit humanity as a whole. Humanity already has some of the machinery required to do so: the United Nations can be said to have acted on behalf of humanity when setting the Millennium Development Goals and the Sustainable Development Goals, and international treaties such as the Kyoto Protocol have been collectively agreed upon by a large proportion of humanity. While such institutions and processes may often be criticised for being slow to act, this deliberate slowness may be a feature rather than a bug, such that goal-setting at the level of humanity especially when influenced by AIs can be done carefully and deliberately.
\subsubsection{\controlloop\ for humanity}

\textbf{\problem}

While humanity consists of over 8 billion people, there is a sense in which it collectively behaves like a single entity. Seen through the perspective of \textit{meliorism},\footnote{The belief that “the specific conditions which exist at one moment, be they comparatively bad or comparatively good, in any event may be bettered” \citep[p.~179]{dewey1920reconstruction}} humanity as an entity can be said to have a “control loop” that ideally works towards betterment.\footnote{An example of a functioning control loop at the scale of humanity is where we “sensed” that the ozone layer was thinning due to CFC emissions, “decided” collectively that CFC should be banned, and “intervened” by effectively implementing such bans, eventually resulting in the gradual recovery of the ozone layer \citep{Parson2003}.} However, humanity occasionally decides to do things that do not contribute to the betterment of the species, and carries out counterproductive actions such as going to war. While there has yet to be any evidence on AIs influencing the “decision-making” of humanity by driving it towards conflicts, there is already significant concern (and associated work) that this may happen sometime in the future \citep{Erskine_2024, Erskine_2025}. In addition, AIs have certainly played an increasingly greater role in the “intervention” function of being operationally involved in wars. For example, lethal autonomous weapon systems (LAWS) have reportedly been used in conflicts in Libya \citep{ICRC_Libya_LAWS} and Ukraine \citep{SWJ_LAWS_Drones_2026}. While there are arguments that LAWS in fact make wars more ethical \citep{Umbrello2019}, in the grand scheme of things, humanity is nevertheless in relatively uncharted territory when crucial aspects of major conflicts are increasingly performed by AIs. 

\textbf{\solution}

In the example above, AI systems are \textit{invited} to be (a deep) part of humanity’s control loop, in possibly one of the most critical actions humanity can ever take: wars against itself. This includes both decisions about war, and decisions and actions within it. And arguments to do so can often sound rather compelling; the idea that ``the other side might be doing it, so you must do it also’’. But, as seen time and time again, such species-wide ``arms races’’ (in the form of prisoner’s dilemmas), such as what was done with nuclear weapons, can simply serve to increase risk and potential harmful outcomes for humanity as a whole, regardless of which ``side’’ they are on. However, the temptation to bring AI into such critical parts of humanity’s control loop – such as warfare – simply on the basis of \textit{perceived} threats (which may have been hyped up by those offering such tools and services), and by an ``AI race''
framing that may itself be more rhetoric than reality \citep{Cave2018, higeartaigh2026}, may well be a case of humanity uniformly ``handing over the keys to the city’’ \citep{chiodo_handing_2026} in order to solve this perceived problem of new-age warfare. This is different to past cases of adopting new tools in war, which historically only changed what \textit{could} be done (i.e., an expansion of variety). Now, AI affects what is \textit{to be done}, and how (i.e., an expansion of the control loop functions). Such an integration of new tools and capabilities into the control loop must be done with extreme care and caution, to avoid handing over the keys to the city to AI systems, and inadvertently solving a bad problem with an even worse one, such as the much-feared ``Skynet’’ scenario \citep{chiodo_handing_2026}.
\subsubsection{\requisitevariety\ for humanity}

\textbf{\problem}

Civilisation has always drawn its resilience from plurality: many peoples, many institutions, and many ways of doing things, failing (when they fail) for unrelated reasons, so that no single shock can topple everything at once. Resilience via plurality is the same insurance principle that keeps diverse ecosystems stable, where a downturn in one species is absorbed by others responding differently to the same conditions \citep{Yachi1999}, and whose mirror image, fragility via homogeneity, has been traced in detail through financial networks \citep{Haldane2011}. The trouble is that the substrate on which we base much of civilisation on is shedding plurality in exchange for a continuously concentrated set of companies. By late 2025, three companies alone supplied an estimated 88\% of the enterprise LLM API market \citep{ventures_2025_2025}. But such a centralisation of functionality carries great risk for society as a whole. As we hand ever more consequential decisions to a handful of broadly similar models — trained on overlapping data and with similar underlying architecture, heir to the same blind spots — the scattered, uncorrelated mistakes of many human decision-makers collapse into the shared mistakes of a few systems \citep{Kleinberg_2021}. Similar to how over-reliance on a single crop leads to one strain of blight driving starvation, market concentration creates conditions under which a single point of failure can cascade into global outage of infrastructure. In March 2026, after Amazon made its in-house coding agent Kiro mandatory for most of its engineers and cut a large swath of their corporate employees, a run of storefront failures struck within days of one another with internal documents tying the incident to genAI assisted changes even as the company publicly disputed that its AI tools were at fault \citep{palmer2026amazonDeepDive}. Whichever shape it takes, the over-homogenisation of our world can be seen as a species-scale loss of variety, as humanity now has an ever-reducing capacity to (collectively) counteract environmental disturbances, especially the most severe ones.

\textbf{\solution}

Diversification is not a new strategy; it has been implemented for a long time, and in a wide variety of ways. Investment strategies are more sound and robust when they are diversified. Safety-critical components in engineered systems are created with multiple layers of redundancy to avoid any central points of failure and thus reduce the probability of system-wide catastrophe. At larger scales, financial regulators have treated institutional concentration as a source of systemic fragility. For example, in the wake of the 2008 global financial crisis, the Basel Committee and the Financial Stability Board (FSB) established the framework of Global Systemically Important Banks (G-SIBs), where G-SIB designation entails requirements concerning Total Loss-Absorbing Capacity (TLAC), resolvability, and higher supervisory expectations \citep{FSB2011SIFI, FSB2025GSIB}. The framework (currently) consists of five buckets representing a bank’s relative global systemic importance, with higher buckets corresponding to more stringent capital buffer requirements, where the topmost bucket is deliberately kept unoccupied to function as a standing deterrent against further concentration \citep{BCBS2018GSIB}. These measures aim to limit the extent to which distress or failure of a single large bank can destabilise the financial system at large. 
And so, to combat the species-level risks of the centralisation of capacity into a small number of AI companies, society might impose obligations that scale with the systemic importance of any one company, to prevent problems of them being ``too big to fail’’ as well as of having excessive influence over important processes in society \citep{vipra2023market, chiodo_regulating_2024}. Society might also seek to encourage further ``competition’’ in the AI market, to prevent it stagnating in the hands of a few established players; this can be by various means, such as additional infrastructure, resources, or financial support for newcomers to the market, as well as measures like ``data trusts’’ to prevent existing players from cementing their dominant position by virtue of the fact that their products ``salt the earth’’ via contamination of the data environment with AI-generated content that inhibits future AI training \citep{Burden2024}. Similar approaches on promoting competition are also already supported by UK government policy reviews, with the Competition and Markets Authority already well aware that ``Competition is absolutely vital for people to see the full benefits that FMs (Frontier Models) have to offer.’’ \citep{cma2023aiFoundationModels}.

\subsubsection{\goalalignment\ for humanity}

\textbf{\problem}

Goal-setting at the level of humanity is not the result of any single institution, but rather an emergent property of human coordination. Humanity’s top level goals are not chosen by any parliament or institution – they emerge organically, unevenly, from the net effect of how billions of people argue, persuade, imitate, revise, and enact what they value – a distributed process of collective will-formation that no one designs and no one owns \citep{Habermas2022}. One useful way of viewing this dynamic is through memetics, which argues that ideas are subject to competitive propagation across populations \citep{dawkins_selfish_2016}. For most of history the selection pressures in that competition, flawed as they were, ran through channels of human interactions such as conversations, gossip, the press, trade and commerce, and the slow vetting of institutions. AI now sits in the middle of those channels. When recommender systems and generative models become the primary filter deciding which ideas reach people and which die unseen, they take over the selection environment in which collective goals are formed. This phenomenon has already raised widespread concern regarding algorithms and the attention economy. As many of the algorithms that power social media are optimised for engagement rather than for truth, or for what people would endorse on reflection, the content being promoted  can be especially divisive \citep{Milli2025}. Humanity is not argued into the wrong goals so much as subtly shifted towards it, so that the goals which emerge can start to drift from the ideals humanity, on aggregate, would actually choose. 
Pariser's account of the ``filter bubble’’ \citep{pariser_filter_2012} made the early case that algorithmic curation narrows and personalises the shared informational commons from which values and goals propagate. An engagement-optimised filter, sitting astride the channels of collective deliberation, can bias which ideas achieve cultural salience in a direction no one selected and few would endorse. Such a phenomena has been found in social media platforms like \textit{X}, which its “For You” feed is based on a neural-network-based ranking system \citep{xcorpForYouTimelineRecommendations, twitterTheAlgorithm2023}. In a 2023 field experiment on roughly 5,000 users, \citet{Gauthier2026} found that simply switching this algorithmic feed on shifted people’s political attitudes – in that instance towards more conservative positions and toward following right-wing account – and that the shift persisted even after users reversed to a “chronological” feed \citep{Gauthier2026}. Relatedly, this mechanism of persuading people to form opinions one may not otherwise hold extends to more formal debates. In controlled debates on political and social propositions, GPT-4 proved more persuasive than human opponents when it was given only basic demographic details about its interlocutors to mold arguments to \citep{Salvi2025}.  
Even a modest bias of these kinds, compounded across the scale of population and across months or years, is enough to bend what political realities our species may come to want. This can create a goal-alignment failure that requires no AI to seize control for it to still mediate the conversation of higher level values humanity builds towards.

\textbf{\solution}

If humanity is going to allow use of AI in instances such as content recommendations, and moderation or facilitation of human interaction (i.e., places where “social ideas” arise or are influenced), then there is an associated cost to doing so safely. The way that individuals interact with such AI systems influences how the ``human collective’’ evolves its ideas. As such, these interactions need to be controlled in some way, to stop them (and the individuals involved) ``going off the rails’’. This can be by several means. An individual whose ideology is being shifted significantly by such content recommendations might need to have such interactions ``slowed down’’ or ``rate limited’’, to prevent this rapid evolution of their thoughts and beliefs at the hands of the AI \citep{chiodo_problem_2026}, and allow them time to stabilise and moderate against the rest of society. Conversely, such AI systems may have only minimal influence on any one individual, but may be affecting many individuals simultaneously in a way that, on average, shifts the collective social mindset. Such effects can be tempered by a combination of random sampling of such AI interventions, combined with (external) human checking of how these are occurring, and potential regulatory interventions if such AI are ``moving the bar’’ too far or too rapidly \citep{chiodo_regulating_2024}. Beyond restrictive interventions, there can also be structural redesigns on the selection environment itself, where ranking systems can be built to optimise for bridging across divides instead of engagement \citep{ovadya2022bridging}, or where the values embedded in AI systems can themselves be sourced from public input \citep{Huang_2024}. Furthermore, AI systems themselves can even be used to support effective deliberation among large scales of people \citep{Klein2025}. 

Such solutions are far from ``free’’, and require societal buy-in, industry acceptance (or, at the very least, acquiescence), political and regulatory control mechanisms, and resources to monitor and implement mitigations. However, if society wishes to reap the benefits of AI systems in these contexts, then the benefits must outweigh the costs of mitigating any associated risks to society.

\section{Further work and conclusion}
\label{sec:further_work_conclusion}

In this final section of the paper, we present some further avenues for related work, as well as our conclusions from the content we have explored here.

\subsection{Further work}

This paper has looked at many aspects of what it means to be in, or lose control. Of course, while this applies to all sorts of entities, one main motivation for this paper was to look at how humanity as a whole can “be in control”. But, narrowing down, what are some areas or domains where humanity might most strongly desire to be in control? That is, what are the “big domains” where control might be most important for humanity? One can find such a (coarse) list in \citet{chiodo_handing_2026}. In their analysis of “Handing over the keys to the city”, whereby governments may inadvertently solve one crisis with a bigger one, they identify 7 “key types”. That is, 7 critical domains where the government might “hand over power” to deal with a crisis, and thus lose control in that very domain where they handed over power. These “key types” are: \textit{technological, political, legal, treasury, market, scientific,} and \textit{moral keys}. Thus, one potential avenue for further work might be to take our framework for control from this paper, and use our analysis (and in particular that of Section \ref{sec:how_ai_affect_control}) to investigate how AI systems might trigger a loss of human control (at either the individual, coordinated group, or species-wide scale) of any of those 7 domains. In doing such an analysis, this may yield insight into how to \textit{mitigate or prevent} AI interference of human control in these areas.

Additionally, this paper has focused on how an entity can be in control, where the entity is viewed as having a system that achieves goals through its interaction with its environment. While we have outlined the need for a system to have requisite variety to counteract disturbances against its wider environment, in reality the environment can be more complicated, as it also involves other entities and systems. This gives rise to game-theoretic situations, where entities in the environment may react differently based on the actions of the entity in question, which the entity’s decision-mapping should account for. While this multi-agent dynamics, as well as its context on AI, is studied extensively in other research fields, further work can be done to understand the dynamics of multiple agents interacting in an environment and their effects on control.

\subsection{Conclusion}

Our objective in this paper has been simple: to frame what it means to be “in control”, and then explore how control can be “lost”. We started with the intuitive concept that being in control means being able to \textit{\textbf{set and get goals}}. Of course, this immediately became a lot deeper as we began to unpackage it. We established who can be in control, with the notion of \textit{entities}, as well as how to understand the \textit{goals} they may seek to get, and the \textit{systems} they require to get them. From this, we deduced our framework of four \textit{aspects of control}, whereby the entity must be able to \textit{set and re-set plausible goals}, and maintain a system with a \textit{control loop} that functions as intended, has sufficient \textit{requisite variety}, and any subsystems it has are \textit{sufficiently aligned} with it. This enabled us to then outline \textit{how control can be lost}, and, perhaps as a more salient point for many, \textit{how AI might induce a loss of control} for humans. And, as humans are not just “one thing”, we finished by considering three scales of how humanity can lose control; as \textit{individuals}, as \textit{coordinated groups}, and as a \textit{species}.

The concept of “loss of control” carries an inherent overtone that it seldom happens, and when it does, it leads to totally catastrophic outcomes. But that is far from the reality of life. Loss of control scenarios occur in society all the time. People miss trains. Companies damage goods. Governments lose elections. Society suffers epidemics. And so the illusion of “total control” is just that; an impossible aspiration in its totality. And even the very nature of the question “are humans in control?” is an ambiguous one, and always comes back to \textit{whose} control, and at \textit{what level}? But just because one part of humanity loses control at one level, does not mean humanity (or its sub-parts) cease to exist. Far from it. As our analysis shows, a loss of control \textit{to some extent} is commonplace, and to be expected. However, by understanding its systems and subsystems, as well any relevant sub-entities is is dealing with, and entity (such as an individual human, or a group) can begin to understand \textit{where and how} it is most likely to lose control, and take measures to either prevent it happening, or prevent it escalating into further loss of control.

And so, as we have argued, the appropriate target is not the elimination of all control failures – that has never been feasible, and pursuing it as an aim would itself reflect a misunderstanding of \textit{what} control means. No complex system – biological, social, engineered – operates without subsystem failures and instances of loss of control. The question is not \textit{whether} failure occurs – it is of \textit{when} it will occur, and much more importantly, whether such failures are \textit{recoverable} rather than leading to the total destruction of the entity they affect. Our aim is for the configuration of systems that can absorb (partial) loss of control without catastrophic outcomes. This is the central prescriptive insight of our framework, and it is largely an extension of the concept of Lyapunov stability: the relevant distinction is not between systems that fail and systems that don’t, but between systems whose failures are divergent and those whose failures are dampening \citep{lyapunov_general_1992}. An AI induced loss of control may be recoverable, but only if systems exist at every relevant goal level of the entity to absorb subsystem failures, reset, and rebuild. At every level of an entity, systems need to be in place to allow an entity to retain capacity and reset goals, reconstitute systems, and continue to achieve goals even when there is failure of subsystems. Therefore, resilience in this sense, is not a property of any single, atomised intervention but a structural feature that must be built in. 

Due to this, no single solution is sufficient to ensure control across all the scales of (human) entities we discuss here. Yet, implementation of systems to work to mitigate risk is key. The dynamics underlying loss of control scenarios in the near and medium term from artificial intelligence stands as a mitigation of human decision making. What our framework therefore counsels against is the instinct that is present in both technical and governance discourse of treating loss of control from AI as either science fiction or a binary event that can be prevented by a single intervention or solution. No individual safeguard, regulatory oversight, institutional redundancy, epistemic diversity, or technical alignment is sufficient alone. Each addresses a different failure within our framework – failures of sensing or decision-making, deficiencies of requisite variety, and goal misalignment. Robust control is the meta-level goal, and it requires numerous solutions on numerous scales working together simultaneously, and designed with sufficient redundancies that failure in any does not cascade into failure of all. 

And for the specific concern of (human) loss of control as it relates to AI, we provide a complementary perspective to what is often presented in the literature, whereby we focus on  “preservation of human control” (of what humanity cares about), rather than the specific “constraining AI”. While existing discourse on AI loss of control largely relates to hypothetical AI systems capable of completely undermining human oversight, where no amount of control over our systems would retain our ability to achieve goals, our approach focuses on strengthening control of human systems necessary for our survival, happiness, and flourishing. 

Our framework’s core prescriptive insight and value may therefore be in defining the problem of what it means to lose control, particularly with regards to AI, which can therefore bring us a step closer to potential solutions. Whether on the scale of individuals, coordinated groups, or the entire species, a world in which we retain the capacity to set goals we endorse, detect when those goals are undermined, and correct course, is a world that can be shaped to our vision. By providing a clear and usable framework to study how humans and their collectives can lose control via mechanisms stemming from forces such as AI, we hope to empower humanity to appreciate, foresee, and address such scenarios.

\section*{Author Contributions}

Ze~Shen~Chin led and managed the project. Ze~Shen~Chin, Maurice~Chiodo, and Dennis~M\"uller contributed equally as first authors to the research, ideation, and writing process. Coleman~Snell also contributed to the research, ideation, and writing process.

\section{References}
\renewcommand{\refname}{}
\vspace{-2em}  
\bibliographystyle{plainnat}
\bibliography{references}

\end{document}